\documentclass[twoside,twocolumn,aps,prb,superscriptaddress,longbibliography]{revtex4-2}
\usepackage[latin9]{inputenc}
\usepackage{amsmath}
\usepackage{amssymb}
\usepackage{graphicx}
\PassOptionsToPackage{version=3}{mhchem}
\usepackage{mhchem}

\makeatletter

\usepackage{float}
\usepackage{soul}
\usepackage{changes}
\usepackage{mathrsfs}
\usepackage{comment}
\usepackage{multirow}
\usepackage{xcolor}

\definecolor{kc}{rgb}{0.6,0,0.6}

\makeatother

\begin{document}
\title{\textit{Ab initio} Wannier-representation-based calculations of photocurrent in semiconductors and metals}
\setcounter{page}{1}
\date{\today}
\author{Junqing Xu}
\email{jqxu@hfut.edu.cn}
\affiliation{Department of Physics, Hefei University of Technology, Hefei, Anhui, China}
\author{Haixiao Xiao}
\email{xiaohx@hfut.edu.cn}
\affiliation{Department of Physics, Hefei University of Technology, Hefei, Anhui, China}
\begin{abstract}
We present a general \textit{ab initio} method based on Wannier functions
using the covariant derivative for simulating the photocurrent in
solids. The method is widely applicable to charge/spin DC and AC photocurrent
at any perturbation levels in both semiconductors and metals for both
linearly and circularly polarized light. This is because the method
is theoretically complete (within the relaxation time approximation),
that is to say, it includes all intraband, interband and their cross
terms. Specifically for the second-order DC photocurrent, it includes
all of the following contributions - shift current, gyration current,
(magnetic) injection current, Berry curvature dipole and other Fermi
surface contributions, instead of only a part of them as in most previous
\textit{ab initio} methods. It is also free from the degeneracy issue,
i.e., applicable to arbitrary band structures with arbitrary numbers
of degenerate bands. We apply the method to various semiconductors
and metals, including GaAs, graphene-hBN heterostructure, monolayer
WS$_{2}$, a 2D ferroelectric material - monolayer GeS, bilayer anti-ferromagnetic
MnBi$_{2}$Te$_{4}$ and topological Weyl semimetal RhSi, to simulate
their charge and/or spin, DC and/or AC photocurrent. Our theoretical
results are in agreement with previous theoretical works. Our numerical
tests of GaAs, WS$_{2}$ and GeS suggest setting the degeneracy threshold
in the conventional method as $\hbar\Gamma^{\left(2\right)}$, with
$\Gamma^{\left(2\right)}$ the relaxation rate of the off-diagonal
elements of the density matrix between two states with close energies.
We find that compared with the conventional Wannier-function-based
method using non-dgenerate perturbation theory, the numerical errors
of optical susceptibilities of bilayer anti-ferromagnetic MnBi$_{2}$Te$_{4}$
with the $\mathcal{PT}$ symmetry can be reduced by 1-2 orders of
magnitude by our method for circularly polarized light. Our method
provides a universal computational tool for reliable and accurate
predictions of abundant weak-field photocurrent phenomena in disparate
materials.
\end{abstract}
\maketitle

\section{Introduction}

The electric current generation under uniform light illumination,
known as photocurrent, has been extensively studied in opto-electronic
physics.\citep{fridkin2001bulk,dai2023recent,spanier2016power,de2017quantized,rees2020helicity,yuan2014generation,fei2021p,hong2013optical,silva2019high,chan2021giant,de2020difference}
Recently, the photocurrent phenomena in solids, such as bulk photovoltaic
effect (BPVE, also called photogalvanic effect - PGE), second- and
third-harmonic generation (SHG and THG) and sum-/difference-frequency
generation, have drawn much attention in the research fields of condensed
matter physics, opto-electronics, opto-spintronics, material science,
etc. For example, quantized circular photogalvanic effect (CPGE),
whereby circularly polarized light generates the helicity-dependent
photocurrent, was predicted\citep{de2017quantized} in Weyl semimetals
and later observed\citep{rees2020helicity}. Spin--valley-coupled
CPGE and its electric control were realized in WSe$_{2}$.\citep{yuan2014generation}
Robust pure spin photocurrent was predicted in several materials.\citep{fei2021p,xu2021pure}
Electrically and broadband tunable third-harmonic generation was realized
in graphene.\citep{soavi2018broadband}

In this work we focus on the photocurrent under weak fields, since
the weak-field condition is typically satisfied in the studies of
BPVE and low-order harmonic generation (LHG) and is preferred for
related low-power optical devices. Moreover, the weak-field photocurrent
measurements are invaluable in detecting materials' properties such
as topological and spin ones. This is because the weak-field photocurrent
results from the product of electric fields and optical susceptibilities,
and the latter are a class of materials' properties determined by
band structure, Berry connection, spin-orbit coupling, scattering
strength, etc.

Predictive \textit{ab initio} theories based on density functional
theory (DFT) emerged since the late 1990s for SHG\citep{levine1991calculation,silva2019high}
and since the early 2010s for BPVE\citep{young2012first,dai2023recent,xu2021pure}.
These theories are invaluable for the understandings of experimental
findings and the predictions of new materials with excellent properties.
The main challenge of \textit{ab initio} calculations is the need
for a large number of k-points to achieve the convergence. For instance,
up to 10$^{6}$ k-points may be necessary for topological semimetals,
which makes the calculations computationally expensive. The k-point
convergence issue becomes more serious if \textit{ab initio} sophisticated
forms of the scattering/relaxation processes beyond the simple relaxation
time approximation (RTA) are employed.

To resolve this issue, \textit{ab initio} methods using maximally
localized Wannier functions have been employed for efficient calculations
of the shift current, injection current and Berry curvature dipole
contributions to the photocurrent.\citep{ibanez2018ab,wang2017first,zhang2019switchable,le2020ab,zhang2018berry}
However, the current Wannier-function-based methods have two main
problems: (i) Some important contributions to BPVE, e.g., Fermi surface
ones and the gyration current one proposed by Watanabe and Yanase\citep{watanabe2021chiral},
are not considered and the implementations for LHG are even less complete.
Therefore, large errors may occur in the studies of metallic systems
and/or under circularly polarized light. (ii) The Wannier interpolation
of the Berry connection and its derivatives (needed in their methods)
uses the non-degenerate perturbation theory, for which the degenerate
bands are not treated properly.

The above problems can be removed by employing a technique originally
developed for theoretical simulations of high harmonic generation
(HHG) under strong fields in Ref.~\citenum{silva2019high}, where
the laser term of the electron dynamics is first expressed in the
smooth ``Wannier'' representation and then transformed to the eigenstate
representation. In ``Wannier'' representation, the basis functions
are smooth Bloch-like functions of ${\bf k}$, so that the laser term
(in the length gauge) is well defined and can be easily computed by
finite differences. Therefore, the degeneracy issue is bypassed. Moreover,
with the accurate expression of the laser term, all contributions
to the photocurrent within the RTA, including the shift current, gyration
current, (magnetic) injection current, Berry curvature dipole and
other Fermi surface contributions, can be considered.

Therefore, it is promising to apply the technique of Ref. \citenum{silva2019high}
to other photocurrent properties besides HHG. Here, we thus have developed
an \textit{ab initio} method of the photocurrent, including BPVE and
LHG, based on Wannier functions and within the RTA. The method is
applicable to both semiconductors and metals for both linearly and
circularly polarized light, corresponding to linear PGE (LPGE) and
CPGE respectively.

This article is organized as follows. In Sec. II, we derive the formulae
of optical susceptibilities via perturbative treatment of the density-matrix
(DM) master equation in the length gauge within the RTA. We then relate
the second-order optical susceptibilities to LPGE and CPGE. We further
discuss different contributions to BPVE based on the separation of
intraband and interband parts of perturbative density matrices, and
compare our method with the conventional Wannier-function-based method
using non-degenerate perturbation theory\citep{ibanez2018ab,wang2017first}
for BPVE. In Sec. III, we give the computational setups of our DFT
and photocurrent calculations. In Sec. IV, we apply our method to
various semiconductors and metals, including GaAs, graphene-hBN heterostructure,
monolayer WS$_{2}$, a 2D ferroelectric material - monolayer GeS,
bilayer anti-ferromagnetic MnBi$_{2}$Te$_{4}$ and topological Weyl
semimetal RhSi to simulate their charge and/or spin BPVE and LHG.
In Sec. V, a summary and outlooks are given.

\section{Methods}

Theoretically, the photocurrent formulae can be expressed in both
the length and velocity gauges for the laser.\citep{ventura2017gauge}
In most theoretical works, the length gauge was employed. This is
because that although the velocity gauge has simpler formulae, it
suffers from several issues: (i) Large number of bands are required
to converge the results if additional calculations are not carried
out;\citep{passos2018nonlinear,xu2021pure} (ii) The dephasing processes
and the general scattering term of the master equation beyond the
RTA are hard to be included;\citep{yue2020structure} (iii) The numerical
results may diverge at low-$\omega$ (photon frequency) limit.\citep{ventura2017gauge}

In length gauge, the current density ${\bf J}^{c}\left(t\right)$
and spin-current density ${\bf J}^{s_{\gamma}}\left(t\right)$ are
\begin{align}
{\bf J}^{c/s_{\gamma}}\left(t\right)= & V_{\mathrm{cell}}^{-1}\mathrm{Tr}\left[{\bf j}^{c/s_{\gamma}}\rho\left(t\right)\right],\\
{\bf j}^{c}= & -e{\bf v},\label{eq:jc_operator}\\
{\bf v}= & \frac{-i}{\hbar}\left[{\bf r},H^{0}\right],\label{eq:v_from_rH}\\
{\bf j}^{s_{\gamma}}= & 0.5\times\left(s_{\gamma}{\bf v}+{\bf v}s_{\gamma}\right),\label{eq:js_operator}
\end{align}

where $\rho$ is the DM operator of Bloch electrons. $V_{\mathrm{cell}}$
is the unitcell volume/area of the crystal for 3D/2D systems. ${\bf j}^{c}$
is the charge current operator. ${\bf v}$ is the velocity operator.
${\bf r}$ is the position operator, $H$ is Hamiltonian operator
and $H^{0}$ is unperturbed Hamiltonian operator. In the eigenbasis,
$H_{kab}^{0}=\epsilon_{ka}\delta_{ab}$ with $\epsilon$ the eigenvalue,
$k$ the k-point index and $a$ ($b$) the band index. $s_{\gamma}$
is the spin operator along $\gamma$ direction. ${\bf j}^{s_{\gamma}}$
is the conventional spin-current operator\citep{sun2008persistent}.

In general, the DM operator $\rho$ is expressed in the eigenbasis
representation as
\begin{align}
\rho= & \sum_{kk'ab}\rho_{ka,k'b}\left|ka\right\rangle \left\langle k'b\right|,\\
\rho_{ka,k'b}= & \left\langle ka\right|\rho\left|k'b\right\rangle .
\end{align}

In this work, since we focus on circumstances in which translational
symmetry is not broken, $\rho_{ka,k'b}$ is always diagonal in Bloch-state
wavevector. Thus, $\rho_{ka,k'b}$ is simplified as $\rho_{kab}$
and $\rho_{k}$ is used as the band matrix of $\rho$ at ${\bf k}$.

In this work, except in Sec. \ref{subsec:Wannier}, we usually omit
the subscript $k$ for the matrices at ${\bf k}$ for simplicity,
but still keep it along with band indices for matrix elements.

\subsection{DM master equation in the length gauge\label{subsec:DM-master-equation}}

We solve the quantum master equation of the single-particle $\rho\left(t\right)$
in the Schr{\"o}dinger picture as\citep{ventura2017gauge,xu2021ab,xu2023ab}
\begin{align}
\frac{d\rho\left(t\right)}{dt}= & -\frac{i}{\hbar}\left[H^{0},\rho\left(t\right)\right]+D^{E}\left[\rho\right]+C\left[\rho\right],\label{eq:master}\\
D^{E}\left[\rho\right]= & \frac{e}{\hbar}{\bf E}\left(t\right)\cdot\frac{D\rho}{D{\bf k}},\label{eq:Efield_term}
\end{align}

where $D^{E}\left[\rho\right]$ and $C\left[\rho\right]$ is the laser
and collision terms of the DM dynamics respectively. ${\bf E}\left(t\right)$
is the time-dependent electric field of a laser. $\frac{D\rho}{D{\bf k}}$
is the covariant derivative of $\rho$.

For a laser with photon frequency $\omega$,
\begin{align}
{\bf E}\left(t\right)= & {\bf E}\left(\omega\right)e^{i\omega t}+{\bf E}\left(-\omega\right)e^{-i\omega t}\label{eq:Et}
\end{align}

with ${\bf E}\left(-\omega\right)\equiv{\bf E}^{*}\left(\omega\right)$
being the constant amplitude.

The covariant derivative of an arbitrary matrix $A$ - $\frac{DA}{D{\bf k}}$
is defined as\citep{ventura2017gauge}
\begin{align}
\frac{DA}{D{\bf k}}= & \frac{dA}{d{\bf k}}-i\left[\xi,A\right],\label{eq:cov-der}\\
\xi_{kab}= & i\left\langle u_{ka}|\frac{du_{kb}}{d{\bf k}}\right\rangle ,
\end{align}

where $\frac{d}{d{\bf k}}$ is the gradient, $\xi$ is the Berry connection
and $u$ is the basis function or the periodic part of Bloch wavefunction.
Note that Eq. \ref{eq:cov-der} above is the same as Eq. 34 of Ref.
\citenum{ventura2017gauge}, but different notations are used. The
operator $\frac{D}{D{\bf k}}$ is directly related to the position
operator ${\bf r}$ as follows:
\begin{align}
\frac{DA}{D{\bf k}}= & -i\left[{\bf r},A\right].\label{eq:D-r-relation}
\end{align}

Therefore, the laser or electric-field term of the master equation
$D^{E}\left[\rho\right]$ can be expressed as
\begin{align}
D^{E}\left[\rho\right]= & -\frac{i}{\hbar}\left[H^{E},\rho\right],\\
H^{E}= & e{\bf E}\left(t\right)\cdot{\bf r}.
\end{align}

The computation of $\frac{D\rho}{D{\bf k}}$ via Eq. \ref{eq:cov-der}
is non-trivial due to the following issues: First, the basis functions
$u$ are usually obtained by diagonalizing $H^{0}$ at different ${\bf k}$
independently, so that the basis functions contain arbitrary phase
factors and are arbitrary in degenerate subspaces. Therefore, the
basis functions $u$ are in general not smooth over ${\bf k}$, which
makes $\frac{d\rho}{d{\bf k}}$ not well-defined (except when $\rho=f^{\mathrm{eq}}$).
Second, the computation of $\xi$ may suffer from the degeneracy issue,
as discussed later in Sec. \ref{subsec:wannier-interpolation}. The
above issues are bypassed through the use of a Wannier-function-based
technique given below in Sec. \ref{subsec:DrhoDk_method}.

The collision term $C\left[\rho\right]$ of Eq. \ref{eq:master} describes
the decay of $\rho$ to its equilibrium due to various processes such
as the electron-phonon scattering, the electron-hole recombination,
etc. In this work, we approximate $C\left[\rho\right]$ within the
RTA as
\begin{align}
C\left[\rho\right]= & -\Gamma\odot\left\{ \rho-f^{\mathrm{eq}}\right\} ,
\end{align}

where $f^{\mathrm{eq}}$ is the equilibrium part of $\rho$ and a
diagonal matrix whose elements are Fermi-Dirac functions. $\Gamma$
is the relaxation rate matrix and $\Gamma_{kaa}$ is the relaxation
rate of the electronic state $\left(k,a\right)$ - $\Gamma_{ka}$.
Hadamard product $A\odot B$ means the elementwise multiplication
of matrices $A$ and $B$.

Suppose
\begin{align}
\rho= & f^{\mathrm{eq}}+\rho^{E}.
\end{align}

Considering that $\frac{df^{\mathrm{eq}}}{dt}=0$ and $\left[\epsilon,f^{\mathrm{eq}}\right]=0$,
within the RTA, Eq. \ref{eq:master} becomes
\begin{align}
i\hbar\frac{d\rho^{E}}{dt}+\left(i\hbar\Gamma-\Delta\right)\odot\rho^{E}= & ie{\bf E}\left(t\right)\cdot\frac{D\rho}{D{\bf k}},\label{eq:master_rta}\\
\Delta_{kab}= & \epsilon_{ka}-\epsilon_{kb}.
\end{align}

\subsection{Perturbative solution of $\rho^{E}$ and optical susceptibilities
for charge and spin current\label{subsec:Perturbative-solution}}

At weak fields, $\rho$ can be expanded as $\rho=\sum_{n}\rho^{\left(n\right)}$
with $\rho^{\left(n\right)}=O\left(\left|{\bf E}\left(\omega\right)\right|^{n}\right)$
and $\rho^{\left(0\right)}\equiv f^{\mathrm{eq}}$. Therefore, the
nth-order master equation is
\begin{align}
i\hbar\frac{d\rho^{E,\left(n\right)}}{dt}+\left(i\hbar\Gamma-\Delta\right)\odot\rho^{E,\left(n\right)}= & ie{\bf E}\left(t\right)\cdot\frac{D\rho^{\left(n-1\right)}}{D{\bf k}},\label{eq:master_rta_pert}\\
\rho^{E,\left(n\right)}= & \left(1-\delta_{n0}\right)\rho^{\left(n\right)}.
\end{align}

The above equation is a first-order (for the time derivative) ordinary
differential equation. If the minimum element of $\Gamma$ - $\Gamma_{\mathrm{min}}$
is positive. At $t\gg\Gamma_{\mathrm{min}}^{-1}$, Eq. \ref{eq:master_rta_pert}
has a stationary solution
\begin{align}
\rho^{E,\left(n\right)}\left(t\right)= & \sum_{m}\rho^{E,\left(n\right)}\left(m\omega\right)e^{im\omega t},\label{eq:rhot_expansion}
\end{align}

where $\rho^{E,\left(n\right)}\left(m\omega\right)$ is time-independent.
Therefore, at $t\gg\Gamma_{\mathrm{min}}^{-1}$, from Eq. \ref{eq:master_rta_pert},
\begin{align}
 & \left(-m\hbar\omega-\Delta+i\hbar\Gamma\right)\odot\rho^{E,\left(n\right)}\left(m\omega\right)\nonumber \\
= & ie\sum_{\pm}{\bf E}\left(\pm\omega\right)\cdot\frac{D\rho^{\left(n-1\right)}\left(\left(m\mp1\right)\omega\right)}{D{\bf k}}.
\end{align}

Define
\begin{align}
d_{kab}^{\Gamma}\left(\omega\right)= & \frac{1}{-\hbar\omega-\Delta_{kab}+i\hbar\Gamma_{kab}},\label{eq:dw_Gamma}
\end{align}

we have,
\begin{align}
\rho^{E,\left(n\right)}\left(m\omega\right)= & ie\sum_{\pm}{\bf E}\left(\pm\omega\right)\cdot\frac{D\rho^{\left(n-1\right)}\left(\left(m\mp1\right)\omega\right)}{D{\bf k}}\nonumber \\
 & \odot d^{\Gamma}\left(m\omega\right),
\end{align}

From the above equation, we can define a nth-order DM from the following
iterative formulae,
\begin{align}
\rho_{\alpha_{1}...\alpha_{n}}^{E,\left(n\right)}\left(\omega_{1},...,\omega_{n}\right)= & \left\{ \begin{array}{c}
ieE_{\alpha_{1}}\left(\omega_{1}\right)\frac{D\rho_{\alpha_{2}...\alpha_{n}}^{\left(n-1\right)}\left(\omega_{2},...,\omega_{n}\right)}{Dk_{\alpha_{1}}}\\
\odot d^{\Gamma}\left(\sum_{j=1}^{n}\omega_{j}\right)
\end{array}\right\} ,\label{eq:nth-order-rho}\\
\rho_{\alpha_{n}}^{E,\left(1\right)}\left(\omega_{n}\right)= & ieE_{\alpha_{n}}\left(\omega_{n}\right)\frac{D\rho^{\left(0\right)}}{Dk_{\alpha_{n}}}\odot d^{\Gamma}\left(\omega_{n}\right),\label{eq:first-order-rho}\\
\omega_{j}= & \pm\omega.
\end{align}

We further define the normalized nth-order DM as
\begin{align}
\widetilde{\rho}_{\alpha_{1}...\alpha_{n}}^{E,\left(n\right)}\left(\omega_{1},...,\omega_{n}\right)= & \frac{\rho_{\alpha_{1}...\alpha_{n}}^{E,\left(n\right)}\left(\omega_{1},...,\omega_{n}\right)}{\prod_{i}^{n}E_{\alpha_{i}}\left(\omega_{i}\right)}.\label{eq:normailized-nth-order-rho}
\end{align}

Therefore, the nth-order current and spin-current densities can be
expressed as
\begin{align}
{\bf J}^{c/s_{\gamma},\left(n\right)}\left(t\right)= & \sum_{m}{\bf J}^{c/s_{\gamma},\left(n\right)}\left(m\omega\right)e^{im\omega t},\\
{\bf J}^{c/s_{\gamma},\left(n\right)}\left(\sum_{j=1}^{n}\omega_{j}\right)= & V_{\mathrm{cell}}^{-1}\prod_{i}^{n}E_{\alpha_{i}}\left(\omega_{i}\right)\\
 & \times\mathrm{Tr}\left[{\bf j}^{c/s_{\gamma}}\widetilde{\rho}_{\alpha_{1}...\alpha_{n}}^{E,\left(n\right)}\left(\omega_{1},...,\omega_{n}\right)\right].
\end{align}

Since nth-order optical susceptibilities for charge and spin current
are defined as
\begin{align}
J_{\beta}^{c/s_{\gamma},\left(n\right)}\left(\sum_{j=1}^{n}\omega_{j}\right)= & \sum_{\alpha_{1}...\alpha_{n}}\prod_{i}^{n}E_{\alpha_{i}}\left(\omega_{i}\right)\nonumber \\
 & \times\sigma_{\alpha_{1}...\alpha_{n}}^{c/s_{\gamma},\beta}\left(\omega_{1},...,\omega_{n}\right),\label{eq:J-sigma-relation}
\end{align}

we have
\begin{align}
\sigma_{\alpha_{1}...\alpha_{n}}^{c/s_{\gamma},\beta}\left(\omega_{1},...,\omega_{n}\right)= & V_{\mathrm{cell}}^{-1}\mathrm{Tr}\left[j_{\beta}^{c/s_{\gamma}}\widetilde{\rho}_{\alpha_{1}...\alpha_{n}}^{E,\left(n\right)}\left(\omega_{1},...,\omega_{n}\right)\right].\label{eq:nth-order-sigma}
\end{align}

From the above equations, we can obtain arbitrary-order perturbative
optical susceptibilities and photocurrent. Considering that BPVE,
SHG and THG, three of the most important photocurrent phenomena, are
determined by second- and third-order optical susceptibilities, we
present the detailed formulae of optical susceptibilities in first
three orders as follows.

\subsubsection{First-order}

From Eq. \ref{eq:first-order-rho}, \ref{eq:normailized-nth-order-rho}
and \ref{eq:nth-order-sigma}, we have
\begin{align}
\widetilde{\rho}_{\alpha}^{E,\left(1\right)}\left(\omega\right)= & ie\frac{Df^{\mathrm{eq}}}{Dk_{\alpha}}\odot d^{\Gamma}\left(\omega\right),\label{eq:rho1}\\
\sigma_{\alpha}^{c/s_{\gamma},\beta}\left(\omega\right)= & ieV_{\mathrm{cell}}^{-1}\mathrm{Tr}\left[j_{\beta}^{c/s_{\gamma}}\frac{Df^{\mathrm{eq}}}{Dk_{\alpha}}\odot d^{\Gamma}\left(\omega\right)\right].
\end{align}

$\sigma_{\alpha}^{c/s_{\gamma},\beta}\left(\omega\right)$ determines
the optical conductivity.

\subsubsection{Second-order}

From Eq. \ref{eq:nth-order-rho}, \ref{eq:first-order-rho}, \ref{eq:normailized-nth-order-rho}
and \ref{eq:nth-order-sigma}, we have the DC component
\begin{align}
\widetilde{\rho}_{\alpha_{1}\alpha_{2}}^{E,\left(2\right)}\left(-\omega,\omega\right)= & ie\frac{D\widetilde{\rho}_{\alpha_{2}}^{E,\left(1\right)}\left(\omega\right)}{Dk_{\alpha_{1}}}\odot d^{\Gamma}\left(0\right),\label{eq:rho2-+}\\
\sigma_{\alpha_{1}\alpha_{2}}^{c/s_{\gamma},\beta}\left(-\omega,\omega\right)= & V_{\mathrm{cell}}^{-1}\mathrm{Tr}\left[j_{\beta}^{c/s_{\gamma}}\widetilde{\rho}_{\alpha_{1}\alpha_{2}}^{E,\left(2\right)}\left(-\omega,\omega\right)\right].\label{eq:sigma2-+}
\end{align}

and the $2\omega$ AC component
\begin{align}
\widetilde{\rho}_{\alpha_{1}\alpha_{2}}^{E,\left(2\right)}\left(\omega,\omega\right)= & ie\frac{D\widetilde{\rho}_{\alpha_{2}}^{E,\left(1\right)}\left(\omega\right)}{Dk_{\alpha_{1}}}\odot d^{\Gamma}\left(2\omega\right),\label{eq:rho2++}\\
\sigma_{\alpha_{1}\alpha_{2}}^{c/s_{\gamma},\beta}\left(\omega,\omega\right)= & V_{\mathrm{cell}}^{-1}\mathrm{Tr}\left[j_{\beta}^{c/s_{\gamma}}\widetilde{\rho}_{\alpha_{1}\alpha_{2}}^{E,\left(2\right)}\left(\omega,\omega\right)\right].\label{eq:sigma2++}
\end{align}

$\sigma_{\alpha_{1}\alpha_{2}}^{c/s_{\gamma},\beta}\left(-\omega,\omega\right)$
and $\sigma_{\alpha_{1}\alpha_{2}}^{c/s_{\gamma},\beta}\left(\omega,\omega\right)$
determine BPVE and SHG respectively.

\subsubsection{Third-order}

From Eq. \ref{eq:nth-order-rho}, \ref{eq:first-order-rho}, \ref{eq:normailized-nth-order-rho}
and \ref{eq:nth-order-sigma}, we have the $3\omega$ AC component
\begin{align}
\widetilde{\rho}_{\alpha_{1}\alpha_{2}\alpha_{3}}^{E,\left(3\right)}\left(\omega,\omega,\omega\right)= & ie\frac{D\widetilde{\rho}_{\alpha_{2}\alpha_{3}}^{E,\left(2\right)}\left(\omega,\omega\right)}{Dk_{\alpha_{1}}}\odot d^{\Gamma}\left(3\omega\right),\label{eq:rho3}\\
\sigma_{\alpha_{1}\alpha_{2}\alpha_{3}}^{c/s_{\gamma},\beta}\left(\omega,\omega,\omega\right)= & V_{\mathrm{cell}}^{-1}\mathrm{Tr}\left[j_{\beta}^{c/s_{\gamma}}\widetilde{\rho}_{\alpha_{1}\alpha_{2}\alpha_{3}}^{E,\left(3\right)}\left(\omega,\omega,\omega\right)\right].\label{eq:sigma3}
\end{align}

$\sigma_{\alpha_{1}\alpha_{2}\alpha_{3}}^{c/s_{\gamma},\beta}\left(\omega,\omega,\omega\right)$
determines THG.

Under weak fields, the photocurrent mechanisms can be separated into
two classes\citep{dai2023recent} - (i) One is described using the
single-particle electronic quantities and with the scattering in Born
approximation. The scattering is usually further simplified within
the RTA. (ii) Another is due to the asymmetric scattering beyond Born
approximation and is called ballistic current. Since the former class
seems more important in most cases and \textit{ab initio} simulations
of ballistic current are numerically difficult\citep{dai2021phonon},
most \textit{ab initio} works only consider the former class of the
mechanisms\citep{bhalla2023quantum,watanabe2021chiral}.

The former class can be further separated into various types of contributions
depending on whether intraband or interband parts of perturbative
density matrices $\rho^{\left(n\right)}$ are considered.\citep{bhalla2023quantum}
See the discussions in Sec. \ref{subsec:different-parts-of-BPVE}
below and Appendix A. The following contributions to BPVE were identified
previously\citep{bhalla2023quantum,watanabe2021chiral}: the shift
current, gyration current, (magnetic) injection current, Berry curvature
dipole and other Fermi surface contributions. Since our method includes
both intraband and interband parts of all $\rho^{\left(n\right)}$
matrices, all types of contributions belonging to the former class
are considered.

\subsection{LPGE and CPGE}

From Eq. \ref{eq:Et}, the electric field amplitudes are ${\bf E}\left(\pm\omega\right)$.
For linearly polarized light, $E_{\alpha_{1}}\left(-\omega\right)E_{\alpha_{2}}\left(\omega\right)\equiv E_{\alpha_{1}}^{*}\left(\omega\right)E_{\alpha_{2}}\left(\omega\right)$
is real for any $\alpha_{1}$ and $\alpha_{2}$ and ${\bf E}\left(-\omega\right)\times{\bf E}\left(\omega\right)\equiv{\bf E}^{*}\left(\omega\right)\times{\bf E}\left(\omega\right)=0$
is always satisfied. While for circularly polarized light, ${\bf E}\left(-\omega\right)\times{\bf E}\left(\omega\right)\equiv{\bf E}^{*}\left(\omega\right)\times{\bf E}\left(\omega\right)$
is always purely imaginary. Therefore, we introduce the following
definitions for LPGE and CPGE:
\begin{align}
L_{\alpha_{1}\alpha_{2}}\left(\omega\right)= & \mathrm{Re}\left\{ E_{\alpha_{1}}^{*}\left(\omega\right)E_{\alpha_{2}}\left(\omega\right)\right\} ,\\
{\bf F}\left(\omega\right)= & \frac{1}{2}i{\bf E}^{*}\left(\omega\right)\times{\bf E}\left(\omega\right),\\
\sigma_{\alpha_{1}\alpha_{2}}^{\mathrm{DC},c/s_{\gamma},\beta}\left(\omega\right)= & \frac{1}{2}\left(\sigma_{\alpha_{1}\alpha_{2}}^{c/s_{\gamma},\beta}\left(-\omega,\omega\right)+\sigma_{\alpha_{2}\alpha_{1}}^{c/s_{\gamma},\beta}\left(\omega,-\omega\right)\right),\\
\eta_{\alpha_{1}\alpha_{2}}^{c/s_{\gamma},\beta}\left(\omega\right)= & \mathrm{Re}\left[\sigma_{\alpha_{1}\alpha_{2}}^{\mathrm{DC},c/s_{\gamma},\beta}\left(\omega\right)\right],\\
\kappa_{\lambda}^{c/s_{\gamma},\beta}\left(\omega\right)= & \epsilon_{\alpha_{1}\alpha_{2}\lambda}\mathrm{Im}\left[\sigma_{\alpha_{1}\alpha_{2}}^{\mathrm{DC},c/s_{\gamma},\beta}\left(\omega\right)\right],
\end{align}

where $\epsilon_{\alpha_{1}\alpha_{2}\lambda}$ is Levi-Civita symbol.
We note that $L_{\alpha_{1}\alpha_{2}}\left(\omega\right)\equiv L_{\alpha_{2}\alpha_{1}}\left(\omega\right)$
and $\eta_{\alpha_{1}\alpha_{2}}^{c/s_{\gamma},\beta}\equiv\eta_{\alpha_{2}\alpha_{1}}^{c/s_{\gamma},\beta}$.
Here we call $\eta_{\alpha_{1}\alpha_{2}}^{c/s_{\gamma},\beta}$ ($\kappa_{\lambda}^{c/s_{\gamma},\beta}$)
LPGE (CPGE) coefficient or susceptibility.

For LPGE, ${\bf F}\left(\omega\right)=0$, so that only the real parts
of $\sigma_{\alpha_{1}\alpha_{2}}^{c/s_{\gamma},\beta}\left(\mp\omega,\pm\omega\right)$
contribute. For CPGE, both the imaginary and real parts of $\sigma_{\alpha_{1}\alpha_{2}}^{c/s_{\gamma},\beta}\left(\mp\omega,\pm\omega\right)$
can contribute, as ${\bf F}\left(\omega\right)\neq0$ and some of
$L_{\alpha_{1}\alpha_{2}}\left(\omega\right)$ can be nonzero.

Using the relation
\begin{align}
\sigma_{\alpha_{1}\alpha_{2}}^{c/s_{\gamma},\beta}\left(-\omega,\omega\right)= & \left[\sigma_{\alpha_{1}\alpha_{2}}^{c/s_{\gamma},\beta}\left(\omega,-\omega\right)\right]^{*},
\end{align}

the second-order dc current density for photon-frequency $\omega$
can be expressed as
\begin{align}
J_{\beta}^{s_{\gamma}}\left(0\right)= & \sum_{\alpha_{1}\alpha_{2}\pm}E_{\alpha_{1}}\left(\mp\omega\right)E_{\alpha_{2}}\left(\pm\omega\right)\sigma_{\alpha_{1}\alpha_{2}}^{c/s_{\gamma},\beta}\left(\mp\omega,\pm\omega\right)\nonumber \\
= & 2\left\{ \begin{array}{c}
\sum_{\alpha_{1}\alpha_{2}}L_{\alpha_{1}\alpha_{2}}\left(\omega\right)\eta_{\alpha_{1}\alpha_{2}}^{c/s_{\gamma},\beta}\left(\omega\right)\\
+\sum_{\lambda}F_{\lambda}\left(\omega\right)\kappa_{\lambda}^{c/s_{\gamma},\beta}\left(\omega\right)
\end{array}\right\} .
\end{align}

For CPGE, in many cases, $\kappa_{\lambda}^{c/s_{\gamma},\beta}$
are found much larger than $\eta_{\alpha_{1}\alpha_{2}}^{c/s_{\gamma},\beta}$,
so that $\eta_{\alpha_{1}\alpha_{2}}^{c/s_{\gamma},\beta}$ are often
not considered.

Let's consider a few special cases below:

(i) Suppose ${\bf E}\left(\omega\right)=E\left(1,1,0\right)$ with
$E$ a real value for a linearly polarized light. As ${\bf F}\left(\omega\right)=0$,
we have
\begin{align}
J_{\beta}^{s_{\gamma}}\left(0\right)= & 2\sum_{\alpha_{1},\alpha_{2}=x,y}\eta_{\alpha_{1}\alpha_{2}}^{c/s_{\gamma},\beta}L_{\alpha_{1}\alpha_{2}}\left(\omega\right).
\end{align}

(ii) Suppose ${\bf E}\left(\omega\right)=E\left(1,i,0\right)$ with
$E$ a real value for a circularly polarized light. As $L_{xy}\left(\omega\right)=0$
and ${\bf F}\left(\omega\right)=E^{2}\left(0,0,-1\right)$, we have
\begin{align}
J_{\beta}^{s_{\gamma}}\left(0\right)= & 2\left(\sum_{\alpha=x,y}L_{\alpha\alpha}\left(\omega\right)\eta_{\alpha\alpha}^{c/s_{\gamma},\beta}+F_{z}\left(\omega\right)\kappa_{z}^{c/s_{\gamma},\beta}\right).
\end{align}

\subsection{Wannier interpolation and the computation of the covariant derivative
$\frac{D\rho_{k}}{D{\bf k}}$\label{subsec:Wannier}}

The Wannier interpolation based on maximally localized Wannier functions
of electronic quantities has been widely employed to simulate various
physical properties.\citep{marzari2012maximally,wang2006ab,xu2024spin,xu2020spin,xu2021giant,xu2023substrate}
The Wannier interpolation contains four steps: (i) The electronic
quantities are first calculated on coarse k meshes, e.g., 6$\times$6$\times$6
and 12$\times$12 for 3D and 2D systems respectively. (ii) Secondly,
they are transformed to the corresponding real-space matrix elements
with (real-space) localized Wannier functions (WFs) as basis. (iii)
Thirdly, electronic quantities are transformed back to the reciprocal
space. At this step, quantities on very fine k meshes (e.g., 2000$\times$2000)
or at many arbitrary k-points in Wannier representation are obtained.
In Wannier representation, the basis functions are the smooth Bloch-like
functions (see Eq. \ref{eq:uW} below). (iv) Finally, Wannier representation
is replaced by the eigenbasis representation where the basis functions
are the Bloch eigenstates of the Wannier-interpolated Hamiltonian.
Thus, physical properties can be conveniently calculated with electronic
quantities in the eigenbasis representation with converged number
of k-points.

In this subsection, different representations are used to express
electronic quantities. Therefore, for clarity, no additional notation
or superscript is used for the eigenbasis representation, while superscript
$W$ is used for the Wannier representation. Note that all equations
above this subsection use the eigenbasis representation.

\subsubsection{Wannier representation and Wannier interpolation\label{subsec:wannier-interpolation}}

The WFs are noted as $\left|{\bf R}a\right\rangle $, where $a$ is
the index of a WF in the unitcell and ${\bf R}$ labels the unitcell.
The smooth Bloch-like functions are given by the phased sum of WFs
\begin{align}
\left|u_{ka}^{W}\right\rangle = & \sum_{{\bf R}}e^{-i{\bf k}\cdot\left(\widehat{{\bf r}}-{\bf R}\right)}\left|{\bf R}a\right\rangle ,\label{eq:uW}
\end{align}

which span the actual Bloch eigenstates $\left|u_{ka}\right\rangle $
at each ${\bf k}$. $\widehat{{\bf r}}$ is position operator. Here
a hat is used to emphasize that it is an operator instead of a coordinate
of electron position.

Define
\begin{align}
\widehat{H_{k}^{0}}= & e^{-i{\bf k}\cdot\widehat{{\bf r}}}\widehat{H^{0}}e^{i{\bf k}\cdot\widehat{{\bf r}}}
\end{align}

with $\widehat{H^{0}}$ the unperturbed Hamiltonian operator.

It follows that, if we construct the Hamiltonian in the Wannier representation
\begin{align}
H_{kab}^{W}= & \left\langle u_{ka}^{W}\right|\widehat{H_{k}^{0}}\left|u_{kb}^{W}\right\rangle 
\end{align}

and diagonalize it as
\begin{align}
U_{k}^{\dagger}H_{k}^{W}U_{k}= & \epsilon_{k},
\end{align}

where $U_{k}$ are the eigenstate matrix and $\epsilon_{k}$ is the
diagonal matrix of eigenvalues. The corresponding Bloch eigenstates
are
\begin{align}
\left|u_{ka}\right\rangle = & \sum_{b}\left|u_{kb}^{W}\right\rangle U_{kba}.
\end{align}

Similar to $H_{k}^{W}$, the velocity and spin matrices are well defined
in Wannier representation and are noted as ${\bf v}_{k}^{W}$ and
${\bf s}_{k}^{W}$ respectively. The computations of $H_{k}^{W}$,
${\bf v}_{k}^{W}$ and ${\bf s}_{k}^{W}$ are efficient and done through
standard techniques developed in Ref. \citenum{wang2006ab}. With
$U_{k}$, the velocity and spin matrices in the eigenbasis representation
read
\begin{align}
{\bf v}_{k}= & U_{k}^{\dagger}{\bf v}_{k}^{W}U_{k},\\
{\bf s}_{k}= & U_{k}^{\dagger}{\bf s}_{k}^{W}U_{k}.
\end{align}

Having ${\bf v}_{k}$ and ${\bf s}_{k}$, $j_{\beta,k}^{c/s_{\gamma}}$
are obtained straightforwardly from Eq. \ref{eq:jc_operator} and
\ref{eq:js_operator}. As the basis size of Wannier representation
is usually small (same as the eigenbasis representation), the computations
of $\epsilon_{k}$, $U_{k}$, ${\bf v}_{k}$ and ${\bf s}_{k}$ are
all efficient. The computational technique of Berry connection in
Wannier representation ${\bf \xi}_{k}^{W}$ is slightly different
from that of $H_{k}^{W}$ and is also efficient.\citep{wang2006ab}

In the conventional Wannier-function-based \textit{ab initio} methods
(using the length gauge) of the photocurrent,\citep{ibanez2018ab,wang2017first,zhang2019switchable,le2020ab,zhang2018berry}
it is necessary to compute $\xi_{k}$ and the derivative of its off-diagonal
part $\xi_{k}^{o}$. $\xi_{k}$ is expressed as
\begin{align}
\xi_{k}= & i{\bf D}_{k}+\overline{\xi}_{k},\label{eq:xi}\\
{\bf D}_{k}= & U_{k}^{\dagger}\frac{dU_{k}}{d{\bf k}},\label{eq:D_for_xi}\\
\overline{\xi}_{k}= & U_{k}^{\dagger}\xi_{k}^{W}U_{k},\label{eq:xibar}\\
{\bf \xi}_{kab}^{W}= & i\left\langle u_{ka}^{W}|\frac{du_{kb}^{W}}{d{\bf k}}\right\rangle .\label{eq:xiW}
\end{align}

However, computing ${\bf D}_{k}$ directly via Eq. \ref{eq:D_for_xi}
is non-trivial and usually done using non-degenerate perturbation
theory,\citep{wang2006ab,ibanez2018ab}
\begin{align}
{\bf D}_{kab}\approx & {\bf D}_{kab}^{\mathrm{pert}},\\
{\bf D}_{kab}^{\mathrm{pert}}= & \left\{ \begin{array}{c}
\frac{\left(U_{k}^{\dagger}\frac{dH_{k}^{W}}{d{\bf k}}U_{k}\right)_{ab}}{\epsilon_{kb}-\epsilon_{ka}},\text{ if }\epsilon_{ka}\neq\epsilon_{kb}\\
0,\text{ if }\epsilon_{ka}=\epsilon_{kb}
\end{array}.\right.
\end{align}

Obviously, ${\bf D}_{kab}^{\mathrm{pert}}$ are problematic for degenerate
bands. This issue may be removed for two-fold degeneracy by choosing
a specific gauge of $U_{k}$,\citep{chen2022basic} but computing
${\bf D}_{kab}$ for arbitrarily degenerate bands without an approximation
is still difficult. Similarly, in the conventional method, the expression
of the derivative of $\xi_{k}^{o}$ contains $U_{k}^{\dagger}\frac{d^{2}U_{k}}{d{\bf k}^{2}}$,
which is also done using non-degenerate perturbation theory\citep{wang2017first},
so that the derivative of $\xi_{k}^{o}$ may have some random errors
for degenerate bands. The degeneracy issue is completely removed in
our method, since ${\bf D}_{k}$ and $U_{k}^{\dagger}\frac{d^{2}U_{k}}{d{\bf k}^{2}}$
are absent in the computation of $\frac{D\rho_{k}}{D{\bf k}}$, as
present clearly in the next subsection.

\subsubsection{The computation of the covariant derivative $\frac{D\rho_{k}}{D{\bf k}}$\label{subsec:DrhoDk_method}}

$\frac{D\rho_{k}}{D{\bf k}}$ is called the covariant derivative because
it satisfies the following relation for arbitrary $U_{k}$:
\begin{align}
\frac{D\rho_{k}}{D{\bf k}}= & U_{k}^{\dagger}\frac{D\rho_{k}^{W}}{D{\bf k}}U_{k},\label{eq:cov-der-between-rho-rhoW}
\end{align}

where $\rho_{k}^{W}=U_{k}\rho_{k}U_{k}^{\dagger}$. The proof is given
in Appendix B. Since
\begin{align}
\frac{D\rho_{k}^{W}}{D{\bf k}}= & \frac{d\rho_{k}^{W}}{d{\bf k}}-i\left[\xi_{k}^{W},\rho_{k}^{W}\right],
\end{align}

we further have
\begin{align}
\frac{D\rho_{k}}{D{\bf k}}= & U_{k}^{\dagger}\frac{d\rho_{k}^{W}}{d{\bf k}}U_{k}-i\left[\overline{\xi}_{k},\rho_{k}\right].\label{eq:DrhoDk_from_drhoWdk}
\end{align}

Since the basis function of Wannier representation $u_{ka}^{W}$ is
smooth over ${\bf k}$ for each index $a$, the matrix derivative
$\frac{d\rho_{k}^{W}}{d{\bf k}}$ is well defined and can be computed
numerically by finite differences. We use the central difference here.
Define ${\bf k}_{p}={\bf k}+d{\bf k}$ and ${\bf k}_{m}={\bf k}-d{\bf k}$
so that
\begin{align}
U_{k}^{\dagger}\frac{d\rho_{k}^{W}}{d{\bf k}}U_{k}= & U_{k}^{\dagger}\frac{\rho_{k_{p}}^{W}-\rho_{k_{m}}^{W}}{2d{\bf k}}U_{k}\nonumber \\
= & \frac{\left(\begin{array}{c}
o_{kk_{p}}\rho_{k_{p}}o_{kk_{p}}^{\dagger}-o_{kk_{m}}\rho_{k_{m}}o_{kk_{m}}^{\dagger}\end{array}\right)}{2d{\bf k}}\label{eq:UHdU}
\end{align}

with $o_{k_{1}k_{2}}$ the overlap matrix
\begin{align}
o_{k_{1}k_{2}}= & U_{k_{1}}^{\dagger}U_{k_{2}}.
\end{align}

Due to the use of WFs, electronic quantities including $\rho_{k}^{\left(n\right)}$
can be computed at arbitrary ${\bf k}$. Therefore, $\left|d{\bf k}\right|$
can be arbitrarily small and is typically chosen as $10^{-8}$, which
guarantees the accuracy of finite-difference computations. For the
finite-difference computation of $\frac{D\rho_{k_{0}}^{\left(n\right)}}{D{\bf k}}$,
$H_{k}^{W}$, ${\bf v}_{k}^{W}$ and $\xi_{k}^{W}$ matrices at a
set of k-points surround the central k-point ${\bf k}_{0}$ are needed.
Such Wannier-representation matrices at these k-points can be computed
either directly via Wannier interpolation at each k-point or via Taylor
expansions of $H_{k}^{W}$, ${\bf v}_{k}^{W}$ and $\xi_{k}^{W}$
matrices around ${\bf k}_{0}$, whose (few-order) derivatives can
be computed efficiently\citep{wang2006ab}.

With Eq. \ref{eq:UHdU} and $\overline{\xi}_{k}$ computed by Eq.
\ref{eq:xibar}, $\frac{D\rho_{k}}{D{\bf k}}$ is then obtained from
Eq. \ref{eq:DrhoDk_from_drhoWdk}. For numerical implementation of
$\frac{D\rho_{k}}{D{\bf k}}$, helpful techniques are employed as
described in Appendix C and D.

\subsection{Different contributions to BPVE\label{subsec:different-parts-of-BPVE}}

For further discussions, we first introduce the following separation
of an arbitrary matrix $A$:
\begin{align}
A= & A^{d}+A^{o},
\end{align}

where the subscripts $d$ and $o$ stand for the diagonal and off-diagonal
parts.

The second-order optical susceptibilities for BPVE (corresponding
to charge photocurrent) can be separated into four parts:
\begin{align}
\sigma_{\alpha_{1}\alpha_{2}}^{\mathrm{DC},c,\beta}\left(\omega\right)= & \sigma_{\alpha_{1}\alpha_{2}}^{\mathrm{DC},c,\beta,dd}\left(\omega\right)+\sigma_{\alpha_{1}\alpha_{2}}^{\mathrm{DC},c,\beta,od}\left(\omega\right)\nonumber \\
 & +\sigma_{\alpha_{1}\alpha_{2}}^{\mathrm{DC},c,\beta,do}\left(\omega\right)+\sigma_{\alpha_{1}\alpha_{2}}^{\mathrm{DC},c,\beta,oo}\left(\omega\right).
\end{align}

In the double subscripts $dd$, $od$, $do$ and $oo$, the first
letter indicates the diagonal and off-diagonal parts of $\rho^{\left(2\right)}$
and the latter letter corresponds to the dependence of the relevant
parts of $\rho^{\left(1\right)}$ on $\rho^{\left(2\right)}$.

We present the forms of the above four parts in this subsection but
leave the detailed derivations in Appendix A:

(1) The $dd$ intraband-intraband part.

This is a Fermi-surface term - Drude term and reads
\begin{align}
\sigma_{\alpha_{1}\alpha_{2}}^{\mathrm{DC},c,\beta,dd}\left(\omega\right)= & \frac{-e^{3}V_{\mathrm{cell}}^{-1}N_{k}^{-1}}{\hbar^{2}\left(\omega^{2}+\Gamma^{2}\right)}\sum_{ka}v_{\beta,kaa}\frac{d^{2}f_{ka}^{\mathrm{eq}}}{dk_{\alpha_{1}}dk_{\alpha_{2}}}.
\end{align}

(2) The $od$ interband-intraband part.

This is another Fermi-surface term, which corresponds to the Berry
curvature dipole term when $\Gamma=0$, and reads
\begin{align}
\sigma_{\alpha_{1}\alpha_{2}}^{\mathrm{DC},c,\beta,od}\left(\omega\right)= & \frac{e^{3}V_{\mathrm{cell}}^{-1}N_{k}^{-1}}{2\left(-\hbar\omega+i\hbar\Gamma\right)}\nonumber \\
 & \times\sum_{k,ab}\xi_{\beta,kba}^{o}\xi_{\alpha_{1},kab}^{o}\frac{df_{kab}^{\mathrm{eq}}}{dk_{\alpha_{2}}}\Delta_{kab}d_{kab}^{\Gamma}\left(0\right)\nonumber \\
 & +\left[\left(\alpha_{1},-\omega\right)\leftrightarrow\left(\alpha_{2},\omega\right)\right].
\end{align}

(3) The $do$ intraband-interband part.

This is the injection current term and reads
\begin{align}
\sigma_{\alpha_{1}\alpha_{2}}^{\mathrm{DC},c,\beta,do}\left(\omega\right)= & \frac{e^{3}\pi V_{\mathrm{cell}}^{-1}N_{k}^{-1}}{\hbar\Gamma}\sum_{kab}\xi_{\alpha_{2},kab}\xi_{\alpha_{1},kba}\nonumber \\
 & \times\left(v_{\beta,aa}-v_{\beta,bb}\right)f_{kab}^{\mathrm{eq}}\delta^{\Gamma}\left(\hbar\omega+\Delta_{kab}\right),\\
\delta^{\Gamma}\left(\hbar\omega\right)= & \frac{1}{\pi}\frac{\hbar\Gamma}{\left(\hbar\omega\right)^{2}+\left(\hbar\Gamma\right)^{2}}.
\end{align}

This is the same as Ref. \citenum{bhalla2023quantum}.

(4) The $oo$ interband-interband part.

This part contains the shift current for linearly polarized light
and its counterpart for circularly polarized light - the gyration
current, and also several other contributions including the Fermi
surface ones.\citep{bhalla2023quantum,watanabe2021chiral} According
to Appendix A, this part reads
\begin{align}
\sigma_{\alpha_{1}\alpha_{2}}^{\mathrm{DC},c,\beta,oo}\left(-\omega,\omega\right)= & \frac{e^{3}V_{\mathrm{cell}}^{-1}N_{k}^{-1}}{\hbar}\nonumber \\
 & \times\sum_{kab}\left(\frac{D\xi_{\beta,k}^{\Gamma,o}}{Dk_{\alpha_{1}}}\right)_{ba}\xi_{\alpha_{2},kab}^{o}f_{kab}^{\mathrm{eq}}d_{kab}^{\Gamma}\left(\omega\right)\nonumber \\
 & +\left[\left(\alpha_{1},-\omega\right)\leftrightarrow\left(\alpha_{2},\omega\right)\right],\label{eq:sigmaDCoo}
\end{align}

where
\begin{align}
\xi_{\beta,kab}^{\Gamma,o}= & -i\hbar v_{\beta,kab}^{o}d_{kba}^{\Gamma}\left(0\right),\nonumber \\
= & \xi_{\beta,kab}^{o}\frac{\Delta_{kab}}{\Delta_{kab}+i\hbar\Gamma}.\label{eq:xio_Gamma}
\end{align}

In Eq. \ref{eq:sigmaDCoo}, relaxation rate $\Gamma$ appears in two
places - one in $d_{k}^{\Gamma}\left(\pm\omega\right)=1/\left(\mp\hbar\omega-\Delta_{k}+i\hbar\Gamma\right)$
of $\widetilde{\rho}_{\alpha}^{E,\left(1\right)}\left(\pm\omega\right)$
and another in $d_{k}^{\Gamma}\left(0\right)=1/\left(-\Delta_{k}+i\hbar\Gamma\right)$
(see the above Eq. \ref{eq:xio_Gamma}) of $\widetilde{\rho}_{\alpha_{1}\alpha_{2}}^{E,\left(2\right)}\left(\mp\omega,\pm\omega\right)$
(Eq. \ref{eq:sigma2-+}). These two $\Gamma$s play very different
roles in the shift and gyration current, as discussed below. Thus,
we name $\Gamma$ in $d_{k}^{\Gamma}\left(0\right)$ as $\Gamma^{\left(2\right)}$
below for the $oo$ part.

To obtain the standard formulae of the shift and gyration current\citep{watanabe2021chiral},
we need to take the weak-scattering limit $\Gamma^{\left(2\right)}\rightarrow0$,
so that Eq. \ref{eq:xio_Gamma} is approximated as
\begin{align}
\xi_{\beta}^{\Gamma^{\left(2\right)},o}\approx & \xi_{\beta}^{o}.\label{eq:scattering-free-limit}
\end{align}

In realistic samples and/or at finite temperatures, we need to consider
that $\Gamma^{\left(2\right)}$ is finite. When $\left|\Delta_{kab}\right|\gg\hbar\Gamma^{\left(2\right)}$,
the above equation is still accurate for matrix elements $\xi_{\beta,kab}^{\Gamma^{\left(2\right)},o}$.
However, when $\left|\Delta_{kab}\right|\lesssim\hbar\Gamma^{\left(2\right)}$
($\lesssim$ means smaller than or comparable to), i.e., for two degenerate
or near-degenerate states $\left|{\bf k}a\right\rangle $ and $\left|{\bf k}b\right\rangle $
($a\neq b$), this equation may introduce notable errors of $\xi_{\beta,kab}^{\Gamma^{\left(2\right)},o}$.
Therefore, $\Gamma^{\left(2\right)}$ plays a role of a ``smooth''
degeneracy threshold for matrix elements $\xi_{\beta,kab}^{o}$, i.e.,
$\xi_{\beta,kab}^{o}$ are kept and neglected when $\left|\Delta_{kab}\right|\gg\hbar\Gamma^{\left(2\right)}$
and $\left|\Delta_{kab}\right|\ll\hbar\Gamma^{\left(2\right)}$ respectively.
Practically, for $\rho_{kab}$ with small $\Delta_{kab}$, i.e., the
off-diagonal elements of the density matrix between two states with
close energies, $\hbar\Gamma^{\left(2\right)}$ are often relatively
large and of order 0.01 to 0.1 eV due to the electron-phonon and electron-impurity
scattering processes, but can be small in certain cases, e.g., at
low temperatures in clean samples, or for states close to Dirac cones.

Using Eq. \ref{eq:scattering-free-limit} above and Eq. \ref{eq:relation_DxiDk_xigd}
in Appendix A, Eq. \ref{eq:sigmaDCoo} becomes
\begin{align}
\sigma_{\alpha_{1}\alpha_{2}}^{\mathrm{DC},c,\beta,oo}\left(\omega\right)= & \frac{-e^{3}V_{\mathrm{cell}}^{-1}N_{k}^{-1}}{2\hbar}\nonumber \\
 & \times\sum_{kab}\xi_{\alpha_{1};\beta,kab}^{o}\xi_{\alpha_{2},kba}^{o}f_{kab}^{\mathrm{eq}}d_{kba}^{\Gamma}\left(\omega\right)\nonumber \\
 & +\left[\left(\alpha_{1},-\omega\right)\leftrightarrow\left(\alpha_{2},\omega\right)\right],
\end{align}

where $\xi_{\alpha;\beta}^{o}$ is the so-called ``generalized derivative''\citep{sipe2000second,wang2017first,ibanez2018ab,watanabe2021chiral}
of matrix $\xi_{\alpha}^{o}$ along direction $\beta$:
\begin{align}
\xi_{\alpha;\beta}^{o}= & \frac{d\xi_{\alpha}^{o}}{dk_{\beta}}-i\left[\xi_{\beta}^{d},\xi_{\alpha}^{o}\right].
\end{align}

Since $d_{kba}^{\Gamma}\left(\omega\right)$ can be separated into
the principal and Dirac-delta parts
\begin{align}
d_{kab}^{\Gamma}\left(\omega\right)= & -\mathrm{P}^{\Gamma}\frac{1}{\hbar\omega+\Delta_{kab}}-i\pi\delta^{\Gamma}\left(\hbar\omega+\Delta_{kab}\right),\\
\mathrm{P}^{\Gamma}\frac{1}{\hbar\omega}= & \frac{\hbar\omega}{\left(\hbar\omega\right)^{2}+\left(\hbar\Gamma\right)^{2}},
\end{align}

$\sigma_{\alpha_{1}\alpha_{2}}^{\mathrm{DC},c,\beta,oo}\left(\omega\right)$
can also be separated into the principal and Dirac-delta parts. From
symmetry analysis, the principal part is purely imaginary under $\mathcal{T}$
(time-reversal) symmetry, so that it is absent for linearly polarized
light under $\mathcal{T}$ symmetry, and it is purely real under $\mathcal{PT}$
symmetry ($\mathcal{P}$ is parity or spatial inversion operation).

Next we focus on the Dirac-delta part, which reads
\begin{align}
\sigma_{\alpha_{1}\alpha_{2}}^{\mathrm{DC},c,\beta,oo,\delta}\left(\omega\right)= & \sigma_{\alpha_{1}\alpha_{2}}^{\mathrm{shift},c,\beta}\left(\omega\right)+i\sigma_{\alpha_{1}\alpha_{2}}^{\mathrm{gyr},c,\beta}\left(\omega\right),
\end{align}

where $\sigma_{\alpha_{1}\alpha_{2}}^{\mathrm{shift},c,\beta}\left(\omega\right)$
is the well-known shift current term and $\sigma_{\alpha_{1}\alpha_{2}}^{\mathrm{gyr},c,\beta}\left(\omega\right)$
is the gyration current term:
\begin{align}
\sigma_{\alpha_{1}\alpha_{2}}^{\mathrm{shift},c,\beta}\left(\omega\right)= & \frac{-\pi e^{3}V_{\mathrm{cell}}^{-1}N_{k}^{-1}}{4\hbar}\sum_{kab}f_{kab}^{\mathrm{eq}}\delta^{\Gamma}\left(\hbar\omega-\Delta_{kab}\right)\nonumber \\
 & \times\mathrm{Im}\left\{ \xi_{\alpha_{1};\beta,kab}^{o}\xi_{\alpha_{2},kba}^{o}+\xi_{\alpha_{2};\beta,kab}^{o}\xi_{\alpha_{1},kba}^{o}\right\} ,\label{eq:conventional_shift}\\
\sigma_{\alpha_{1}\alpha_{2}}^{\mathrm{gyr},c,\beta}\left(\omega\right)= & \frac{\pi e^{3}V_{\mathrm{cell}}^{-1}N_{k}^{-1}}{4\hbar}\sum_{kab}f_{kab}^{\mathrm{eq}}\delta^{\Gamma}\left(\hbar\omega-\Delta_{kab}\right)\nonumber \\
 & \times\mathrm{Re}\left\{ \xi_{\alpha_{1};\beta,kab}^{o}\xi_{\alpha_{2},kba}^{o}-\xi_{\alpha_{2};\beta,kab}^{o}\xi_{\alpha_{1},kba}^{o}\right\} .\label{eq:gyration_current}
\end{align}

Other equivalent forms of $\sigma_{\alpha_{1}\alpha_{2}}^{\mathrm{shift},c,\beta}\left(\omega\right)$
and $\sigma_{\alpha_{1}\alpha_{2}}^{\mathrm{gyr},c,\beta}\left(\omega\right)$
are given in Appendix A. From symmetry analysis, the gyration/shift
current is absent under $\mathcal{T}$/$\mathcal{PT}$ symmetry.

From Eq. \ref{eq:conventional_shift} and \ref{eq:gyration_current},
the role of $\Gamma$ in $d_{k}^{\Gamma}\left(\pm\omega\right)$ of
$\widetilde{\rho}_{\alpha}^{E,\left(1\right)}\left(\pm\omega\right)$
is simply introducing Lorentzian smearing to the BPVE spectra due
to the shift and gyration current, which is completely different from
that of $\Gamma^{\left(2\right)}$.

\subsection{Comparison with the conventional Wannier-function-based method of
BPVE\label{subsec:compare_with_conventional}}

According to Sec. \ref{subsec:different-parts-of-BPVE}, within the
Wannier-function approach, the implementations of $dd$, $od$ and
$do$ parts of $\sigma_{\alpha_{1}\alpha_{2}}^{\mathrm{DC},c,\beta}\left(\omega\right)$
are all trivial, since the required electronic quantities are all
well-defined. For the $oo$ part, in the conventional Wannier-function-based
method, the generalized derivative $\xi_{\alpha;\beta}^{o}$ is computed
using non-degenerate perturbation theory\citep{wang2017first,ibanez2018ab},
which may introduce gauge-dependent errors. In this work, we implemented
the conventional method of the shift current based on JDFTx package,
the same as our method using the covariant derivative.

In the conventional method, a parameter called degeneracy threshold
$t^{\mathrm{deg}}$ is necessary for dealing with $1/\left(\epsilon_{a}-\epsilon_{b}\right)$
and can be defined in two ways: (i) $1/\left(\epsilon_{a}-\epsilon_{b}\right)$
is set zero if $\left|\epsilon_{a}-\epsilon_{b}\right|<t^{\mathrm{deg}}$.
This is a natural choice of the standard conventional method of the
shift and gyration current. This treatment has been widely employed,
e.g., in Ref. \citenum{wang2017first}; (ii) $1/\left(\epsilon_{a}-\epsilon_{b}\right)$
is regularized as its principal value broadened by $t^{\mathrm{deg}}$
- $\left(\epsilon_{a}-\epsilon_{b}\right)/\left[\left(\epsilon_{a}-\epsilon_{b}\right)^{2}+\left(t^{\mathrm{deg}}\right)^{2}\right]$,
which was suggested in Ref. \citenum{ibanez2018ab}. Here we use the
latter but we have seen that two treatments lead to quite similar
results with the same $t^{\mathrm{deg}}$.

The standard formulae of the conventional method are derived in the
weak-scattering limit using the non-degenerate perturbation theory,
so that theoretically $t^{\mathrm{deg}}\rightarrow0$ limit needs
to be taken and small $t^{\mathrm{deg}}$ should be preferred in numerical
simulations. On the other hand, according to the above discussions
about the $oo$ part of BPVE in Sec. \ref{subsec:different-parts-of-BPVE},
the quantity $\Gamma^{\left(2\right)}$ in the BPVE formulae of our
method, which is relaxation rate of $\rho_{kab}$ with small $\Delta_{kab}$,
can be physically regarded as a ``smooth'' degeneracy threshold
(see discussions below Eq. \ref{eq:scattering-free-limit}). Therefore,
in this aspect, $t^{\mathrm{deg}}=\hbar\Gamma^{\left(2\right)}$ is
probably a proper choice, although $t^{\mathrm{deg}}$ and $\hbar\Gamma^{\left(2\right)}$
are not the same in theory and $t^{\mathrm{deg}}$ seems to appear
in more places in the BPVE formulae\citep{wang2017first,ibanez2018ab}
of the conventional method than $\Gamma^{\left(2\right)}$. Numerically,
we indeed find that if $t^{\mathrm{deg}}=\hbar\Gamma^{\left(2\right)}$
is satisfied, theoretical results of BPVE coefficients by our and
the conventional methods are quite similar for GaAs and GeS (but not
for bilayer AFM MBT, which will be discussed later).

For BPVE, our method requires the computations of $H^{W}$, ${\bf v}^{W}$,
$\xi^{W}$, $\frac{dH^{W}}{d{\bf k}}$ and $\frac{d{\bf v}^{W}}{d{\bf k}}$
on uniform k meshes. Similarly, the conventional method requires\citep{wang2017first,ibanez2018ab}
$H^{W}$, $\xi^{W}$, $\frac{dH^{W}}{d{\bf k}}$, $\frac{d\xi^{W}}{d{\bf k}}$
and $\frac{d^{2}H^{W}}{d{\bf k}^{2}}$ on the same k meshes. Therefore,
the computational complexities of our method and the conventional
method are similar. However, our method has the following advantages:
(i) It includes all charge/spin photocurrent mechanisms in general
cases (with finite and state-resolved $\Gamma$ in both semiconductor
and metals). (ii) It is free from the degeneracy issue. (iii) It applies
well in other types of weak-field charge/spin photocurrent such as
THG.

\begin{figure*}
\includegraphics[scale=0.45]{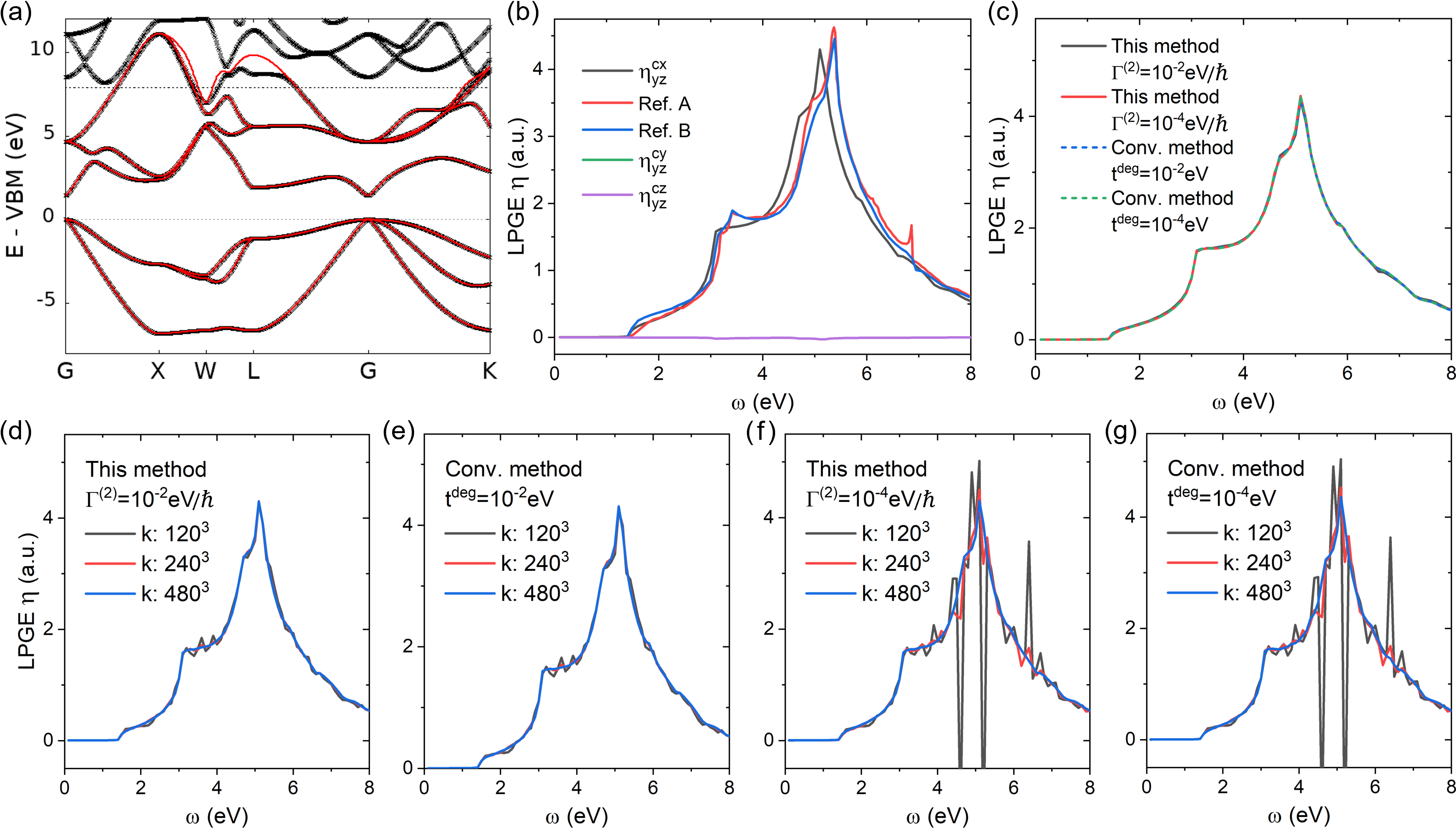}

\caption{Theoretical results of GaAs. (a) DFT and Wannier band structures.
(b) LPGE coefficients $\eta_{\alpha_{1}\alpha_{2}}^{c\beta}$ by our
method using the covariant derivative compared with $\eta_{yz}^{cx}$
from previous theoretical results - ``Ref. A'' and ``Ref. B''.
(c) $\eta_{yz}^{cx}$ by our method (labelled as ``this method'')
with $\hbar\Gamma^{\left(2\right)}$=0.01 eV and $\hbar\Gamma^{\left(2\right)}$=10$^{-4}$
eV, compared with $\eta_{yz}^{cx}$ by the conventional Wannier-function-based
method (labelled as ``conv. method'') using non-degenerate perturbation
theory with different degeneracy thresholds $t^{\mathrm{deg}}$. Subfigures
(b) and (c) use $480\times480\times480$ k meshes. (d) and (f) are
$\eta_{yz}^{cx}$ by our method with $\hbar\Gamma^{\left(2\right)}$=0.01
eV and $\hbar\Gamma^{\left(2\right)}$=10$^{-4}$ eV respectively.
(e) and (g) are $\eta_{yz}^{cx}$ by the conventional methods with
$t^{\mathrm{deg}}$ of $10^{-2}$ and $10^{-4}$ eV respectively.
The implementation details of the conventional method are given in
Sec. \ref{subsec:compare_with_conventional}. $\Gamma^{\left(2\right)}$
is relaxation rate $\Gamma$ of the off-diagonal elements of the density
matrix between two states with close energies and appearing in $d_{k}^{\Gamma}\left(0\right)=1/\left(-\Delta_{k}+i\hbar\Gamma\right)$
of the second-order perturbative density matrix $\widetilde{\rho}_{\alpha_{1}\alpha_{2}}^{E,\left(2\right)}\left(\mp\omega,\pm\omega\right)$
(Eq. \ref{eq:sigma2-+}). See detailed discussions of $\Gamma^{\left(2\right)}$
in Sec. \ref{subsec:different-parts-of-BPVE} below Eqs. \ref{eq:xio_Gamma}
and \ref{eq:scattering-free-limit}. ``Ref. A'' and ``Ref. B''
correspond to theoretical results of Refs. \citenum{ibanez2018ab}
and \citenum{nastos2006optical} respectively. A scissor correction
is included using the same method as Refs. \citenum{ibanez2018ab}
and \citenum{nastos2005scissors} to enlarge the theoretical band
gap to the experimental value 1.43 eV. The relaxation rate $\Gamma$=0.01
eV/$\hbar$, which is also the Lorentzian smearing parameter of the
conventional method.\label{fig:gaas}}
\end{figure*}

\section{Computational details}

The ground-state electronic structure is first calculated using DFT
with relatively coarse k meshes. The DFT calculations use $12\times12\times12$,
$12\times12$, $12\times12$, $12\times12$, $18\times18$ and $6\times6\times6$
k meshes for GaAs, graphenen-hBN, WS$_{2}$, GeS, MBT and RhSi respectively.
We use Perdew-Burke-Ernzerhof exchange-correlation functional\citep{perdew1996generalized}.
For bilayer AFM MBT, the DFT+U method is adopted to treat the $d$
orbitals of Mn atoms with Hubbard $U$ parameter 4.0 eV, and its lattice
structures and internal geometries are fully relaxed using the DFT+D3
correction method\citep{grimme2010consistent} for dispersion interactions.
For graphene-hBN, the DFT+D2 correction method\citep{grimme2006semiempirical}
with scale factor $s_{6}=0.5$ is used to be consistent with our previous
work\citep{habib2022electric}. For monolayer materials studied here,
we have not considered van der Waals corrections, as their effects
are found weak (within 1\%) on lattice constants. For GaAs, we use
the experimental lattice constant of 5.653 \AA as in our previous
work\citep{xu2021ab}. For RhSi, we use experimental lattice constant\citep{geller1954crystal}
of 4.67 \AA. For WS$_{2}$ and GeS, we use the fully relaxed lattice
constants. We use Optimized Norm-Conserving Vanderbilt (ONCV) pseudopotentials\citep{hamann2013optimized,van2018pseudodojo}.
The plane-wave cutoff energies are 76, 74, 62, 78, 82 and 80 Ry for
GaAs, graphene-hBN, WS$_{2}$, GeS, MBT and RhSi respectively. For
all 2D systems, the Coulomb truncation technique\citep{sundararaman2013regularization}
developed by R. Sundararaman and T. A. Arias is employed to accelerate
convergence with vacuum sizes and the vacuum sizes are 20 bohr (additional
to the thickness of the 2D systems), which leads to quite similar
optimized lattice constants and band structures compared with 60 bohr
vacuum size.

We then transform all quantities from the plane-wave basis to the
basis of maximally localized Wannier functions, and interpolate them
to substantially finer k meshes.\citep{wang2006ab,marzari2012maximally}
For the photocurrent calculations, the fine k meshes are $480\times480\times480$,
$12000\times12000$, $2400\times2400$, $2400\times2400$, $1080\times1080$
and $240\times240\times240$ for GaAs, graphene-hBN, WS$_{2}$, GeS,
MBT and RhSi respectively. For simplicity, the elements of the relaxation
rate matrices $\Gamma_{k}$ are all set to the same constant, which
unless specified is 0.001 eV/$\hbar$ for graphene-hBN and 0.01 eV/$\hbar$
for other materials, corresponding to a relaxation time of 666 and
66 fs respectively. All calculations are done based on the open-source
plane-wave DFT code JDFTx\citep{sundararaman2017jdftx,brown2016nonradiative,habib2018hot,kumar2022fermi}.

\begin{figure*}
\includegraphics[scale=0.45]{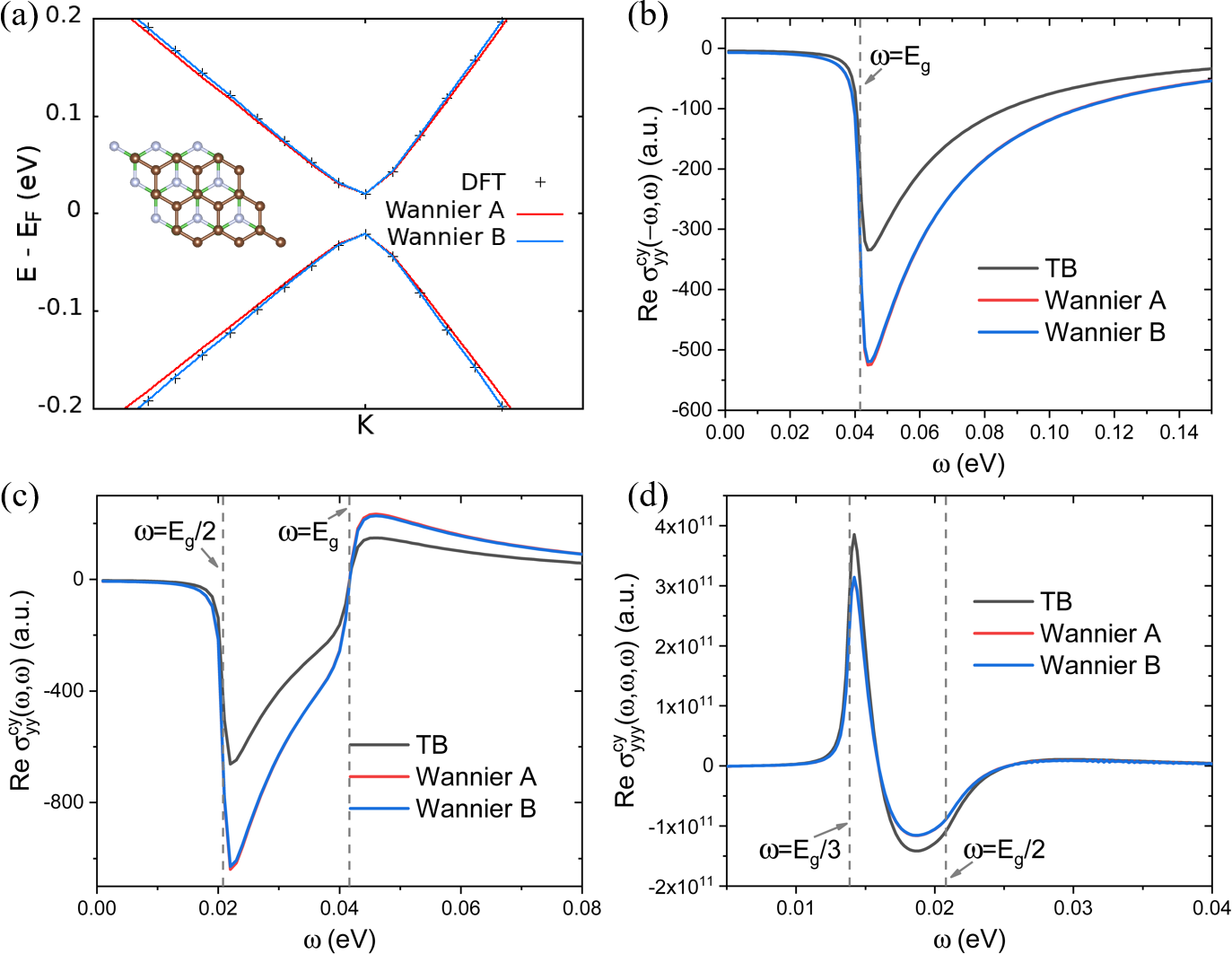}

\caption{Theoretical results of graphene-hBN.\label{fig:gr-hbn} (a) DFT and
Wannier band structures. ``Wannier A'' and ``Wannier B'' correspond
to two types of Wannier-interpolated Hamiltonians with 2 and 20 WFs
(in the unitcell) respectively. (b), (c) and (d) are optical susceptibilities
of LPGE (Re$\sigma_{yyy}^{cy}\left(\omega,\omega,\omega\right)$),
SHG (Re$\sigma_{yy}^{cy}\left(\omega,\omega\right)$) and THG (Re$\sigma_{yyy}^{cy}\left(\omega,\omega,\omega\right)$)
respectively calculated using three types of Hamiltonians - a two-band
tight-binding (TB) Hamiltonian, ``Wannier A'' and ``Wannier B''.
See more details of TB, ``Wannier A'' and ``Wannier B'' in the
text of Sec. \ref{subsec:gr-hbn}. Three special photon energies satisfying
$\omega,2\omega,3\omega=E_{g}$, corresponding to one-, two-, three-photon
processes respectively, are labeled in (b), (c) and (d) using vertical
dashed lines. $\Gamma$=0.001 eV/$\hbar$.}
\end{figure*}

\section{Results and discussions}

Before presenting our theoretical results, we would like to clarify
two points: (i) We simulate the photocurrent within the RTA, so that
the ballistic current is absent. (ii) The signs of optical susceptibilities
and LPGE/CPGE coefficients depend on the definitions of $x$, $y$,
$z$ directions. For CPGE, the sign of $\kappa_{\lambda}^{c/s_{\gamma},\beta}$
depends on the definition of ${\bf F}\left(\omega\right)$. Therefore,
when comparing with other theoretical works, we should be careful
about these definitions.

\subsection{First benchmark: GaAs}

GaAs is a typical semiconductor with broken inversion symmetry, which
allows the presence of the second-order photocurrent - BPVE and SHG.
Due to its symmetry, LPGE of GaAs is allowed while CPGE not. Without
considering the ballistic current, LPGE of GaAs is determined by shift
current. GaAs was the first piezoelectric crystal whose shift-current
spectrum was evaluated using modern band structure methods\citep{sipe2000second}
and later simulated in other method papers\citep{nastos2006optical,ibanez2018ab}.
Therefore, we first carry out benchmark calculations of LPGE coefficients
of GaAs.

In Fig. \ref{fig:gaas}(a) and (b), we compare DFT and Wannier band
structures and find that they agree perfectly. This ensures the accuracy
of the photocurrent calculation based on Wannier functions. From symmetry
analysis, it is known that LPGE coefficients $\eta_{\alpha_{1}\alpha_{2}}^{c\beta}$
are only non-zero for permutations $\beta\alpha_{1}\alpha_{2}$ of
$xyz$. Indeed, in Fig. \ref{fig:gaas}(b), numerically we find that
$\eta_{yz}^{cx}\neq0$ while $\eta_{yz}^{cy}$ and $\eta_{yz}^{cz}$
almost vanish. Our calculated $\eta_{yz}^{cx}$ are in agreement with
previous theoretical results\citep{nastos2006optical,ibanez2018ab},
which indicates the reliability of the implementation of our method.

In Fig. \ref{fig:gaas}(c), we compare LPGE coefficients by our method
using the covariant derivative (labelled as ``this method'') with
different $\Gamma^{\left(2\right)}$, which is relaxation rate of
$\rho_{kab}$ with small $\Delta_{kab}$ and appearing in $d_{k}^{\Gamma}\left(0\right)=1/\left(-\Delta_{k}+i\hbar\Gamma\right)$
of the second-order perturbative density matrix $\widetilde{\rho}_{\alpha_{1}\alpha_{2}}^{E,\left(2\right)}\left(-\omega,\omega\right)$
(Eq. \ref{eq:sigma2-+}), and the conventional Wannier-function-based
method using non-degenerate perturbation theory (labelled as ``conv.
method'') with different degeneracy thresholds $t^{\mathrm{deg}}$.
From Fig. 1(c), we find that theoretical results by our and the conventional
methods are identical with converged k meshes for different $\Gamma^{\left(2\right)}$
and $t^{\mathrm{deg}}$. Note that in all simulations, relaxation
rate $\Gamma$ (in $d_{k}^{\Gamma}\left(\pm\omega\right)=1/\left(\mp\hbar\omega-\Delta_{k}+i\hbar\Gamma\right)$
of $\widetilde{\rho}_{\alpha}^{E,\left(1\right)}\left(\pm\omega\right)$)
is fixed as 0.01 eV/$\hbar$, which is also the Lorentzian smearing
parameter of the conventional method.

Moreover, we examine theoretical results by both methods with different
k meshes in Fig. \ref{fig:gaas}(d)-(g). We find that with larger
$\hbar\Gamma^{\left(2\right)}$ and $t^{\mathrm{deg}}$ of 10$^{-2}$
eV, results by both methods show fast k-point convergence - results
with $120\times120\times120$ k meshes are already close to the converged
ones. On the other hand, with smaller $\hbar\Gamma^{\left(2\right)}$
and $t^{\mathrm{deg}}$ of 10$^{-4}$ eV, results by both methods
show slower k-point convergence - $480\times480\times480$ k meshes
are required to converge the LPGE spectrum and suspicious peaks and
dips can appear with not converged k meshes. Similar phenomena are
observed for theoretical results with another Wannierization setup
(see Fig. \ref{fig:gaas-diffwann}), which has more WFs. The $\hbar\Gamma^{\left(2\right)}$,
$t^{\mathrm{deg}}$ and Wannierization dependences of theoretical
results indicate that: When $\hbar\Gamma^{\left(2\right)}$ and $t^{\mathrm{deg}}$
are not large, such as 10$^{-4}$ eV, there seem random-like errors
around the degeneracy and near-degeneracy regions (of electronic states)
partly due to Wannier interpolation errors. For LPGE of GaAs, such
errors are avoided using relatively large $\hbar\Gamma^{\left(2\right)}$
and $t^{\mathrm{deg}}$ such as $10^{-2}$ eV and tend to be cancelled
out by increasing k meshes.

More importantly, from Fig. \ref{fig:gaas}(d)-(g) and Fig. \ref{fig:gaas-diffwann}(c)-(f),
we find that when $t^{\mathrm{deg}}=\hbar\Gamma^{\left(2\right)}$
is satisfied, LPGE coefficients by two methods are always quite similar
for all sets of k meshes. For example, when $t^{\mathrm{deg}}=\hbar\Gamma^{\left(2\right)}$=10$^{-4}$
eV and k meshes are $120\times120\times120$, LPGE spectra including
the suspicious peaks and dips by two different methods are rather
similar. We also find that when $t^{\mathrm{deg}}=\hbar\Gamma^{\left(2\right)}$,
two methods predict consistent LPGE coefficients of monolayer GeS.
For monolayer WS$_{2}$, theoretical results by both methods are also
found identical and independent of $\hbar\Gamma^{\left(2\right)}$
and $t^{\mathrm{deg}}$. These all suggest setting $t^{\mathrm{deg}}$
as $\hbar\Gamma^{\left(2\right)}$, which leads to consistent results
between two methods.

However, for bilayer AFM MBT with the so-called $\mathcal{PT}$ symmetry,
we find that random-like errors of CPGE spectra are introduced by
the conventional method and the errors cannot be removed by changing
$t^{\mathrm{deg}}$ or increasing k meshes. Note that theoretically,
even for non-magnetic systems, it is not guaranteed that setting $t^{\mathrm{deg}}=\hbar\Gamma^{\left(2\right)}$
always makes results by two methods consistent. Therefore, our method
is numerically better than the conventional method, since it completely
avoids the degeneracy issue. Additionally, the degeneracy issue in
the conventional Wannier-function-based method has not been well examined
for other types of (spin-)photocurrent, future theoretical studies
are needed to achieve better understandings.

\begin{figure*}
\includegraphics[scale=0.45]{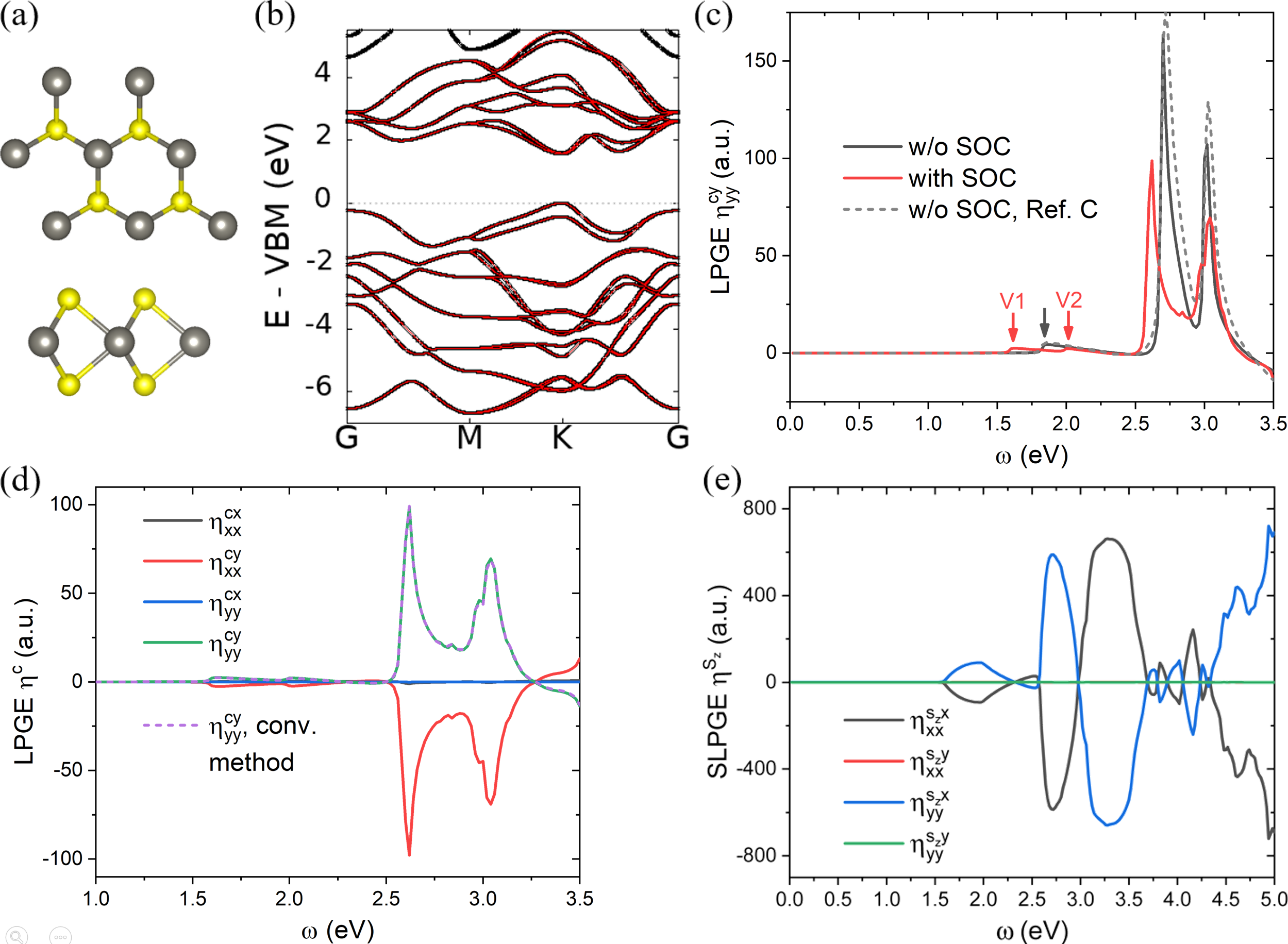}

\caption{Theoretical results of monolayer WS$_{2}$. (a) Top and side views
of the structure. (b) DFT (black lines) and Wannier (red lines) band
structures. (c) LPGE coefficients $\eta_{yy}^{cy}$ with and without
spin-orbit coupling (SOC) compared with previous theoretical results
without SOC. ``Ref. C'' is Ref. \citenum{wang2017first}. (d) LPGE
coefficients by our method with SOC, compared with $\eta_{yy}^{cy}$
by the conventional method (labelled as ``conv. method'') implemented
by us with $t^{\mathrm{deg}}=10^{-2}$ eV. (e) Spin LPGE (SLPGE) coefficients
with SOC. $\Gamma$=0.01 eV/$\hbar$.\label{fig:ws2}}
\end{figure*}

\subsection{A 2D material: LPGE and LHG of graphene-hBN\label{subsec:gr-hbn}}

Since the discovery of graphene, low-frequency nonlinear optical response
of graphene has attracted a lot of attention of both theorists and
experimentalists.\citep{mikhailov2007non,higuchi2017light,hong2013optical,luo2017opto,soavi2018broadband,ventura2020study}
Here we simulate low-order optical susceptibilities of graphene-hBN.
The hBN substrate is introduced to break the inversion symmetry to
allow non-zero LPGE and SHG.

Three types of photocurrent simulations using three types of electronic
Hamiltonians are carried out:

(i) ``TB'': Minimum tight-binding Hamiltonian (as in Ref. \citenum{ventura2020study})
with two atomic orbitals (in the unitcell) and two energy parameters
- gap $E_{g}$=0.0416 eV (same as our DFT value) and a nearest-neighbor
hopping parameter $t$=2.8 eV. A $t$ around 2.8 eV has been commonly
used to model graphene. Note that if choosing $E_{g}$=0.03 eV as
in Ref. \citenum{ventura2020study}, we can reproduce their LPGE and
LHG spectra;

(ii) ``Wannier A'': Minimum \textit{ab initio} Wannier-interpolated
Hamiltonian with two WFs (in the unitcell), which reproduces DFT eigenvalues
within the energy window {[}$E_{F}$-1 eV, $E_{F}$+3 eV{]}. From
Fig. \ref{fig:gr-hbn}(a), it can be seen that ``Wannier A'' nicely
reproduces DFT bands with tiny errors around Dirac cones;

(iii) ``Wannier B'': \textit{ab initio} Wannier-interpolated Hamiltonian
with 20 WFs, which reproduces DFT eigenvalues within {[}$E_{F}$-6
eV, $E_{F}$+7.7 eV{]}. From Fig. \ref{fig:gr-hbn}(a), it can be
seen that ``Wannier B'' perfectly reproduces DFT bands.

From Fig. \ref{fig:gr-hbn}(b)-(d), we find that ``Wannier A'' and
``Wannier B'' results of LPGE, SHG and THG coefficients (susceptibilities)
are identical, and TB leads to qualitatively similar results. The
curves of TB results have the same shapes as \textit{ab initio} results
based on Wannier functions, but there are quantitative differences
and the ratios of TB results to \textit{ab initio} results range from
63\% to 123\%. Therefore, our results indicate that a minimum TB
model and a minimum Wannierization setup are good for qualitative
studies and quantitative simulations respectively of LPGE and LHG
(within the RTA) of graphene-hBN. This conclusion however may not
applicable if band structures are complicated and/or spin-orbit coupling
plays a crucial role, in which cases sophisticated \textit{ab initio}
Wannierization setups are required. Additionally, it is found that
TB results are insensitive to the nearest-neighbor hoping parameter
$t$ (not shown), so that to cure the differences between TB and \textit{ab
initio} results, the so-called ``${\bf \widehat{r}}$-hopping''
corrections\citep{ibanez2022assessing} and/or farther-neighbor hoppings
are probably needed.

We next investigate the response of several different photon processes
for low-frequency LPGE and LHG of the semiconducting graphene-hBN.
From Fig. \ref{fig:gr-hbn}(b), the LPGE spectrum shows a one-photon
resonant peak - a peak right above $\omega=E_{g}$. This is consistent
with the fact that the formula of LPGE coefficients contains a delta-like
function $d_{kab}^{\Gamma}\left(1\omega\right)$ (according to Eq.
\ref{eq:rho1}, \ref{eq:rho2-+} and \ref{eq:sigma2-+}), which has
a resonant energy at $\omega=\Delta_{kab}$. For SHG shown in Fig.
\ref{fig:gr-hbn}(c), it is found that its spectrum shows three peaks
- two one-photon resonant peaks around $\omega=E_{g}$ and one two-photon
resonant peak right above $2\omega=E_{g}$. This is because the formula
of SHG coefficients (Eq. \ref{eq:sigma2++}, \ref{eq:rho2++} and
\ref{eq:rho1}) contains both $d^{\Gamma}\left(2\omega\right)$ and
$d^{\Gamma}\left(1\omega\right)$. For THG, its spectrum (Fig. \ref{fig:gr-hbn}(d))
has a sharp three-photon resonant peak right above $3\omega=E_{g}$
corresponding to $d^{\Gamma}\left(3\omega\right)$ in the formula
of THG (Eq. \ref{eq:sigma3}, \ref{eq:rho3}, \ref{eq:rho2++} and
\ref{eq:rho1}). On the other hand, the THG spectrum shows less clear
features for two-photon processes and no obvious features for one-photon
processes: (i) The second peak of the THG spectrum is a bit away from
$2\omega=E_{g}$ and relatively broad; (ii) THG coefficients around
$\omega=E_{g}$ are much weaker than its maximum value.

\begin{figure*}
\includegraphics[scale=0.26]{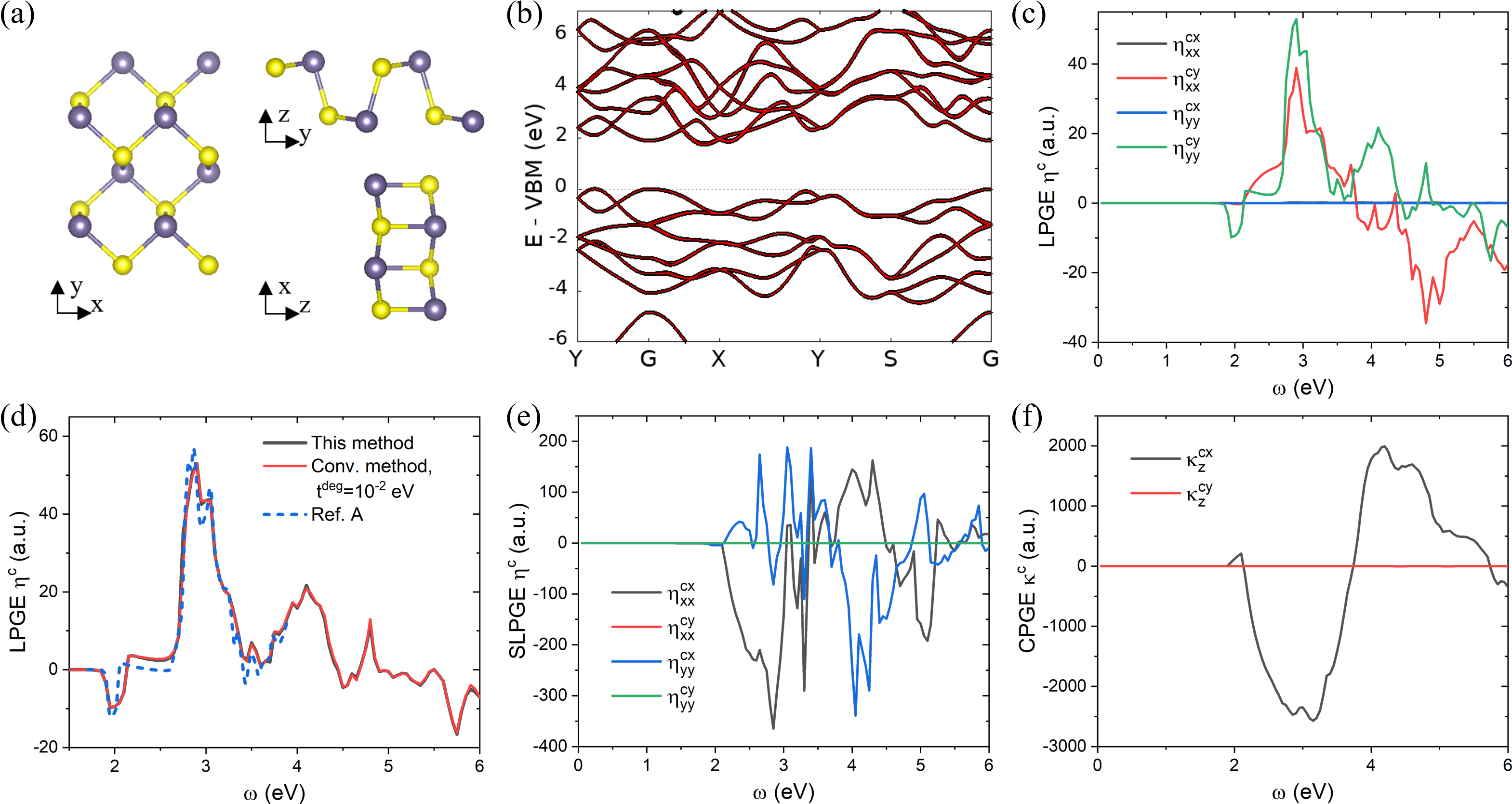}

\caption{Theoretical results of monolayer GeS. (a) Top and side views of the
structure. (b) DFT (black lines) and Wannier (red lines) band structures.
(c) LPGE coefficients. (d) $\eta_{yy}^{cy}$ by our method and the
conventional method with $t^{\mathrm{deg}}=10^{-2}$ eV, compared
with previous theoretical results. Ref. A is Ref. \citenum{ibanez2018ab}.
(e) and (f) are calculated SLPGE and CPGE coefficients respectively.
$\Gamma$=0.01 eV/$\hbar$.\label{fig:ges}}
\end{figure*}

\begin{figure*}
\includegraphics[scale=0.24]{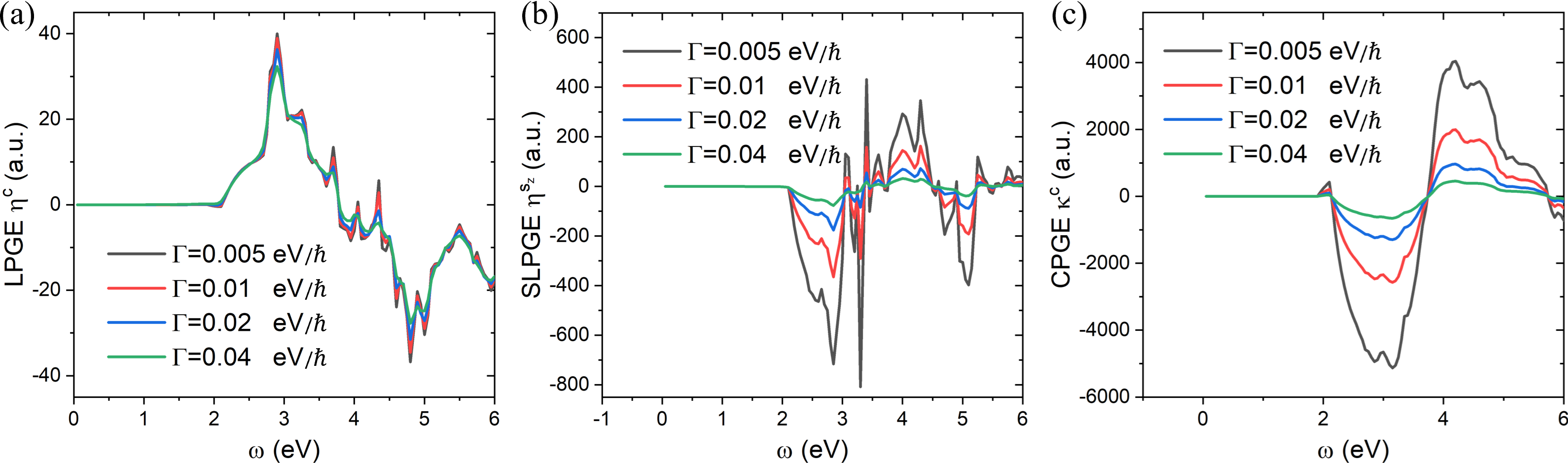}

\caption{Relaxation rate $\Gamma$ dependence of calculated (a) LPGE, (b) SLPGE
and (d) CPGE coefficients of monolayer GeS.\label{fig:ges-gamma}}
\end{figure*}

\begin{figure*}
\includegraphics[scale=0.4]{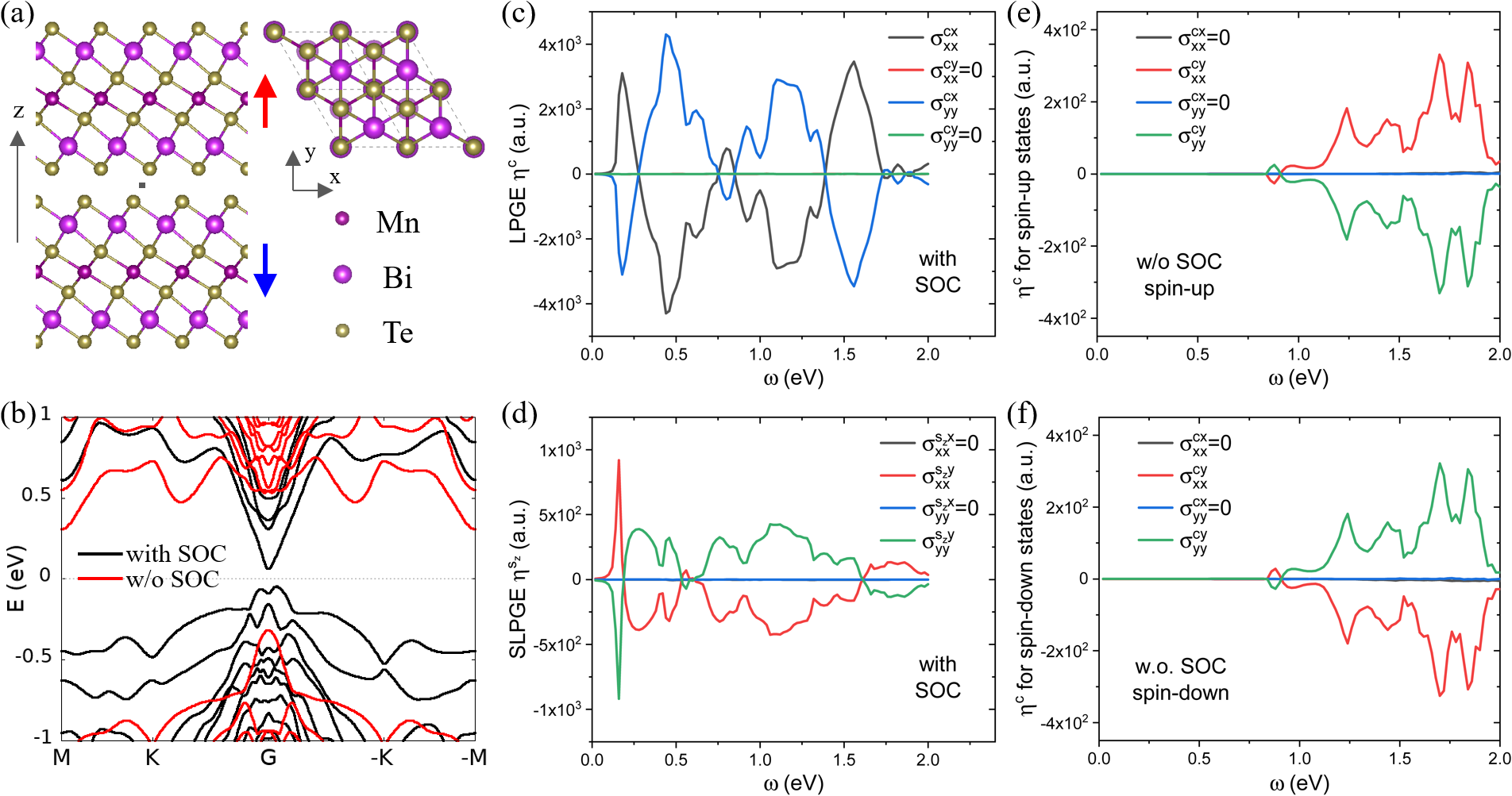}

\caption{Theoretical results of bilayer AFM MBT. (a) Side and top views of
its crystal structure. The system has the so-called $\mathcal{PT}$
symmetry, i.e., the system is invariant if inversion operation $\mathcal{P}$
and time-reversal operation $\mathcal{T}$ are applied together. The
inversion center (without considering magnetic moments) is located
between two layers (black square). Red and blue arrows indicate the
magnetic moment directions of the top and bottom MBT layers respectively.
(b) Band structures with (black lines) and without (red lines) SOC.
(c) and (d) are calculated LPGE and SLPGE coefficients with SOC respectively.
(e) and (f) are calculated LPGE coefficients for spin-up and spin-down
states without SOC respectively. $\Gamma$=0.01 eV/$\hbar$.\label{fig:mbt}}
\end{figure*}

\begin{figure*}
\includegraphics[scale=0.6]{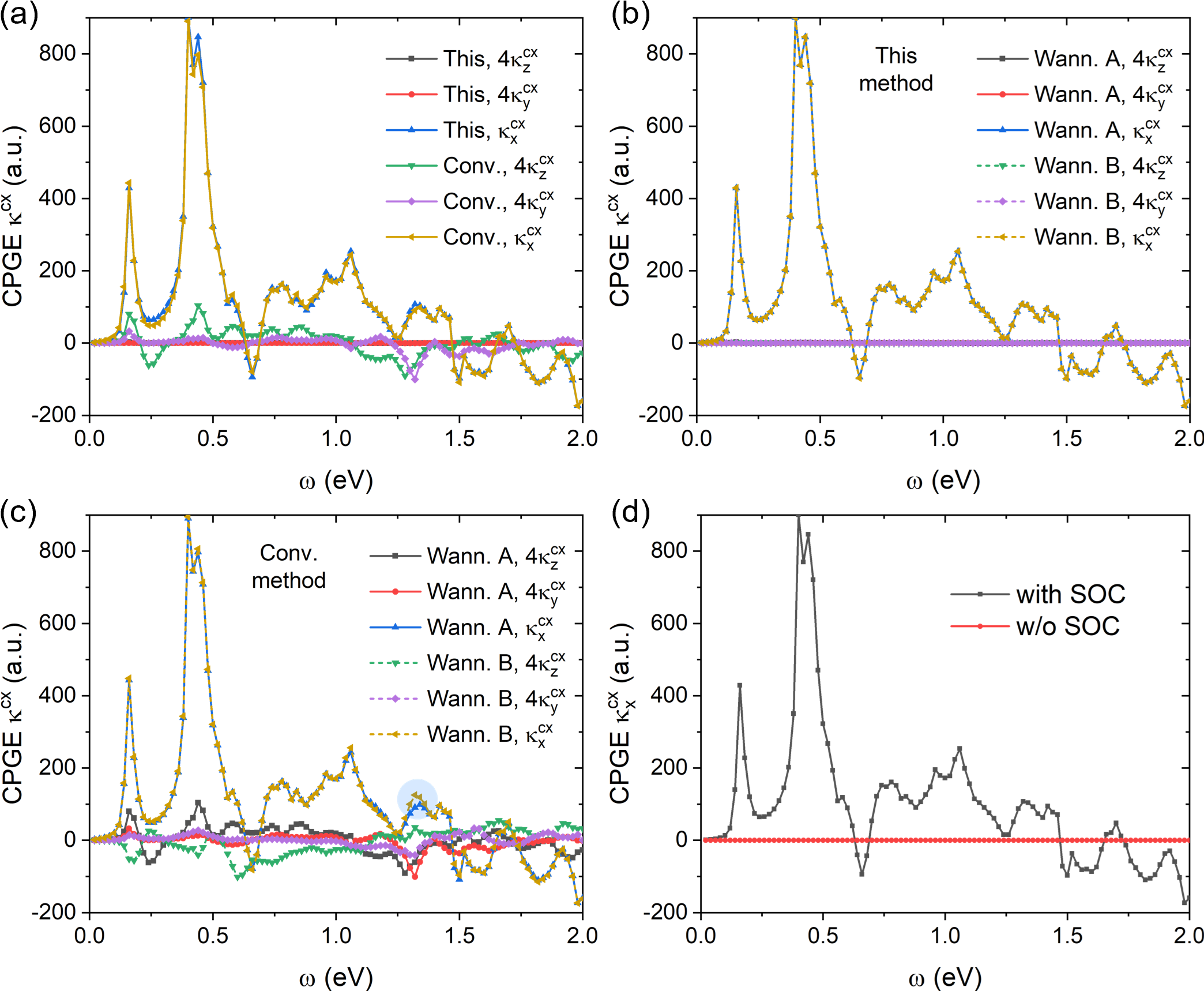}

\caption{Calculated CPGE coefficients $\kappa^{cx}$ of bilayer AFM MBT. (a)
$\kappa^{cx}$ by our method (labelled as ``This'' or ``This method'')
and the conventional method (labelled as ``Conv.'' or ``Conv. method'').
(b) and (c) $\kappa^{cx}$ by our method and the conventional method
respectively with two Wannierization setups - ``Wann. A'' and ``Wann.
B''. (d) $\kappa_{x}^{cx}$ by our method with and without SOC. ``Wann.
A'' and ``B'' have the same number of WFs and energy window but
use different random initial WFs. $\kappa_{z}^{cx}$ and $\kappa_{y}^{cx}$
are multiplied by 4 to make their corresponding numerical errors clearer.
Note that both $\kappa_{z}^{cx}$ and $\kappa_{y}^{cx}$ are expected
zero due to the combined $\mathcal{P}\mathcal{M}_{x}$ symmetry of
bilayer AFM MBT\citep{xu2021pure}. $\mathcal{M}_{x}$ means an in-plane
mirror symmetry.\label{fig:mbt_cpge}}
\end{figure*}

\subsection{A 2D material: (Spin) LPGE of monolayer WS$_{2}$}

Besides graphene, transition metal dichalcogenides (TMDs) are another
important class of 2D materials. Optical (spin-)current generation
is critical to the TMD-based electronic and spintronic applications
and has been extensively studied experimentally and theoretically.\citep{buscema2015photocurrent,eginligil2015dichroic,xie2016manipulating,wang2017first,xu2021pure}

Here we study both LPGE and spin LPGE (SLPGE) of monolayer 2H WS$_{2}$.
As shown in Fig. \ref{fig:ws2}(b), a high-quality Wannierization
is achieved, which ensures the accuracy of our \textit{ab initio}
simulations. We first investigate the effects of SOC on LPGE coefficients.
From Fig. \ref{fig:ws2}(c), our calculated LPGE spectrum is in agreement
with previous theoretical results\citep{wang2017first} and the SOC
effects are found significant. Most importantly, the first peak near
$\omega=E_{g}$ of the LPGE spectrum without SOC is splitted to two
by SOC (labeled as ``V1'' and ``V2'' in Fig. \ref{fig:ws2}(c)),
and the splitting of two peaks is close to the SOC band splitting
between two highest valence bands at ${\bf K}$, $\sim$0.43 eV.

Therefore, we include SOC in further \textit{ab initio} simulations
of LPGE and SLPGE coefficients of monolayer WS$_{2}$, shown in Fig.
\ref{fig:ws2}(d) and (e). Note that spin current is only present
when SOC is turned on. We find that charge and spin currents are perpendicular
to each other - charge current is along $y$ direction while spin
current is along $x$ direction under linearly polarized light. This
means a pure spin current (along $x$ direction) is generated by SLPGE.
This phenomenon is due to the different selection rules on charge
and spin currents in the presence of the in-plane mirror symmetry
$\mathcal{M}_{x}$: $k_{x}\rightarrow-k_{x}$, which leads to the
absence of $\eta_{xx}^{cx}$, $\eta_{yy}^{cx}$, $\eta_{xx}^{s_{z}y}$
and $\eta_{yy}^{s_{z}y}$.\citep{xu2021pure}

Additionally, in Fig. \ref{fig:ws2}(d), we compare LPGE coefficients,
which is fully determined by the shift current for undoped WS$_{2}$,
calculated by both our and the conventional methods with the same
computational setups. It is found that two methods predict identical
results. Different from GaAs, we find that theoretical results are
almost independent of $\Gamma^{\left(2\right)}$ and $t^{\mathrm{deg}}$,
indicating that the degeneracy errors are negligible for WS$_{2}$.

\subsection{A 2D material: LPGE and CPGE of 2D ferroelectric GeS}

Recently, ferroelectric group-IV monochalcogenide monolayers have
attracted growing interests due to their exciting properties, such
as selective valley excitations, valley Hall effects, and persistent
spin helix behavior.\citep{barraza2021colloquium} They also show
interesting nonlinear optical properties including an unusual strong
SHG intensity and large BPVE.\citep{barraza2021colloquium}

Here we study LPGE, SLPGE and CPGE of a group-IV monochalcogenide
monolayer - monolayer GeS. Our results shown in Fig. \ref{fig:ges}
are in agreement with previous theoretical ones\citep{ibanez2018ab,mu2021pure,panday2019injection},
e.g., our calculated LPGE is in agreement with that of Ref. \citenum{ibanez2018ab}
(Fig. \ref{fig:ges}(d)). We also compare LPGE by our method and the
conventional method (Fig. \ref{fig:ges}(d)). It is found that with
$t^{\mathrm{deg}}=\hbar\Gamma^{\left(2\right)}=\hbar\Gamma$=0.01
eV, two methods predict very similar results. This observation is
similar to GaAs case. Additionally, we find that with $t^{\mathrm{deg}}=\hbar\Gamma^{\left(2\right)}=10^{-4}$
eV and $\hbar\Gamma$=0.01 eV, theoretical results by two methods
are also consistent but are slightly different from those with $t^{\mathrm{deg}}=\hbar\Gamma^{\left(2\right)}=\hbar\Gamma$=0.01
eV when k meshes are converged, as we have checked. Therefore, calculated
LPGE coefficients by our method and the conventional method are $\Gamma^{\left(2\right)}$-
and $t^{\mathrm{deg}}$-dependent respectively, which indicates that
the treatment of near-degenerate subspaces has important effects on
LPGE coefficients.

Similar to monolayer WS$_{2}$, from Fig. \ref{fig:ges}(c) and (e),
we find pure spin currents perpendicular to charge currents, which
is again due to the presence of in-plane mirror symmetry $\mathcal{M}_{x}$:
$k_{x}\rightarrow-k_{x}$. Further, we observe strong CPGE (Fig. \ref{fig:ges}(f)),
40 times stronger than LPGE. According to previous theoretical works\citep{watanabe2021chiral,dai2023recent,panday2019injection},
as monolayer GeS is nonmagnetic, its CPGE is mainly attributed to
the injection current, which is the intraband-interband contribution
to BPVE (see Sec. \ref{subsec:different-parts-of-BPVE}). Strong CPGE
due to injection current has been predicted in ferroelectric group-IV
monochalcogenide monolayers including GeS, GeSe, SnS and SnSe, and
it is attributed to various factors in these materials such as anisotropy,
in-plane polarization and wave function delocalization.\citep{panday2019injection}

It is well known that optical susceptibilities due to injection current
are proportion to relaxation time $\tau=1/\Gamma$ if $\Gamma>0$.\citep{watanabe2021chiral}
Therefore, we next examine the $\Gamma$ dependence of CPGE coefficients
as well as LPGE and SLPGE. From Fig. \ref{fig:ges-gamma}(a), it is
found that LPGE is independent of $\Gamma$. This is because LPGE
of a nonmagnetic semiconducting material such as monolayer GeS should
be dominated by the shift current contribution, which is known independent
of $\Gamma$.\citep{watanabe2021chiral,dai2023recent} Calculated
CPGE coefficients $\kappa^{c}$ are found proportional to $1/\Gamma$
(Fig. \ref{fig:ges-gamma}(c)) as expected. Calculated SLPGE coefficients
$\eta^{s_{z}}$ are found proportional to $1/\Gamma$ (Fig. \ref{fig:ges-gamma}(d)),
which is probably because SLPGE is also dominated by the injection
current (the same as CPGE) as discussed in Ref. \citenum{lihm2022comprehensive}.

\begin{figure*}
\includegraphics[scale=0.263]{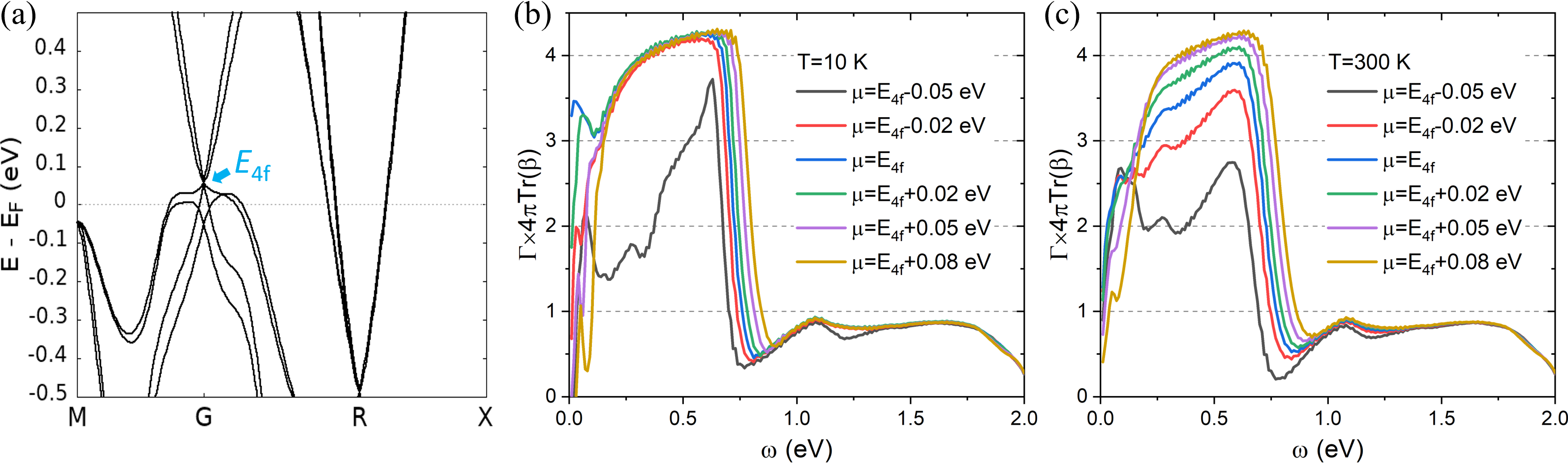}

\caption{Theoretical results of RhSi with SOC. (a) Band structure. $E_{\mathrm{4f}}$
is the energy of the four-fold degenerate point. (b) and (c) are traces
of CPGE coefficients $\kappa^{c}$ multiplied by a factor of $4\pi\Gamma$
at different chemical potentials ($\mu$) at 10 K and 300 K respectively.
$\Gamma$=0.01 eV/$\hbar$.\label{fig:rhsi}}
\end{figure*}

\subsection{A magnet: Bilayer AFM MBT}

Recently, various exotic BPVE properties have been predicted for AFM
systems with the so-called $\mathcal{PT}$ symmetry, which means the
systems are invariant if inversion operation $\mathcal{P}$ and time-reversal
operation $\mathcal{T}$ are applied together.\citep{watanabe2021chiral,fei2021p,xu2021pure}
Here we apply our method to simulate LPGE and SLPGE of bilayer AFM
MBT, which has the $\mathcal{PT}$ symmetry. Similar to Ref. \citenum{xu2021pure},
we have the following observations from Fig. \ref{fig:mbt}: (i) SOC
affects both band structure and BPVE significantly; (ii) Pure spin
currents are present regardless of SOC. This is due to the different
selection rules on charge and spin-$z$ currents in the presence of
the $\mathcal{PM}_{x}$ symmetry.\citep{xu2021pure} (iii) With SOC,
the charge current is present (see Fig. \ref{fig:mbt}(b)) and perpendicular
to spin-$z$ current (see Fig. \ref{fig:mbt}(d)). However, without
SOC, the charge current, which is the sum of the spin-up and spin-down
currents, is absent (see Fig. \ref{fig:mbt}(e) and (f)). This is
because the so-called inversion-spin-rotation $\mathcal{PS}$ symmetry
is satisfied without SOC but is broken if SOC is turned on, according
to Ref. \citenum{xu2021pure}.

In Fig. \ref{fig:mbt_cpge}, we show CPGE coefficients $\kappa^{cx}$,
due to the gyration current (Eq. \ref{eq:gyration_current}), calculated
by our and the conventional methods. Due to the $\mathcal{PM}_{x}$
symmetry, $\kappa_{z}^{cx}$ and $\kappa_{y}^{cx}$ should be zero
and $\kappa_{x}^{cx}$ can be nonzero\citep{xu2021pure}. From Fig.
\ref{fig:mbt_cpge}(a), we first find that calculated $\kappa_{x}^{cx}$
are quite large and the differences of $\kappa_{x}^{cx}$ by two methods
are small compared with the magnitudes of $\kappa_{x}^{cx}$. However,
for $\kappa_{z}^{cx}$ and $\kappa_{y}^{cx}$, we find that theoretical
results by two methods have non-negligible differences. Results by
our method are found zero as expected (with maximum error $\sim$0.5
a.u.). On the other hand, results by the conventional method show
non-negligible errors (with maximum error $\sim$25 a.u.) and the
errors cannot be removed by changing $t^{\mathrm{deg}}$ or increasing
k meshes, as we have checked. Moreover, from Fig. \ref{fig:mbt_cpge}(b)
and (c), it is found that $\kappa^{cx}$ by our method are independent
of the Wannierization setups. However, for the conventional method,
$\kappa_{x}^{cx}$ with different Wannierization setups have non-negligible
differences around $\omega=$1.3 eV, and $\kappa_{z}^{cx}$ ($\kappa_{y}^{cx}$)
with different Wannierization setups are completely different.

Considering that all bands of bilayer AFM MBT are Kramers degenerate
due to the $\mathcal{PT}$ symmetry, the $t^{\mathrm{deg}}$-independence
and the Wannierization-setup-dependence of the CPGE results by the
conventional method probably indicate gauge-dependent (here gauge
means the choice of eigenstates) numerical errors within the Kramers-degenerate
subspaces. Similar conclusion was previously obtained for bilayer
AFM CrI$_{3}$ with the $\mathcal{PT}$ symmetry in Ref. \citenum{chen2022basic}.
Our results indicate that compared with the conventional method, our
method may significantly reduce numerical errors of the BPVE simulations
of magnetic systems with the $\mathcal{PT}$ symmetry.

Additionally, we show the SOC effects in Fig. \ref{fig:mbt_cpge}(d).
It can be seen that $\kappa_{x}^{cx}$ becomes zero if SOC is turned
off, similar to the SOC effects of LPGE discussed above.

\subsection{A topological Weyl semimetal: Quantized CPGE in RhSi}

CPGE serves as an invaluable tool to detect the chirality, topological
charge, symmetries and other properties of topological Weyl semimetals.\citep{de2017quantized,le2020ab,de2020difference,rees2020helicity}
Previously, quantized CPGE has been studied theoretically considering
only the injection current contribution, and via model Hamiltonians\citep{de2017quantized}
or Wannier-function-based \textit{ab initio} methods with the so-called
diagonal tight-binding approximation (DTBA)\citep{le2020ab,de2020difference},
in which $\xi^{W}$ is treated approximately\citep{ibanez2018ab}.

The so-called quantized-CPGE suggests that the relation: $\mathrm{Tr}\left[\beta\left(\omega\right)\right]=i\pi\frac{e^{3}}{h^{2}}C_{L}$
with $C_{L}$ topological charge, is satisfied in a certain photon-frequency
range.\citep{de2017quantized,le2020ab} For the injection current
(which dominates quantized CPGE) with a finite relaxation time $\tau=1/\Gamma$,
there is $\beta\left(\omega\right)=i\Gamma\kappa^{c}\left(\omega\right)$.\citep{watanabe2021chiral,dai2023recent}
Therefore, the quantized-CPGE relation becomes: $4\pi\mathrm{Tr}\left[\Gamma\kappa^{c}\right]=C_{L}$
in atomic units.

In this work, we apply our \textit{ab initio} method to simulate CPGE
of RhSi at various temperatures and chemical potentials. We have gone
beyond DTBA and considered photocurrent contributions beyond just
the injection current. From our calculated CPGE spectra in Fig. \ref{fig:rhsi},
we observe that $4\pi\mathrm{Tr}\left[\Gamma\kappa^{c}\left(\omega\right)\right]\approx4$
is satisfied in a relatively wide photon-energy range {[}0.3, 0.6{]}
eV at both low (Fig. \ref{fig:rhsi}(b)) and high (Fig. \ref{fig:rhsi}(c))
temperatures, if the chemical potential is not too low. Our results
suggest that it seems easier to observe quantized CPGE at lower temperatures
and higher chemical potentials.

\section{Summary and outlooks}

We have developed an \textit{ab initio} method based on Wannier functions
for simulating weak-field BPVE and LHG in solids. The method is of
great predictive power and widely applicable to semiconductors and
metals with arbitrary band structures for both linearly and circularly
polarized light. We have demonstrated its power through its applications
into the simulations of (S)LPGE, (S)CPGE and LHG in various types
of systems.

This method has the potential to be greatly improved in various directions,
such as: (i) By introducing a static electric field, which can be
done straightforwardly, the so-called jerk current\citep{fregoso2018jerk}
can be simulated. (ii) The scattering term within the RTA with a global
constant relaxation time $\tau=1/\Gamma$ can be replaced by the fully
\textit{ab initio} sophisticated scattering term developed in our
previous works\citep{xu2023ab,xu2021ab}, so that the energy-, ${\bf k}$-
and transition-resolved relaxation and decoherence are properly considered.
This generation may have important effects on quantized CPGE, which
is predicted within the RTA with a global constant $\tau$. (iii)
By solving the density matrix non-perturbatively via real-time dynamics,
the photocurrent at stronger fields can be simulated.

\section*{Acknowledgments}

Junqing Xu thanks Ravishankar Sundararaman for helpful discussions.
This work is supported by National Natural Science Foundation of China
(Grant No. 12304214 and No. 12104122), Fundamental Research Funds
for Central Universities (Grant No. JZ2023HGPA0291), Anhui Provincial
Natural Science Foundation (Grant No. 1908085QA16). This research
used resources of the HPC Platform of Hefei University of Technology.

\section*{Appendices}

\subsection*{Appendix A: The derivations of the explicit forms of different contributions
to BPVE}

From Eq. \ref{eq:rho1} and Appendix C, we have
\begin{align}
\widetilde{\rho}_{\alpha,kaa}^{E,\left(1\right),d}\left(\omega\right)= & \frac{ie}{-\hbar\omega+i\hbar\Gamma}\frac{df_{ka}^{\mathrm{eq}}}{dk_{\alpha}},\label{eq:rho1_diag}\\
\widetilde{\rho}_{\alpha,kab}^{E,\left(1\right),o}\left(\omega\right)= & -e\xi_{\alpha,kab}f_{kab}^{\mathrm{eq}}d_{kab}^{\Gamma}\left(\omega\right),\label{eq:rho1_off}\\
f_{kab}^{\mathrm{eq}}= & f_{ka}^{\mathrm{eq}}-f_{kb}^{\mathrm{eq}},
\end{align}

where we have assumed that $\Gamma_{kab}$ are all equal to the same
constant $\Gamma$ for simplicity. Note that $\rho_{\alpha,kab}^{\left(1\right),o}\left(\omega\right)=0$
when $\epsilon_{ka}=\epsilon_{kb}$.

Then, we can obtain different parts of $\sigma_{\alpha_{1}\alpha_{2}}^{\mathrm{DC},c,\beta}\left(\omega\right)$:

(1) The $dd$ intraband-intraband part.

From Eq. \ref{eq:sigma2-+} and \ref{eq:rho2-+}, this part reads

\begin{align}
\sigma_{\alpha_{1}\alpha_{2}}^{c,\beta,dd}\left(-\omega,\omega\right)= & -eV_{\mathrm{cell}}^{-1}\mathrm{Tr}\left\{ v_{\beta}^{d}\widetilde{\rho}_{\alpha_{1}\alpha_{2}}^{E,\left(2\right),dd}\left(-\omega,\omega\right)\right\} \nonumber \\
= & -ie^{2}V_{\mathrm{cell}}^{-1}\nonumber \\
 & \times\mathrm{Tr}\left\{ v_{\beta}^{d}\left[\frac{D\widetilde{\rho}_{\alpha_{2}}^{E,\left(1\right),d}\left(\omega\right)}{Dk_{\alpha_{1}}}\odot d^{\Gamma}\left(0\right)\right]\right\} .
\end{align}

Then, from \ref{eq:rho1_diag} and \ref{eq:cov-der}, we obtain
\begin{align}
\sigma_{\alpha_{1}\alpha_{2}}^{c,\beta,dd}\left(-\omega,\omega\right)= & \frac{e^{3}V_{\mathrm{cell}}^{-1}N_{k}^{-1}}{\hbar^{2}i\Gamma\left(-\omega+i\Gamma\right)}\sum_{ka}v_{\beta,kaa}\frac{d^{2}f_{ka}^{\mathrm{eq}}}{dk_{\alpha_{1}}dk_{\alpha_{2}}},
\end{align}

thus,
\begin{align}
\sigma_{\alpha_{1}\alpha_{2}}^{\mathrm{DC},c,\beta,dd}\left(\omega\right)= & \frac{1}{2}\left(\sigma_{\alpha_{1}\alpha_{2}}^{c,\beta,dd}\left(-\omega,\omega\right)+\sigma_{\alpha_{2}\alpha_{1}}^{c,\beta,dd}\left(\omega,-\omega\right)\right),\nonumber \\
= & \frac{-e^{3}V_{\mathrm{cell}}^{-1}N_{k}^{-1}}{\hbar^{2}\left(\omega^{2}+\Gamma^{2}\right)}\sum_{ka}v_{\beta,kaa}\frac{d^{2}f_{ka}^{\mathrm{eq}}}{dk_{\alpha_{1}}dk_{\alpha_{2}}}.
\end{align}

(2) The $od$ interband-intraband part.

This part reads

\begin{align}
\sigma_{\alpha_{1}\alpha_{2}}^{c,\beta,od}\left(-\omega,\omega\right)= & -ie^{2}V_{\mathrm{cell}}^{-1}\nonumber \\
 & \times\mathrm{Tr}\left\{ v_{\beta}^{o}\left(\frac{D\widetilde{\rho}_{\alpha_{2}}^{E,\left(1\right),d}\left(\omega\right)}{Dk_{\alpha_{1}}}\odot d^{\Gamma}\left(0\right)\right)\right\} .
\end{align}

From \ref{eq:rho1_diag}, \ref{eq:cov-der} and \ref{eq:relation_xi_and_v},
we then obtain
\begin{align}
\sigma_{\alpha_{1}\alpha_{2}}^{c,\beta,od}\left(-\omega,\omega\right)= & -e^{2}V_{\mathrm{cell}}^{-1}\nonumber \\
 & \times\mathrm{Tr}\left\{ v_{\beta}^{o}\left(\left[\xi_{\alpha_{1}},\widetilde{\rho}_{\alpha_{2}}^{E,\left(1\right),d}\left(\omega\right)\right]\odot d^{\Gamma}\left(0\right)\right)\right\} \nonumber \\
= & \frac{-e^{3}V_{\mathrm{cell}}^{-1}}{-\hbar\omega+i\hbar\Gamma}\nonumber \\
 & \times\mathrm{Tr}\left\{ \left(\xi_{\beta}^{o}\odot\Delta\right)\left(\left[\frac{df^{\mathrm{eq}}}{dk_{\alpha}},\xi_{\alpha_{1}}\right]\odot d^{\Gamma}\left(0\right)\right)\right\} \nonumber \\
= & \frac{e^{3}V_{\mathrm{cell}}^{-1}N_{k}^{-1}}{-\hbar\omega+i\hbar\Gamma}\nonumber \\
 & \times\sum_{k,ab}\xi_{\beta,kba}^{o}\xi_{\alpha_{1},kab}^{o}\frac{df_{kab}^{\mathrm{eq}}}{dk_{\alpha_{2}}}\Delta_{kab}d_{kab}^{\Gamma}\left(0\right).
\end{align}

(3) The $do$ intraband-interband part.

This part reads

\begin{align}
\sigma_{\alpha_{1}\alpha_{2}}^{c,\beta,do}\left(-\omega,\omega\right)= & -ie^{2}V_{\mathrm{cell}}^{-1}\nonumber \\
 & \times\mathrm{Tr}\left\{ v_{\beta}^{d}\left(\frac{D\widetilde{\rho}_{\alpha_{2}}^{E,\left(1\right),o}\left(\omega\right)}{Dk_{\alpha_{1}}}\odot d^{\Gamma}\left(0\right)\right)\right\} .
\end{align}

From Eq. \ref{eq:rho1_off} and \ref{eq:cov-der}, we then obtain
\begin{align}
\sigma_{\alpha_{1}\alpha_{2}}^{c,\beta,do}\left(-\omega,\omega\right)= & \frac{-e^{2}V_{\mathrm{cell}}^{-1}}{i\hbar\Gamma}\mathrm{Tr}\left\{ v_{\beta}^{d}\left[\xi_{\alpha_{1}},\widetilde{\rho}_{\alpha_{2}}^{E,\left(1\right),o}\left(\omega\right)\right]\right\} \nonumber \\
= & \frac{-e^{2}V_{\mathrm{cell}}^{-1}}{i\hbar\Gamma}\mathrm{Tr}\left\{ \left[v_{\beta}^{d},\xi_{\alpha_{1}}\right]\widetilde{\rho}_{\alpha_{2}}^{E,\left(1\right),o}\left(\omega\right)\right\} \nonumber \\
= & \frac{ie^{3}V_{\mathrm{cell}}^{-1}N_{k}^{-1}}{\hbar\Gamma}\sum_{kab}\xi_{\alpha_{1},kba}\xi_{\alpha_{2},kab}\nonumber \\
 & \times\left(v_{\beta,aa}-v_{\beta,bb}\right)f_{kab}^{\mathrm{eq}}d_{kab}^{\Gamma}\left(\omega\right),
\end{align}

thus,
\begin{align}
\sigma_{\alpha_{1}\alpha_{2}}^{\mathrm{DC},c,\beta,do}\left(\omega\right)= & \frac{ie^{3}V_{\mathrm{cell}}^{-1}N_{k}^{-1}}{2\hbar\Gamma}\sum_{kab}\left(v_{\beta,aa}-v_{\beta,bb}\right)f_{kab}^{\mathrm{eq}}\nonumber \\
 & \times\left\{ \begin{array}{c}
\xi_{\alpha_{2},kab}\xi_{\alpha_{1},kba}d_{kab}^{\Gamma}\left(\omega\right)\\
+\xi_{\alpha_{1},kab}\xi_{\alpha_{2},kba}d_{kab}^{\Gamma}\left(-\omega\right)
\end{array}\right\} \nonumber \\
= & \frac{ie^{3}V_{\mathrm{cell}}^{-1}N_{k}^{-1}}{2\hbar\Gamma}\sum_{kab}\xi_{\alpha_{2},kab}\xi_{\alpha_{1},kba}\nonumber \\
 & \times\left(v_{\beta,aa}-v_{\beta,bb}\right)f_{kab}^{\mathrm{eq}}\nonumber \\
 & \times\left\{ d_{kab}^{\Gamma}\left(\omega\right)+d_{kba}^{\Gamma}\left(-\omega\right)\right\} .
\end{align}

Since
\begin{align}
d_{kab}^{\Gamma}\left(\omega\right)+d_{kba}^{\Gamma}\left(-\omega\right)= & \frac{-2i\hbar\Gamma}{\left(-\hbar\omega-\Delta_{kab}\right)^{2}+\left(\hbar\Gamma\right)^{2}}
\end{align}

and
\begin{align}
\delta^{\Gamma}\left(\hbar\omega\right)= & \frac{1}{\pi}\frac{\hbar\Gamma}{\left(\hbar\omega\right)^{2}+\left(\hbar\Gamma\right)^{2}},
\end{align}

we have
\begin{align}
\sigma_{\alpha_{1}\alpha_{2}}^{\mathrm{DC},c,\beta,do}\left(\omega\right)= & \frac{e^{3}\pi V_{\mathrm{cell}}^{-1}N_{k}^{-1}}{\hbar\Gamma}\sum_{kab}\xi_{\alpha_{2},kab}\xi_{\alpha_{1},kba}\nonumber \\
 & \times\left(v_{\beta,kaa}-v_{\beta,kbb}\right)f_{kab}^{\mathrm{eq}}\delta^{\Gamma}\left(\hbar\omega+\Delta_{kab}\right).
\end{align}

(4) The $oo$ interband-interband part.

This part reads
\begin{align}
\sigma_{\alpha_{1}\alpha_{2}}^{c,\beta,oo}\left(-\omega,\omega\right)= & -ie^{2}V_{\mathrm{cell}}^{-1}\nonumber \\
 & \times\mathrm{Tr}\left\{ v_{\beta}^{o}\left(\frac{D\widetilde{\rho}_{\alpha_{2}}^{E,\left(1\right),o}\left(\omega\right)}{Dk_{\alpha_{1}}}\odot d^{\Gamma}\left(0\right)\right)\right\} \nonumber \\
= & \frac{e^{2}V_{\mathrm{cell}}^{-1}}{\hbar}\mathrm{Tr}\left(\xi_{\beta}^{\Gamma,o}\frac{D\widetilde{\rho}_{\alpha_{2}}^{E,\left(1\right),o}\left(\omega\right)}{Dk_{\alpha_{1}}}\right),
\end{align}

where
\begin{align}
\xi_{\beta,kab}^{\Gamma,o}= & -i\hbar v_{\beta,kab}^{o}d_{kba}^{\Gamma}\left(0\right),\nonumber \\
= & \xi_{\beta,kab}^{o}\frac{\Delta_{kab}}{\Delta_{kab}+i\hbar\Gamma}.
\end{align}

Since the relation $\mathrm{Tr}\left\{ A\frac{DB}{D{\bf k}}\right\} =-\mathrm{Tr}\left\{ \frac{DA}{D{\bf k}}B\right\} $
is satisfied for arbitrary matrices $A$ and $B$, we have
\begin{align}
\sigma_{\alpha_{1}\alpha_{2}}^{c,\beta,oo}\left(-\omega,\omega\right)= & \frac{-e^{2}V_{\mathrm{cell}}^{-1}}{\hbar}\mathrm{Tr}\left\{ \frac{D\xi_{\beta}^{\Gamma,o}}{Dk_{\alpha_{1}}}\widetilde{\rho}_{\alpha_{2}}^{E,\left(1\right),o}\left(\omega\right)\right\} \nonumber \\
= & \frac{e^{3}V_{\mathrm{cell}}^{-1}N_{k}^{-1}}{\hbar}\nonumber \\
 & \times\sum_{kab}\left(\frac{D\xi_{\beta,k}^{\Gamma,o}}{Dk_{\alpha_{1}}}\right)_{ba}\xi_{\alpha_{2},kab}f_{kab}^{\mathrm{eq}}d_{kab}^{\Gamma}\left(\omega\right).
\end{align}

To obtain the standard formulae of the shift and gyration current\citep{watanabe2021chiral},
we need to take the weak-scattering limit $\Gamma\rightarrow0$, so
that Eq. \ref{eq:xio_Gamma} is approximated as
\begin{align}
\xi_{\beta}^{\Gamma,o}\approx & \xi_{\beta}^{o}.
\end{align}

Using Eq. \ref{eq:scattering-free-limit}, we obtain
\begin{align}
\sigma_{\alpha_{1}\alpha_{2}}^{c,\beta,oo}\left(-\omega,\omega\right)= & \frac{-e^{2}V_{\mathrm{cell}}^{-1}}{\hbar}\mathrm{Tr}\left\{ \frac{D\xi_{\beta}^{o}}{Dk_{\alpha_{1}}}\widetilde{\rho}_{\alpha_{2}}^{E,\left(1\right),o}\left(\omega\right)\right\} \nonumber \\
= & \frac{e^{3}V_{\mathrm{cell}}^{-1}N_{k}^{-1}}{\hbar}\nonumber \\
 & \times\sum_{kab}\left(\frac{D\xi_{\beta,k}^{o}}{Dk_{\alpha_{1}}}\right)_{ba}\xi_{\alpha_{2},kab}f_{kab}^{\mathrm{eq}}d_{kab}^{\Gamma}\left(\omega\right).
\end{align}

Using the relation (derived using ${\bf \xi}_{\alpha,kab}=i\left\langle u_{ka}|\frac{du_{kb}}{dk_{\alpha}}\right\rangle $
and $\xi=\xi^{d}+\xi^{o}$)
\begin{align}
\left(\frac{D\xi_{\beta}^{o}}{Dk_{\alpha_{1}}}\right)^{o}= & \xi_{\alpha_{1};\beta}^{o},\label{eq:relation_DxiDk_xigd}
\end{align}

we have
\begin{align}
\sigma_{\alpha_{1}\alpha_{2}}^{c,\beta,oo}\left(-\omega,\omega\right)= & \frac{-e^{3}V_{\mathrm{cell}}^{-1}N_{k}^{-1}}{\hbar}\nonumber \\
 & \times\sum_{kab}\xi_{\alpha_{1};\beta,kab}^{o}\xi_{\alpha_{2},kba}^{o}f_{kab}^{\mathrm{eq}}d_{kba}^{\Gamma}\left(\omega\right).
\end{align}

Then, our obtained $\sigma_{\alpha_{1}\alpha_{2}}^{\mathrm{DC},c,\beta,oo}\left(\omega\right)=\left(\sigma_{\alpha_{1}\alpha_{2}}^{c,\beta,oo}\left(-\omega,\omega\right)+\sigma_{\alpha_{2}\alpha_{1}}^{c,\beta,oo}\left(\omega,-\omega\right)\right)/2$
is exactly the same as Eq. 93 of Ref. \citenum{watanabe2021chiral},
considering that their $q$ is $-e$ and they expand $\rho\left(t\right)=\sum_{m}\rho\left(m\omega\right)e^{-im\omega}$
different from our Eq. \ref{eq:rhot_expansion}.

As mentioned in Sec. \ref{subsec:different-parts-of-BPVE}, $\sigma_{\alpha_{1}\alpha_{2}}^{\mathrm{DC},c,\beta,oo}\left(\omega\right)$
can be separated into the principal and Dirac-delta parts. We focus
on the Dirac-delta part $\sigma_{\alpha_{1}\alpha_{2}}^{\mathrm{DC},c,\beta,oo,\delta}\left(\omega\right)$.
By replacing $d_{kba}^{\Gamma}\left(\omega\right)$ to $-i\pi\delta^{\Gamma}\left(\hbar\omega+\Delta_{kab}\right)$,
we obtain
\begin{align}
\sigma_{\alpha_{1}\alpha_{2}}^{\mathrm{DC},c,\beta,oo,\delta}\left(\omega\right)= & \frac{i\pi e^{3}V_{\mathrm{cell}}^{-1}N_{k}^{-1}}{2\hbar}\nonumber \\
 & \times\sum_{kab}\left\{ \begin{array}{c}
\xi_{\alpha_{1};\beta,kab}^{o}\xi_{\alpha_{2},kba}^{o}\\
\times f_{kab}^{\mathrm{eq}}\delta^{\Gamma}\left(\hbar\omega+\Delta_{kba}\right)\\
+\xi_{\alpha_{2};\beta,kab}^{o}\xi_{\alpha_{1},kba}^{o}\\
\times f_{kab}^{\mathrm{eq}}\delta^{\Gamma}\left(\hbar\omega-\Delta_{kba}\right)
\end{array}\right\} \nonumber \\
= & \frac{i\pi e^{3}V_{\mathrm{cell}}^{-1}N_{k}^{-1}}{2\hbar}\nonumber \\
 & \times\sum_{kab}\left\{ \begin{array}{c}
\xi_{\alpha_{1};\beta,kab}^{o}\xi_{\alpha_{2},kba}^{o}\\
\times f_{kab}^{\mathrm{eq}}\delta^{\Gamma}\left(\hbar\omega-\Delta_{kab}\right)\\
+\xi_{\alpha_{2};\beta,kab}^{o,*}\xi_{\alpha_{1},kba}^{o,*}\\
\times f_{kba}^{\mathrm{eq}}\delta^{\Gamma}\left(\hbar\omega-\Delta_{kab}\right)
\end{array}\right\} \nonumber \\
= & \frac{i\pi e^{3}V_{\mathrm{cell}}^{-1}N_{k}^{-1}}{2\hbar}\sum_{kab}f_{kab}^{\mathrm{eq}}\delta^{\Gamma}\left(\hbar\omega-\Delta_{kab}\right)\nonumber \\
 & \times\left\{ \xi_{\alpha_{1};\beta,kab}^{o}\xi_{\alpha_{2},kba}^{o}-\xi_{\alpha_{2};\beta,kab}^{o,*}\xi_{\alpha_{1},kba}^{o,*}\right\} .
\end{align}

Therefore, the shift current contribution to $\sigma_{\alpha_{1}\alpha_{2}}^{\mathrm{DC},c,\beta}\left(\omega\right)$
is the real part of $\sigma_{\alpha_{1}\alpha_{2}}^{\mathrm{DC},c,\beta,oo,\delta}\left(\omega\right)$:
\begin{align}
\sigma_{\alpha_{1}\alpha_{2}}^{\mathrm{shift},c,\beta}\left(\omega\right)= & \mathrm{Re}\left[\sigma_{\alpha_{1}\alpha_{2}}^{\mathrm{DC},c,\beta,oo,\delta}\left(\omega\right)\right]\nonumber \\
= & \frac{-\pi e^{3}V_{\mathrm{cell}}^{-1}N_{k}^{-1}}{2\hbar}\sum_{kab}f_{kab}^{\mathrm{eq}}\delta^{\Gamma}\left(\hbar\omega-\Delta_{kab}\right)\nonumber \\
 & \times\mathrm{Im}\left\{ \xi_{\alpha_{1};\beta,kab}^{o}\xi_{\alpha_{2},kba}^{o}+\xi_{\alpha_{2};\beta,kab}^{o}\xi_{\alpha_{1},kba}^{o}\right\} ,
\end{align}

This is the same as the shift current formula or Eq. 104 of Ref. \citenum{watanabe2021chiral},
again considering that their $q$ is $-e$ and they expand $\rho\left(t\right)=\sum_{m}\rho\left(m\omega\right)e^{-im}$
different from our Eq. \ref{eq:rhot_expansion}.

Equivalently by swapping $a$ and $b$, we have
\begin{align}
\sigma_{\alpha_{1}\alpha_{2}}^{\mathrm{shift},c,\beta}\left(\omega\right)= & \frac{-\pi e^{3}V_{\mathrm{cell}}^{-1}N_{k}^{-1}}{2\hbar}\sum_{kab}f_{kba}^{\mathrm{eq}}\delta^{\Gamma}\left(\hbar\omega-\Delta_{kba}\right)\nonumber \\
 & \times\mathrm{Im}\left\{ \xi_{\alpha_{1};\beta,kab}^{o,*}\xi_{\alpha_{2},kba}^{o,*}-\xi_{\alpha_{2};\beta,kab}^{o}\xi_{\alpha_{1},kba}^{o}\right\} \nonumber \\
= & \frac{-\pi e^{3}V_{\mathrm{cell}}^{-1}N_{k}^{-1}}{2\hbar}\sum_{kab}f_{kab}^{\mathrm{eq}}\delta^{\Gamma}\left(\hbar\omega+\Delta_{kab}\right)\nonumber \\
 & \times\mathrm{Im}\left\{ \xi_{\alpha_{1};\beta,kab}^{o}\xi_{\alpha_{2},kba}^{o}+\xi_{\alpha_{2};\beta,kab}^{o}\xi_{\alpha_{1},kba}^{o}\right\} .
\end{align}

Then, we also have
\begin{align}
\sigma_{\alpha_{1}\alpha_{2}}^{\mathrm{shift},c,\beta}\left(\omega\right)= & \frac{-\pi e^{3}V_{\mathrm{cell}}^{-1}N_{k}^{-1}}{4\hbar}\nonumber \\
 & \times\sum_{kab}f_{kab}^{\mathrm{eq}}\left\{ \delta^{\Gamma}\left(\hbar\omega+\Delta_{kab}\right)+\delta^{\Gamma}\left(\hbar\omega-\Delta_{kab}\right)\right\} \nonumber \\
 & \times\mathrm{Im}\left\{ \xi_{\alpha_{1};\beta,kab}^{o}\xi_{\alpha_{2},kba}^{o}+\xi_{\alpha_{2};\beta,kab}^{o}\xi_{\alpha_{1},kba}^{o}\right\} .
\end{align}

Similarly, the gyration current contribution to $\sigma_{\alpha_{1}\alpha_{2}}^{\mathrm{DC},c,\beta}\left(\omega\right)$
is the imaginary part of $\sigma_{\alpha_{1}\alpha_{2}}^{\mathrm{DC},c,\beta,oo,\delta}\left(\omega\right)$:
\begin{align}
\sigma_{\alpha_{1}\alpha_{2}}^{\mathrm{gyr},c,\beta}\left(\omega\right)= & \mathrm{Im}\left[\sigma_{\alpha_{1}\alpha_{2}}^{\mathrm{DC},c,\beta,oo,\delta}\left(\omega\right)\right]\nonumber \\
= & \frac{\pi e^{3}V_{\mathrm{cell}}^{-1}N_{k}^{-1}}{2\hbar}\sum_{kab}f_{kab}^{\mathrm{eq}}\delta^{\Gamma}\left(\hbar\omega-\Delta_{kab}\right)\nonumber \\
 & \times\mathrm{Re}\left\{ \xi_{\alpha_{1};\beta,kab}^{o}\xi_{\alpha_{2},kba}^{o}-\xi_{\alpha_{2};\beta,kab}^{o}\xi_{\alpha_{1},kba}^{o}\right\} .
\end{align}

Equivalently by swapping $a$ and $b$, we have
\begin{align}
\sigma_{\alpha_{1}\alpha_{2}}^{\mathrm{gyr},c,\beta}\left(\omega\right)= & \frac{-\pi e^{3}V_{\mathrm{cell}}^{-1}N_{k}^{-1}}{2\hbar}\sum_{kab}f_{kab}^{\mathrm{eq}}\delta^{\Gamma}\left(\hbar\omega+\Delta_{kab}\right)\nonumber \\
 & \times\mathrm{Re}\left\{ \xi_{\alpha_{1};\beta,kab}^{o}\xi_{\alpha_{2},kba}^{o}-\xi_{\alpha_{2};\beta,kab}^{o}\xi_{\alpha_{1},kba}^{o}\right\} .
\end{align}

Then, we also have
\begin{align}
\sigma_{\alpha_{1}\alpha_{2}}^{\mathrm{gyr},c,\beta}\left(\omega\right)= & \frac{-\pi e^{3}V_{\mathrm{cell}}^{-1}N_{k}^{-1}}{4\hbar}\nonumber \\
 & \times\sum_{kab}f_{kab}^{\mathrm{eq}}\left\{ \delta^{\Gamma}\left(\hbar\omega+\Delta_{kab}\right)-\delta^{\Gamma}\left(\hbar\omega-\Delta_{kab}\right)\right\} \nonumber \\
 & \times\mathrm{Re}\left\{ \xi_{\alpha_{1};\beta,kab}^{o}\xi_{\alpha_{2},kba}^{o}-\xi_{\alpha_{2};\beta,kab}^{o}\xi_{\alpha_{1},kba}^{o}\right\} .
\end{align}

\subsection*{Appendix B: The proof of Eq. \ref{eq:cov-der}}

From $\rho^{W}=U\rho U^{\dagger}$ and Eq. \ref{eq:D_for_xi} - ${\bf D}=U^{\dagger}\frac{dU}{d{\bf k}}$,
we have
\begin{align}
\frac{d\rho^{W}}{d{\bf k}}= & \frac{dU\rho U^{\dagger}}{d{\bf k}}\nonumber \\
= & \frac{dU}{d{\bf k}}\rho U^{\dagger}+U\frac{d\rho}{d{\bf k}}U^{\dagger}+U\rho\frac{dU^{\dagger}}{d{\bf k}}.\nonumber \\
= & U\left(\frac{d\rho}{d{\bf k}}+{\bf D}\rho+\rho{\bf D}^{\dagger}\right)U^{\dagger}\nonumber \\
= & U\left\{ \frac{d\rho}{d{\bf k}}+\left[{\bf D},\rho\right]\right\} U^{\dagger}.
\end{align}

With Eq. \ref{eq:xi} - $\xi=i{\bf D}+U^{\dagger}\xi^{W}U$, we then
have,
\begin{align}
\frac{D\rho}{D{\bf k}}= & \frac{d\rho}{d{\bf k}}-i\left[\xi,\rho\right]\nonumber \\
= & \frac{d\rho}{d{\bf k}}-i\left[i{\bf D}+U^{\dagger}\xi^{W}U,\rho\right]\nonumber \\
= & \frac{d\rho}{d{\bf k}}+\left[{\bf D},\rho\right]-i\left[U^{\dagger}\xi^{W}U,\rho\right]\nonumber \\
= & U^{\dagger}\frac{d\rho^{W}}{d{\bf k}}U-i\left[U^{\dagger}\xi^{W}U,U^{\dagger}\rho^{W}U\right]\nonumber \\
= & U^{\dagger}\left(\frac{d\rho^{W}}{d{\bf k}}--i\left[\xi^{W},\rho^{W}\right]\right)U\nonumber \\
= & U^{\dagger}\frac{D\rho^{W}}{D{\bf k}}U.
\end{align}

\begin{figure*}
\includegraphics[scale=0.45]{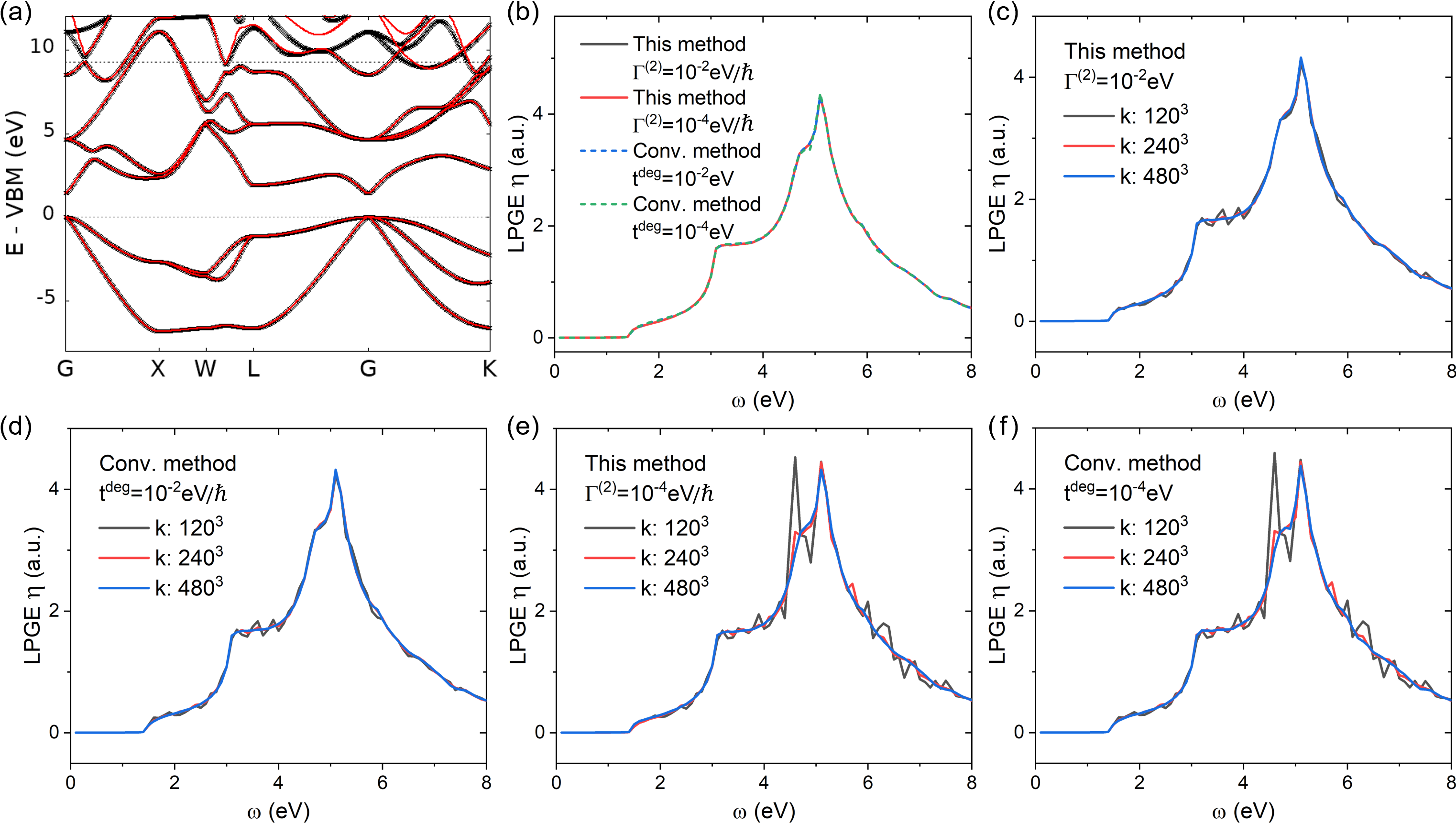}

\caption{Theoretical results of GaAs with a different Wannierization setup
from that for Fig. \ref{fig:gaas}. This Wannierization setup and
that for Fig. \ref{fig:gaas} have 12 and 8 WFs respectively. Their
energy windows (relative to VBM energy) are {[}-8 eV, 9.3 eV{]} and
{[}-8 eV, 7.9 eV{]} respectively. (a) DFT and Wannier band structures.
(b) $\eta_{yz}^{cx}$ by our method with $\hbar\Gamma^{\left(2\right)}$=0.01
eV and $\hbar\Gamma^{\left(2\right)}$=10$^{-4}$ eV, compared with
$\eta_{yz}^{cx}$ by the conventional method with different degeneracy
thresholds $t^{\mathrm{deg}}$. Subfigure (b) uses $480\times480\times480$
k meshes. (c) and (e) are $\eta_{yz}^{cx}$ by our method with $\hbar\Gamma^{\left(2\right)}$=0.01
eV and $\hbar\Gamma^{\left(2\right)}$=10$^{-4}$ eV respectively.
(d) and (f) are $\eta_{yz}^{cx}$ by the conventional method with
$t^{\mathrm{deg}}$ of $10^{-2}$ and $10^{-4}$ eV respectively.
The relaxation rate $\Gamma$=0.01 eV/$\hbar$, which is also the
Lorentzian smearing parameter of the conventional method.\label{fig:gaas-diffwann}}
\end{figure*}

\subsection*{Appendix C: The computation of $\frac{Df^{\mathrm{eq}}}{D{\bf k}}$
without finite differences}

From Eq. \ref{eq:v_from_rH} and \ref{eq:D-r-relation}, we have
\begin{align}
{\bf v}_{kab}= & \frac{1}{\hbar}\left(\frac{DH_{k}^{0}}{D{\bf k}}\right)_{ab}\nonumber \\
= & \frac{1}{\hbar}\frac{d\epsilon_{ka}}{d{\bf k}}\delta_{ab}+\frac{i}{\hbar}\xi_{kab}\Delta_{kab},
\end{align}

so that
\begin{align}
\xi_{ab}= & -i\frac{\hbar{\bf v}_{kab}}{\Delta_{kab}},\text{ if }\Delta_{kab}\neq0.\label{eq:relation_xi_and_v}
\end{align}

Then, we have
\begin{align}
-i\left[\xi_{k},f_{k}^{\mathrm{eq}}\right]_{ab}= & i\xi_{kab}\left(f_{ka}^{\mathrm{eq}}-f_{kb}^{\mathrm{eq}}\right)\nonumber \\
= & i\xi_{kab}\left(f_{ka}^{\mathrm{eq}}-f_{kb}^{\mathrm{eq}}\right)\left(1-\delta_{\epsilon_{ka},\epsilon_{kb}}\right)\nonumber \\
= & \hbar{\bf v}_{kab}\frac{f_{ka}^{\mathrm{eq}}-f_{kb}^{\mathrm{eq}}}{\epsilon_{ka}-\epsilon_{kb}}\left(1-\delta_{\epsilon_{ka},\epsilon_{kb}}\right).
\end{align}

Therefore,
\begin{align}
\left(\frac{Df_{k}^{\mathrm{eq}}}{D{\bf k}}\right)_{ab}= & \left(\frac{\Delta f^{\mathrm{eq}}}{\Delta\epsilon}\right)_{kab}\hbar{\bf v}_{kab},\label{eq:DfDk_from_v}\\
\left(\frac{\Delta f^{\mathrm{eq}}}{\Delta\epsilon}\right)_{kab}= & \left(\frac{df_{ka}^{\mathrm{eq}}}{d\epsilon}\right)\delta_{ab}\\
 & +\frac{f_{ka}^{\mathrm{eq}}-f_{kb}^{\mathrm{eq}}}{\epsilon_{ka}-\epsilon_{kb}}\left(1-\delta_{\epsilon_{ka},\epsilon_{kb}}\right).
\end{align}

As $\frac{df_{ka}^{\mathrm{eq}}}{d\epsilon}=\left(k_{B}T\right)^{-1}f_{ka}^{\mathrm{eq}}\left(f_{ka}^{\mathrm{eq}}-1\right)$
can be evaluated analytically, numerical finite differences are avoided
for the computation of $\frac{Df^{\mathrm{eq}}}{D{\bf k}}$. Numerically,
we have found that computing $\frac{Df^{\mathrm{eq}}}{D{\bf k}}$
via Eq. \ref{eq:UHdU} with finite differences and via Eq. \ref{eq:DfDk_from_v}
lead to almost the same results. Eq. \ref{eq:DfDk_from_v} is preferred
since it is computationally convenient.

\subsection*{Appendix D: The computation of $\overline{\xi}$ (Eq. \ref{eq:xibar})}

The accuracy of $\overline{{\bf \xi}}_{k}$ seems sometimes a bit
worse when DFT meshes are not so dense, compared with $H_{k}$, ${\bf v}_{k}$
and ${\bf s}_{k}$, whose accuracy is good even when DFT coarse k
meshes for constructing WFs are quite coarse\citep{marzari2012maximally},
e.g., 4$\times$4$\times$4.\citep{wang2006ab} This is because: In
usual implementation of ${\bf \xi}_{k}^{W}$ (Eq. \ref{eq:xiW}, which
determines $\overline{\xi}$) using the plane-wave DFT method, finite
differences of $u_{k}^{W}$ on DFT coarse meshes are required. Although
$u_{k}^{W}$ is smooth over ${\bf k}$, too coarse k meshes may still
lead to some errors. This issue can be removed by increasing DFT k
meshes or by using another implementation of ${\bf \xi}_{k}^{W}$
without finite differences.\citep{wang2006ab}

Here we introduce another technique to improve the accuracy of $\overline{\xi}$:

From Eq. \ref{eq:v_from_rH}, \ref{eq:D-r-relation} and \ref{eq:UHdU},
we have
\begin{align}
{\bf v}_{kab}= & \frac{1}{\hbar}\left(\frac{DH_{k}^{0}}{D{\bf k}}\right)_{ab}\nonumber \\
= & \left(\frac{1}{\hbar}U_{k}^{\dagger}\frac{dH_{k}^{W}}{d{\bf k}}U_{k}-\frac{i}{\hbar}\left[\overline{\xi}_{k},\epsilon_{k}\right]\right)_{ab}.
\end{align}

Therefore,
\begin{align}
\overline{\xi}_{kab}= & \frac{\hbar{\bf v}_{kab}-\left(U_{k}^{\dagger}\frac{dH_{k}^{W}}{d{\bf k}}U_{k}\right)_{ab}}{i\Delta_{kab}}\text{ if }\epsilon_{ka}\neq\epsilon_{kb}.\label{eq:xibar_from_v}
\end{align}

Since $\frac{dH^{W}}{d{\bf k}}$ can be computed accurately and efficiently
without finite differences,\citep{wang2006ab} the computation of
$\overline{\xi}_{kab}$ for the elements satisfying $\epsilon_{ka}\neq\epsilon_{kb}$
by Eq. \ref{eq:xibar_from_v} above is also accurate and efficient.

\subsection*{Appendix E: GaAs results with a different Wannierization setup}

In Fig. \ref{fig:gaas-diffwann}, we show theoretical results of GaAs
with a Wannierization setup having more WFs and large energy window
than that for Fig. \ref{fig:gaas}. Similarly to Fig. \ref{fig:gaas},
we find that results by two methods with different $\Gamma^{\left(2\right)}$
and $t^{\mathrm{deg}}$ agree well when k meshes are converged, and
the k-point convergence of our method and the conventional method
with relatively large $\hbar\Gamma^{\left(2\right)}$ and $t^{\mathrm{deg}}$
of $10^{-2}$ eV is much faster than that with smaller $\hbar\Gamma^{\left(2\right)}$
and $t^{\mathrm{deg}}$ of $10^{-4}$ eV. By comparing Fig. \ref{fig:gaas-diffwann}
with Fig. \ref{fig:gaas}, it can be seen that when k meshes are not
converged and with smaller $\hbar\Gamma^{\left(2\right)}$ and $t^{\mathrm{deg}}$
of $10^{-4}$ eV, theoretical results are Wannierization dependent.

\section*{Reference}


\begin{thebibliography}{64}%
\makeatletter
\providecommand \@ifxundefined [1]{%
 \@ifx{#1\undefined}
}%
\providecommand \@ifnum [1]{%
 \ifnum #1\expandafter \@firstoftwo
 \else \expandafter \@secondoftwo
 \fi
}%
\providecommand \@ifx [1]{%
 \ifx #1\expandafter \@firstoftwo
 \else \expandafter \@secondoftwo
 \fi
}%
\providecommand \natexlab [1]{#1}%
\providecommand \enquote  [1]{``#1''}%
\providecommand \bibnamefont  [1]{#1}%
\providecommand \bibfnamefont [1]{#1}%
\providecommand \citenamefont [1]{#1}%
\providecommand \href@noop [0]{\@secondoftwo}%
\providecommand \href [0]{\begingroup \@sanitize@url \@href}%
\providecommand \@href[1]{\@@startlink{#1}\@@href}%
\providecommand \@@href[1]{\endgroup#1\@@endlink}%
\providecommand \@sanitize@url [0]{\catcode `\\12\catcode `\$12\catcode
  `\&12\catcode `\#12\catcode `\^12\catcode `\_12\catcode `\%12\relax}%
\providecommand \@@startlink[1]{}%
\providecommand \@@endlink[0]{}%
\providecommand \url  [0]{\begingroup\@sanitize@url \@url }%
\providecommand \@url [1]{\endgroup\@href {#1}{\urlprefix }}%
\providecommand \urlprefix  [0]{URL }%
\providecommand \Eprint [0]{\href }%
\providecommand \doibase [0]{https://doi.org/}%
\providecommand \selectlanguage [0]{\@gobble}%
\providecommand \bibinfo  [0]{\@secondoftwo}%
\providecommand \bibfield  [0]{\@secondoftwo}%
\providecommand \translation [1]{[#1]}%
\providecommand \BibitemOpen [0]{}%
\providecommand \bibitemStop [0]{}%
\providecommand \bibitemNoStop [0]{.\EOS\space}%
\providecommand \EOS [0]{\spacefactor3000\relax}%
\providecommand \BibitemShut  [1]{\csname bibitem#1\endcsname}%
\let\auto@bib@innerbib\@empty
\bibitem [{\citenamefont {Fridkin}(2001)}]{fridkin2001bulk}%
  \BibitemOpen
  \bibfield  {author} {\bibinfo {author} {\bibfnamefont {V.}~\bibnamefont
  {Fridkin}},\ }\bibfield  {title} {\bibinfo {title} {Bulk photovoltaic effect
  in noncentrosymmetric crystals},\ }\href@noop {} {\bibfield  {journal}
  {\bibinfo  {journal} {Crystallogr. Rep.}\ }\textbf {\bibinfo {volume} {46}},\
  \bibinfo {pages} {654} (\bibinfo {year} {2001})}\BibitemShut {NoStop}%
\bibitem [{\citenamefont {Dai}\ and\ \citenamefont
  {Rappe}(2023)}]{dai2023recent}%
  \BibitemOpen
  \bibfield  {author} {\bibinfo {author} {\bibfnamefont {Z.}~\bibnamefont
  {Dai}}\ and\ \bibinfo {author} {\bibfnamefont {A.~M.}\ \bibnamefont
  {Rappe}},\ }\bibfield  {title} {\bibinfo {title} {Recent progress in the
  theory of bulk photovoltaic effect},\ }\href@noop {} {\bibfield  {journal}
  {\bibinfo  {journal} {Chem. Phys. Rev.}\ }\textbf {\bibinfo {volume} {4}}
  (\bibinfo {year} {2023})}\BibitemShut {NoStop}%
\bibitem [{\citenamefont {Spanier}\ \emph {et~al.}(2016)\citenamefont
  {Spanier}, \citenamefont {Fridkin}, \citenamefont {Rappe}, \citenamefont
  {Akbashev}, \citenamefont {Polemi}, \citenamefont {Qi}, \citenamefont {Gu},
  \citenamefont {Young}, \citenamefont {Hawley}, \citenamefont {Imbrenda} \emph
  {et~al.}}]{spanier2016power}%
  \BibitemOpen
  \bibfield  {author} {\bibinfo {author} {\bibfnamefont {J.~E.}\ \bibnamefont
  {Spanier}}, \bibinfo {author} {\bibfnamefont {V.~M.}\ \bibnamefont
  {Fridkin}}, \bibinfo {author} {\bibfnamefont {A.~M.}\ \bibnamefont {Rappe}},
  \bibinfo {author} {\bibfnamefont {A.~R.}\ \bibnamefont {Akbashev}}, \bibinfo
  {author} {\bibfnamefont {A.}~\bibnamefont {Polemi}}, \bibinfo {author}
  {\bibfnamefont {Y.}~\bibnamefont {Qi}}, \bibinfo {author} {\bibfnamefont
  {Z.}~\bibnamefont {Gu}}, \bibinfo {author} {\bibfnamefont {S.~M.}\
  \bibnamefont {Young}}, \bibinfo {author} {\bibfnamefont {C.~J.}\ \bibnamefont
  {Hawley}}, \bibinfo {author} {\bibfnamefont {D.}~\bibnamefont {Imbrenda}},
  \emph {et~al.},\ }\bibfield  {title} {\bibinfo {title} {Power conversion
  efficiency exceeding the shockley--queisser limit in a ferroelectric
  insulator},\ }\href@noop {} {\bibfield  {journal} {\bibinfo  {journal} {Nat.
  Photonics}\ }\textbf {\bibinfo {volume} {10}},\ \bibinfo {pages} {611}
  (\bibinfo {year} {2016})}\BibitemShut {NoStop}%
\bibitem [{\citenamefont {De~Juan}\ \emph {et~al.}(2017)\citenamefont
  {De~Juan}, \citenamefont {Grushin}, \citenamefont {Morimoto},\ and\
  \citenamefont {Moore}}]{de2017quantized}%
  \BibitemOpen
  \bibfield  {author} {\bibinfo {author} {\bibfnamefont {F.}~\bibnamefont
  {De~Juan}}, \bibinfo {author} {\bibfnamefont {A.~G.}\ \bibnamefont
  {Grushin}}, \bibinfo {author} {\bibfnamefont {T.}~\bibnamefont {Morimoto}},\
  and\ \bibinfo {author} {\bibfnamefont {J.~E.}\ \bibnamefont {Moore}},\
  }\bibfield  {title} {\bibinfo {title} {Quantized circular photogalvanic
  effect in weyl semimetals},\ }\href@noop {} {\bibfield  {journal} {\bibinfo
  {journal} {Nat. Commun.}\ }\textbf {\bibinfo {volume} {8}},\ \bibinfo {pages}
  {15995} (\bibinfo {year} {2017})}\BibitemShut {NoStop}%
\bibitem [{\citenamefont {Rees}\ \emph {et~al.}(2020)\citenamefont {Rees},
  \citenamefont {Manna}, \citenamefont {Lu}, \citenamefont {Morimoto},
  \citenamefont {Borrmann}, \citenamefont {Felser}, \citenamefont {Moore},
  \citenamefont {Torchinsky},\ and\ \citenamefont
  {Orenstein}}]{rees2020helicity}%
  \BibitemOpen
  \bibfield  {author} {\bibinfo {author} {\bibfnamefont {D.}~\bibnamefont
  {Rees}}, \bibinfo {author} {\bibfnamefont {K.}~\bibnamefont {Manna}},
  \bibinfo {author} {\bibfnamefont {B.}~\bibnamefont {Lu}}, \bibinfo {author}
  {\bibfnamefont {T.}~\bibnamefont {Morimoto}}, \bibinfo {author}
  {\bibfnamefont {H.}~\bibnamefont {Borrmann}}, \bibinfo {author}
  {\bibfnamefont {C.}~\bibnamefont {Felser}}, \bibinfo {author} {\bibfnamefont
  {J.}~\bibnamefont {Moore}}, \bibinfo {author} {\bibfnamefont {D.~H.}\
  \bibnamefont {Torchinsky}},\ and\ \bibinfo {author} {\bibfnamefont
  {J.}~\bibnamefont {Orenstein}},\ }\bibfield  {title} {\bibinfo {title}
  {Helicity-dependent photocurrents in the chiral weyl semimetal rhsi},\
  }\href@noop {} {\bibfield  {journal} {\bibinfo  {journal} {Sci. Adv.}\
  }\textbf {\bibinfo {volume} {6}},\ \bibinfo {pages} {eaba0509} (\bibinfo
  {year} {2020})}\BibitemShut {NoStop}%
\bibitem [{\citenamefont {Yuan}\ \emph {et~al.}(2014)\citenamefont {Yuan},
  \citenamefont {Wang}, \citenamefont {Lian}, \citenamefont {Zhang},
  \citenamefont {Fang}, \citenamefont {Shen}, \citenamefont {Xu}, \citenamefont
  {Xu}, \citenamefont {Zhang}, \citenamefont {Hwang} \emph
  {et~al.}}]{yuan2014generation}%
  \BibitemOpen
  \bibfield  {author} {\bibinfo {author} {\bibfnamefont {H.}~\bibnamefont
  {Yuan}}, \bibinfo {author} {\bibfnamefont {X.}~\bibnamefont {Wang}}, \bibinfo
  {author} {\bibfnamefont {B.}~\bibnamefont {Lian}}, \bibinfo {author}
  {\bibfnamefont {H.}~\bibnamefont {Zhang}}, \bibinfo {author} {\bibfnamefont
  {X.}~\bibnamefont {Fang}}, \bibinfo {author} {\bibfnamefont {B.}~\bibnamefont
  {Shen}}, \bibinfo {author} {\bibfnamefont {G.}~\bibnamefont {Xu}}, \bibinfo
  {author} {\bibfnamefont {Y.}~\bibnamefont {Xu}}, \bibinfo {author}
  {\bibfnamefont {S.-C.}\ \bibnamefont {Zhang}}, \bibinfo {author}
  {\bibfnamefont {H.~Y.}\ \bibnamefont {Hwang}}, \emph {et~al.},\ }\bibfield
  {title} {\bibinfo {title} {Generation and electric control of
  spin--valley-coupled circular photogalvanic current in wse2},\ }\href@noop {}
  {\bibfield  {journal} {\bibinfo  {journal} {Nat. Nanotechnol.}\ }\textbf
  {\bibinfo {volume} {9}},\ \bibinfo {pages} {851} (\bibinfo {year}
  {2014})}\BibitemShut {NoStop}%
\bibitem [{\citenamefont {Fei}\ \emph {et~al.}(2021)\citenamefont {Fei},
  \citenamefont {Song}, \citenamefont {Pusey-Nazzaro},\ and\ \citenamefont
  {Yang}}]{fei2021p}%
  \BibitemOpen
  \bibfield  {author} {\bibinfo {author} {\bibfnamefont {R.}~\bibnamefont
  {Fei}}, \bibinfo {author} {\bibfnamefont {W.}~\bibnamefont {Song}}, \bibinfo
  {author} {\bibfnamefont {L.}~\bibnamefont {Pusey-Nazzaro}},\ and\ \bibinfo
  {author} {\bibfnamefont {L.}~\bibnamefont {Yang}},\ }\bibfield  {title}
  {\bibinfo {title} {P t-symmetry-enabled spin circular photogalvanic effect in
  antiferromagnetic insulators},\ }\href@noop {} {\bibfield  {journal}
  {\bibinfo  {journal} {Phys. Rev. Lett.}\ }\textbf {\bibinfo {volume} {127}},\
  \bibinfo {pages} {207402} (\bibinfo {year} {2021})}\BibitemShut {NoStop}%
\bibitem [{\citenamefont {Hong}\ \emph {et~al.}(2013)\citenamefont {Hong},
  \citenamefont {Dadap}, \citenamefont {Petrone}, \citenamefont {Yeh},
  \citenamefont {Hone},\ and\ \citenamefont {Osgood~Jr}}]{hong2013optical}%
  \BibitemOpen
  \bibfield  {author} {\bibinfo {author} {\bibfnamefont {S.-Y.}\ \bibnamefont
  {Hong}}, \bibinfo {author} {\bibfnamefont {J.~I.}\ \bibnamefont {Dadap}},
  \bibinfo {author} {\bibfnamefont {N.}~\bibnamefont {Petrone}}, \bibinfo
  {author} {\bibfnamefont {P.-C.}\ \bibnamefont {Yeh}}, \bibinfo {author}
  {\bibfnamefont {J.}~\bibnamefont {Hone}},\ and\ \bibinfo {author}
  {\bibfnamefont {R.~M.}\ \bibnamefont {Osgood~Jr}},\ }\bibfield  {title}
  {\bibinfo {title} {Optical third-harmonic generation in graphene},\
  }\href@noop {} {\bibfield  {journal} {\bibinfo  {journal} {Phys. Rev. X}\
  }\textbf {\bibinfo {volume} {3}},\ \bibinfo {pages} {021014} (\bibinfo {year}
  {2013})}\BibitemShut {NoStop}%
\bibitem [{\citenamefont {Silva}\ \emph {et~al.}(2019)\citenamefont {Silva},
  \citenamefont {Mart{\'\i}n},\ and\ \citenamefont {Ivanov}}]{silva2019high}%
  \BibitemOpen
  \bibfield  {author} {\bibinfo {author} {\bibfnamefont {R.~E.~F.}\
  \bibnamefont {Silva}}, \bibinfo {author} {\bibfnamefont {F.}~\bibnamefont
  {Mart{\'\i}n}},\ and\ \bibinfo {author} {\bibfnamefont {M.}~\bibnamefont
  {Ivanov}},\ }\bibfield  {title} {\bibinfo {title} {High harmonic generation
  in crystals using maximally localized wannier functions},\ }\href@noop {}
  {\bibfield  {journal} {\bibinfo  {journal} {Phys. Rev. B}\ }\textbf {\bibinfo
  {volume} {100}},\ \bibinfo {pages} {195201} (\bibinfo {year}
  {2019})}\BibitemShut {NoStop}%
\bibitem [{\citenamefont {Chan}\ \emph {et~al.}(2021)\citenamefont {Chan},
  \citenamefont {Qiu}, \citenamefont {da~Jornada},\ and\ \citenamefont
  {Louie}}]{chan2021giant}%
  \BibitemOpen
  \bibfield  {author} {\bibinfo {author} {\bibfnamefont {Y.-H.}\ \bibnamefont
  {Chan}}, \bibinfo {author} {\bibfnamefont {D.~Y.}\ \bibnamefont {Qiu}},
  \bibinfo {author} {\bibfnamefont {F.~H.}\ \bibnamefont {da~Jornada}},\ and\
  \bibinfo {author} {\bibfnamefont {S.~G.}\ \bibnamefont {Louie}},\ }\bibfield
  {title} {\bibinfo {title} {Giant exciton-enhanced shift currents and direct
  current conduction with subbandgap photo excitations produced by
  many-electron interactions},\ }\href@noop {} {\bibfield  {journal} {\bibinfo
  {journal} {Proc. Natl. Acad. Sci. USA}\ }\textbf {\bibinfo {volume} {118}},\
  \bibinfo {pages} {e1906938118} (\bibinfo {year} {2021})}\BibitemShut
  {NoStop}%
\bibitem [{\citenamefont {de~Juan}\ \emph {et~al.}(2020)\citenamefont
  {de~Juan}, \citenamefont {Zhang}, \citenamefont {Morimoto}, \citenamefont
  {Sun}, \citenamefont {Moore},\ and\ \citenamefont
  {Grushin}}]{de2020difference}%
  \BibitemOpen
  \bibfield  {author} {\bibinfo {author} {\bibfnamefont {F.}~\bibnamefont
  {de~Juan}}, \bibinfo {author} {\bibfnamefont {Y.}~\bibnamefont {Zhang}},
  \bibinfo {author} {\bibfnamefont {T.}~\bibnamefont {Morimoto}}, \bibinfo
  {author} {\bibfnamefont {Y.}~\bibnamefont {Sun}}, \bibinfo {author}
  {\bibfnamefont {J.~E.}\ \bibnamefont {Moore}},\ and\ \bibinfo {author}
  {\bibfnamefont {A.~G.}\ \bibnamefont {Grushin}},\ }\bibfield  {title}
  {\bibinfo {title} {Difference frequency generation in topological
  semimetals},\ }\href@noop {} {\bibfield  {journal} {\bibinfo  {journal}
  {Phys. Rev. Res.}\ }\textbf {\bibinfo {volume} {2}},\ \bibinfo {pages}
  {012017(R)} (\bibinfo {year} {2020})}\BibitemShut {NoStop}%
\bibitem [{\citenamefont {Xu}\ \emph {et~al.}(2021{\natexlab{a}})\citenamefont
  {Xu}, \citenamefont {Wang}, \citenamefont {Zhou},\ and\ \citenamefont
  {Li}}]{xu2021pure}%
  \BibitemOpen
  \bibfield  {author} {\bibinfo {author} {\bibfnamefont {H.}~\bibnamefont
  {Xu}}, \bibinfo {author} {\bibfnamefont {H.}~\bibnamefont {Wang}}, \bibinfo
  {author} {\bibfnamefont {J.}~\bibnamefont {Zhou}},\ and\ \bibinfo {author}
  {\bibfnamefont {J.}~\bibnamefont {Li}},\ }\bibfield  {title} {\bibinfo
  {title} {Pure spin photocurrent in non-centrosymmetric crystals: bulk spin
  photovoltaic effect},\ }\href@noop {} {\bibfield  {journal} {\bibinfo
  {journal} {Nat. Commun.}\ }\textbf {\bibinfo {volume} {12}},\ \bibinfo
  {pages} {4330} (\bibinfo {year} {2021}{\natexlab{a}})}\BibitemShut {NoStop}%
\bibitem [{\citenamefont {Soavi}\ \emph {et~al.}(2018)\citenamefont {Soavi},
  \citenamefont {Wang}, \citenamefont {Rostami}, \citenamefont {Purdie},
  \citenamefont {De~Fazio}, \citenamefont {Ma}, \citenamefont {Luo},
  \citenamefont {Wang}, \citenamefont {Ott}, \citenamefont {Yoon} \emph
  {et~al.}}]{soavi2018broadband}%
  \BibitemOpen
  \bibfield  {author} {\bibinfo {author} {\bibfnamefont {G.}~\bibnamefont
  {Soavi}}, \bibinfo {author} {\bibfnamefont {G.}~\bibnamefont {Wang}},
  \bibinfo {author} {\bibfnamefont {H.}~\bibnamefont {Rostami}}, \bibinfo
  {author} {\bibfnamefont {D.~G.}\ \bibnamefont {Purdie}}, \bibinfo {author}
  {\bibfnamefont {D.}~\bibnamefont {De~Fazio}}, \bibinfo {author}
  {\bibfnamefont {T.}~\bibnamefont {Ma}}, \bibinfo {author} {\bibfnamefont
  {B.}~\bibnamefont {Luo}}, \bibinfo {author} {\bibfnamefont {J.}~\bibnamefont
  {Wang}}, \bibinfo {author} {\bibfnamefont {A.~K.}\ \bibnamefont {Ott}},
  \bibinfo {author} {\bibfnamefont {D.}~\bibnamefont {Yoon}}, \emph {et~al.},\
  }\bibfield  {title} {\bibinfo {title} {Broadband, electrically tunable
  third-harmonic generation in graphene},\ }\href@noop {} {\bibfield  {journal}
  {\bibinfo  {journal} {Nat. Nanotechnol.}\ }\textbf {\bibinfo {volume} {13}},\
  \bibinfo {pages} {583} (\bibinfo {year} {2018})}\BibitemShut {NoStop}%
\bibitem [{\citenamefont {Levine}\ and\ \citenamefont
  {Allan}(1991)}]{levine1991calculation}%
  \BibitemOpen
  \bibfield  {author} {\bibinfo {author} {\bibfnamefont {Z.~H.}\ \bibnamefont
  {Levine}}\ and\ \bibinfo {author} {\bibfnamefont {D.~C.}\ \bibnamefont
  {Allan}},\ }\bibfield  {title} {\bibinfo {title} {Calculation of the
  nonlinear susceptibility for optical second-harmonic generation in iii-v
  semiconductors},\ }\href@noop {} {\bibfield  {journal} {\bibinfo  {journal}
  {Phys. Rev, Lett.}\ }\textbf {\bibinfo {volume} {66}},\ \bibinfo {pages} {41}
  (\bibinfo {year} {1991})}\BibitemShut {NoStop}%
\bibitem [{\citenamefont {Young}\ and\ \citenamefont
  {Rappe}(2012)}]{young2012first}%
  \BibitemOpen
  \bibfield  {author} {\bibinfo {author} {\bibfnamefont {S.~M.}\ \bibnamefont
  {Young}}\ and\ \bibinfo {author} {\bibfnamefont {A.~M.}\ \bibnamefont
  {Rappe}},\ }\bibfield  {title} {\bibinfo {title} {First principles
  calculation of the shift current photovoltaic effect in ferroelectrics},\
  }\href@noop {} {\bibfield  {journal} {\bibinfo  {journal} {Phys. Rev. Lett.}\
  }\textbf {\bibinfo {volume} {109}},\ \bibinfo {pages} {116601} (\bibinfo
  {year} {2012})}\BibitemShut {NoStop}%
\bibitem [{\citenamefont {Iba{\~n}ez-Azpiroz}\ \emph
  {et~al.}(2018)\citenamefont {Iba{\~n}ez-Azpiroz}, \citenamefont {Tsirkin},\
  and\ \citenamefont {Souza}}]{ibanez2018ab}%
  \BibitemOpen
  \bibfield  {author} {\bibinfo {author} {\bibfnamefont {J.}~\bibnamefont
  {Iba{\~n}ez-Azpiroz}}, \bibinfo {author} {\bibfnamefont {S.~S.}\ \bibnamefont
  {Tsirkin}},\ and\ \bibinfo {author} {\bibfnamefont {I.}~\bibnamefont
  {Souza}},\ }\bibfield  {title} {\bibinfo {title} {Ab initio calculation of
  the shift photocurrent by wannier interpolation},\ }\href@noop {} {\bibfield
  {journal} {\bibinfo  {journal} {Phys. Rev. B}\ }\textbf {\bibinfo {volume}
  {97}},\ \bibinfo {pages} {245143} (\bibinfo {year} {2018})}\BibitemShut
  {NoStop}%
\bibitem [{\citenamefont {Wang}\ \emph {et~al.}(2017)\citenamefont {Wang},
  \citenamefont {Liu}, \citenamefont {Kang}, \citenamefont {Gu}, \citenamefont
  {Xu},\ and\ \citenamefont {Duan}}]{wang2017first}%
  \BibitemOpen
  \bibfield  {author} {\bibinfo {author} {\bibfnamefont {C.}~\bibnamefont
  {Wang}}, \bibinfo {author} {\bibfnamefont {X.}~\bibnamefont {Liu}}, \bibinfo
  {author} {\bibfnamefont {L.}~\bibnamefont {Kang}}, \bibinfo {author}
  {\bibfnamefont {B.-L.}\ \bibnamefont {Gu}}, \bibinfo {author} {\bibfnamefont
  {Y.}~\bibnamefont {Xu}},\ and\ \bibinfo {author} {\bibfnamefont
  {W.}~\bibnamefont {Duan}},\ }\bibfield  {title} {\bibinfo {title}
  {First-principles calculation of nonlinear optical responses by wannier
  interpolation},\ }\href@noop {} {\bibfield  {journal} {\bibinfo  {journal}
  {Phys. Rev. B}\ }\textbf {\bibinfo {volume} {96}},\ \bibinfo {pages} {115147}
  (\bibinfo {year} {2017})}\BibitemShut {NoStop}%
\bibitem [{\citenamefont {Zhang}\ \emph {et~al.}(2019)\citenamefont {Zhang},
  \citenamefont {Holder}, \citenamefont {Ishizuka}, \citenamefont {de~Juan},
  \citenamefont {Nagaosa}, \citenamefont {Felser},\ and\ \citenamefont
  {Yan}}]{zhang2019switchable}%
  \BibitemOpen
  \bibfield  {author} {\bibinfo {author} {\bibfnamefont {Y.}~\bibnamefont
  {Zhang}}, \bibinfo {author} {\bibfnamefont {T.}~\bibnamefont {Holder}},
  \bibinfo {author} {\bibfnamefont {H.}~\bibnamefont {Ishizuka}}, \bibinfo
  {author} {\bibfnamefont {F.}~\bibnamefont {de~Juan}}, \bibinfo {author}
  {\bibfnamefont {N.}~\bibnamefont {Nagaosa}}, \bibinfo {author} {\bibfnamefont
  {C.}~\bibnamefont {Felser}},\ and\ \bibinfo {author} {\bibfnamefont
  {B.}~\bibnamefont {Yan}},\ }\bibfield  {title} {\bibinfo {title} {Switchable
  magnetic bulk photovoltaic effect in the two-dimensional magnet CrI3},\
  }\href@noop {} {\bibfield  {journal} {\bibinfo  {journal} {Nat. Commun.}\
  }\textbf {\bibinfo {volume} {10}},\ \bibinfo {pages} {3783} (\bibinfo {year}
  {2019})}\BibitemShut {NoStop}%
\bibitem [{\citenamefont {Le}\ \emph {et~al.}(2020)\citenamefont {Le},
  \citenamefont {Zhang}, \citenamefont {Felser},\ and\ \citenamefont
  {Sun}}]{le2020ab}%
  \BibitemOpen
  \bibfield  {author} {\bibinfo {author} {\bibfnamefont {C.}~\bibnamefont
  {Le}}, \bibinfo {author} {\bibfnamefont {Y.}~\bibnamefont {Zhang}}, \bibinfo
  {author} {\bibfnamefont {C.}~\bibnamefont {Felser}},\ and\ \bibinfo {author}
  {\bibfnamefont {Y.}~\bibnamefont {Sun}},\ }\bibfield  {title} {\bibinfo
  {title} {Ab initio study of quantized circular photogalvanic effect in chiral
  multifold semimetals},\ }\href@noop {} {\bibfield  {journal} {\bibinfo
  {journal} {Phys. Rev. B}\ }\textbf {\bibinfo {volume} {102}},\ \bibinfo
  {pages} {121111(R)} (\bibinfo {year} {2020})}\BibitemShut {NoStop}%
\bibitem [{\citenamefont {Zhang}\ \emph {et~al.}(2018)\citenamefont {Zhang},
  \citenamefont {Sun},\ and\ \citenamefont {Yan}}]{zhang2018berry}%
  \BibitemOpen
  \bibfield  {author} {\bibinfo {author} {\bibfnamefont {Y.}~\bibnamefont
  {Zhang}}, \bibinfo {author} {\bibfnamefont {Y.}~\bibnamefont {Sun}},\ and\
  \bibinfo {author} {\bibfnamefont {B.}~\bibnamefont {Yan}},\ }\bibfield
  {title} {\bibinfo {title} {Berry curvature dipole in weyl semimetal
  materials: an ab initio study},\ }\href@noop {} {\bibfield  {journal}
  {\bibinfo  {journal} {Phys. Rev. B}\ }\textbf {\bibinfo {volume} {97}},\
  \bibinfo {pages} {041101(R)} (\bibinfo {year} {2018})}\BibitemShut {NoStop}%
\bibitem [{\citenamefont {Watanabe}\ and\ \citenamefont
  {Yanase}(2021)}]{watanabe2021chiral}%
  \BibitemOpen
  \bibfield  {author} {\bibinfo {author} {\bibfnamefont {H.}~\bibnamefont
  {Watanabe}}\ and\ \bibinfo {author} {\bibfnamefont {Y.}~\bibnamefont
  {Yanase}},\ }\bibfield  {title} {\bibinfo {title} {Chiral photocurrent in
  parity-violating magnet and enhanced response in topological
  antiferromagnet},\ }\href@noop {} {\bibfield  {journal} {\bibinfo  {journal}
  {Phys. Rev. X}\ }\textbf {\bibinfo {volume} {11}},\ \bibinfo {pages} {011001}
  (\bibinfo {year} {2021})}\BibitemShut {NoStop}%
\bibitem [{\citenamefont {Ventura}\ \emph {et~al.}(2017)\citenamefont
  {Ventura}, \citenamefont {Passos}, \citenamefont {dos Santos}, \citenamefont
  {Lopes},\ and\ \citenamefont {Peres}}]{ventura2017gauge}%
  \BibitemOpen
  \bibfield  {author} {\bibinfo {author} {\bibfnamefont {G.}~\bibnamefont
  {Ventura}}, \bibinfo {author} {\bibfnamefont {D.}~\bibnamefont {Passos}},
  \bibinfo {author} {\bibfnamefont {J.~L.}\ \bibnamefont {dos Santos}},
  \bibinfo {author} {\bibfnamefont {J.~V.~P.}\ \bibnamefont {Lopes}},\ and\
  \bibinfo {author} {\bibfnamefont {N.}~\bibnamefont {Peres}},\ }\bibfield
  {title} {\bibinfo {title} {Gauge covariances and nonlinear optical
  responses},\ }\href@noop {} {\bibfield  {journal} {\bibinfo  {journal} {Phys.
  Rev. B}\ }\textbf {\bibinfo {volume} {96}},\ \bibinfo {pages} {035431}
  (\bibinfo {year} {2017})}\BibitemShut {NoStop}%
\bibitem [{\citenamefont {Passos}\ \emph {et~al.}(2018)\citenamefont {Passos},
  \citenamefont {Ventura}, \citenamefont {Lopes}, \citenamefont {dos Santos},\
  and\ \citenamefont {Peres}}]{passos2018nonlinear}%
  \BibitemOpen
  \bibfield  {author} {\bibinfo {author} {\bibfnamefont {D.}~\bibnamefont
  {Passos}}, \bibinfo {author} {\bibfnamefont {G.}~\bibnamefont {Ventura}},
  \bibinfo {author} {\bibfnamefont {J.~V.~P.}\ \bibnamefont {Lopes}}, \bibinfo
  {author} {\bibfnamefont {J.~L.}\ \bibnamefont {dos Santos}},\ and\ \bibinfo
  {author} {\bibfnamefont {N.}~\bibnamefont {Peres}},\ }\bibfield  {title}
  {\bibinfo {title} {Nonlinear optical responses of crystalline systems:
  Results from a velocity gauge analysis},\ }\href@noop {} {\bibfield
  {journal} {\bibinfo  {journal} {Phys. Rev. B}\ }\textbf {\bibinfo {volume}
  {97}},\ \bibinfo {pages} {235446} (\bibinfo {year} {2018})}\BibitemShut
  {NoStop}%
\bibitem [{\citenamefont {Yue}\ and\ \citenamefont
  {Gaarde}(2020)}]{yue2020structure}%
  \BibitemOpen
  \bibfield  {author} {\bibinfo {author} {\bibfnamefont {L.}~\bibnamefont
  {Yue}}\ and\ \bibinfo {author} {\bibfnamefont {M.~B.}\ \bibnamefont
  {Gaarde}},\ }\bibfield  {title} {\bibinfo {title} {Structure gauges and laser
  gauges for the semiconductor bloch equations in high-order harmonic
  generation in solids},\ }\href@noop {} {\bibfield  {journal} {\bibinfo
  {journal} {Phys. Rev. A}\ }\textbf {\bibinfo {volume} {101}},\ \bibinfo
  {pages} {053411} (\bibinfo {year} {2020})}\BibitemShut {NoStop}%
\bibitem [{\citenamefont {Sun}\ \emph {et~al.}(2008)\citenamefont {Sun},
  \citenamefont {Xie},\ and\ \citenamefont {Wang}}]{sun2008persistent}%
  \BibitemOpen
  \bibfield  {author} {\bibinfo {author} {\bibfnamefont {Q.}~\bibnamefont
  {Sun}}, \bibinfo {author} {\bibfnamefont {X.}~\bibnamefont {Xie}},\ and\
  \bibinfo {author} {\bibfnamefont {J.}~\bibnamefont {Wang}},\ }\bibfield
  {title} {\bibinfo {title} {Persistent spin current in nanodevices and
  definition of the spin current},\ }\href@noop {} {\bibfield  {journal}
  {\bibinfo  {journal} {Phys. Rev. B}\ }\textbf {\bibinfo {volume} {77}},\
  \bibinfo {pages} {035327} (\bibinfo {year} {2008})}\BibitemShut {NoStop}%
\bibitem [{\citenamefont {Xu}\ \emph {et~al.}(2021{\natexlab{b}})\citenamefont
  {Xu}, \citenamefont {Habib}, \citenamefont {Sundararaman},\ and\
  \citenamefont {Ping}}]{xu2021ab}%
  \BibitemOpen
  \bibfield  {author} {\bibinfo {author} {\bibfnamefont {J.}~\bibnamefont
  {Xu}}, \bibinfo {author} {\bibfnamefont {A.}~\bibnamefont {Habib}}, \bibinfo
  {author} {\bibfnamefont {R.}~\bibnamefont {Sundararaman}},\ and\ \bibinfo
  {author} {\bibfnamefont {Y.}~\bibnamefont {Ping}},\ }\bibfield  {title}
  {\bibinfo {title} {Ab initio ultrafast spin dynamics in solids},\ }\href@noop
  {} {\bibfield  {journal} {\bibinfo  {journal} {Phys. Rev. B}\ }\textbf
  {\bibinfo {volume} {104}},\ \bibinfo {pages} {184418} (\bibinfo {year}
  {2021}{\natexlab{b}})}\BibitemShut {NoStop}%
\bibitem [{\citenamefont {Xu}\ and\ \citenamefont
  {Ping}(2023{\natexlab{a}})}]{xu2023ab}%
  \BibitemOpen
  \bibfield  {author} {\bibinfo {author} {\bibfnamefont {J.}~\bibnamefont
  {Xu}}\ and\ \bibinfo {author} {\bibfnamefont {Y.}~\bibnamefont {Ping}},\
  }\bibfield  {title} {\bibinfo {title} {Ab initio predictions of spin
  relaxation, dephasing, and diffusion in solids},\ }\href@noop {} {\bibfield
  {journal} {\bibinfo  {journal} {J. Chem. Theory Comput.}\ } (\bibinfo {year}
  {2023}{\natexlab{a}})}\BibitemShut {NoStop}%
\bibitem [{\citenamefont {Dai}\ \emph {et~al.}(2021)\citenamefont {Dai},
  \citenamefont {Schankler}, \citenamefont {Gao}, \citenamefont {Tan},\ and\
  \citenamefont {Rappe}}]{dai2021phonon}%
  \BibitemOpen
  \bibfield  {author} {\bibinfo {author} {\bibfnamefont {Z.}~\bibnamefont
  {Dai}}, \bibinfo {author} {\bibfnamefont {A.~M.}\ \bibnamefont {Schankler}},
  \bibinfo {author} {\bibfnamefont {L.}~\bibnamefont {Gao}}, \bibinfo {author}
  {\bibfnamefont {L.~Z.}\ \bibnamefont {Tan}},\ and\ \bibinfo {author}
  {\bibfnamefont {A.~M.}\ \bibnamefont {Rappe}},\ }\bibfield  {title} {\bibinfo
  {title} {Phonon-assisted ballistic current from first-principles
  calculations},\ }\href@noop {} {\bibfield  {journal} {\bibinfo  {journal}
  {Phys. Rev. Lett.}\ }\textbf {\bibinfo {volume} {126}},\ \bibinfo {pages}
  {177403} (\bibinfo {year} {2021})}\BibitemShut {NoStop}%
\bibitem [{\citenamefont {Bhalla}\ \emph {et~al.}(2023)\citenamefont {Bhalla},
  \citenamefont {Das}, \citenamefont {Agarwal},\ and\ \citenamefont
  {Culcer}}]{bhalla2023quantum}%
  \BibitemOpen
  \bibfield  {author} {\bibinfo {author} {\bibfnamefont {P.}~\bibnamefont
  {Bhalla}}, \bibinfo {author} {\bibfnamefont {K.}~\bibnamefont {Das}},
  \bibinfo {author} {\bibfnamefont {A.}~\bibnamefont {Agarwal}},\ and\ \bibinfo
  {author} {\bibfnamefont {D.}~\bibnamefont {Culcer}},\ }\bibfield  {title}
  {\bibinfo {title} {Quantum kinetic theory of nonlinear optical currents:
  Finite fermi surface and fermi sea contributions},\ }\href@noop {} {\bibfield
   {journal} {\bibinfo  {journal} {Phys. Rev. B}\ }\textbf {\bibinfo {volume}
  {107}},\ \bibinfo {pages} {165131} (\bibinfo {year} {2023})}\BibitemShut
  {NoStop}%
\bibitem [{\citenamefont {Marzari}\ \emph {et~al.}(2012)\citenamefont
  {Marzari}, \citenamefont {Mostofi}, \citenamefont {Yates}, \citenamefont
  {Souza},\ and\ \citenamefont {Vanderbilt}}]{marzari2012maximally}%
  \BibitemOpen
  \bibfield  {author} {\bibinfo {author} {\bibfnamefont {N.}~\bibnamefont
  {Marzari}}, \bibinfo {author} {\bibfnamefont {A.~A.}\ \bibnamefont
  {Mostofi}}, \bibinfo {author} {\bibfnamefont {J.~R.}\ \bibnamefont {Yates}},
  \bibinfo {author} {\bibfnamefont {I.}~\bibnamefont {Souza}},\ and\ \bibinfo
  {author} {\bibfnamefont {D.}~\bibnamefont {Vanderbilt}},\ }\bibfield  {title}
  {\bibinfo {title} {Maximally localized wannier functions: Theory and
  applications},\ }\href@noop {} {\bibfield  {journal} {\bibinfo  {journal}
  {Rev. Mod. Phys.}\ }\textbf {\bibinfo {volume} {84}},\ \bibinfo {pages}
  {1419} (\bibinfo {year} {2012})}\BibitemShut {NoStop}%
\bibitem [{\citenamefont {Wang}\ \emph {et~al.}(2006)\citenamefont {Wang},
  \citenamefont {Yates}, \citenamefont {Souza},\ and\ \citenamefont
  {Vanderbilt}}]{wang2006ab}%
  \BibitemOpen
  \bibfield  {author} {\bibinfo {author} {\bibfnamefont {X.}~\bibnamefont
  {Wang}}, \bibinfo {author} {\bibfnamefont {J.~R.}\ \bibnamefont {Yates}},
  \bibinfo {author} {\bibfnamefont {I.}~\bibnamefont {Souza}},\ and\ \bibinfo
  {author} {\bibfnamefont {D.}~\bibnamefont {Vanderbilt}},\ }\bibfield  {title}
  {\bibinfo {title} {Ab initio calculation of the anomalous hall conductivity
  by wannier interpolation},\ }\href@noop {} {\bibfield  {journal} {\bibinfo
  {journal} {Phys. Rev. B}\ }\textbf {\bibinfo {volume} {74}},\ \bibinfo
  {pages} {195118} (\bibinfo {year} {2006})}\BibitemShut {NoStop}%
\bibitem [{\citenamefont {Xu}\ \emph {et~al.}(2024)\citenamefont {Xu},
  \citenamefont {Li}, \citenamefont {Huynh}, \citenamefont {Fadel},
  \citenamefont {Huang}, \citenamefont {Sundararaman}, \citenamefont
  {Vardeny},\ and\ \citenamefont {Ping}}]{xu2024spin}%
  \BibitemOpen
  \bibfield  {author} {\bibinfo {author} {\bibfnamefont {J.}~\bibnamefont
  {Xu}}, \bibinfo {author} {\bibfnamefont {K.}~\bibnamefont {Li}}, \bibinfo
  {author} {\bibfnamefont {U.~N.}\ \bibnamefont {Huynh}}, \bibinfo {author}
  {\bibfnamefont {M.}~\bibnamefont {Fadel}}, \bibinfo {author} {\bibfnamefont
  {J.}~\bibnamefont {Huang}}, \bibinfo {author} {\bibfnamefont
  {R.}~\bibnamefont {Sundararaman}}, \bibinfo {author} {\bibfnamefont
  {V.}~\bibnamefont {Vardeny}},\ and\ \bibinfo {author} {\bibfnamefont
  {Y.}~\bibnamefont {Ping}},\ }\bibfield  {title} {\bibinfo {title} {How spin
  relaxes and dephases in bulk halide perovskites},\ }\href@noop {} {\bibfield
  {journal} {\bibinfo  {journal} {Nat. Commun.}\ }\textbf {\bibinfo {volume}
  {15}},\ \bibinfo {pages} {188} (\bibinfo {year} {2024})}\BibitemShut
  {NoStop}%
\bibitem [{\citenamefont {Xu}\ \emph {et~al.}(2020)\citenamefont {Xu},
  \citenamefont {Habib}, \citenamefont {Kumar}, \citenamefont {Wu},
  \citenamefont {Sundararaman},\ and\ \citenamefont {Ping}}]{xu2020spin}%
  \BibitemOpen
  \bibfield  {author} {\bibinfo {author} {\bibfnamefont {J.}~\bibnamefont
  {Xu}}, \bibinfo {author} {\bibfnamefont {A.}~\bibnamefont {Habib}}, \bibinfo
  {author} {\bibfnamefont {S.}~\bibnamefont {Kumar}}, \bibinfo {author}
  {\bibfnamefont {F.}~\bibnamefont {Wu}}, \bibinfo {author} {\bibfnamefont
  {R.}~\bibnamefont {Sundararaman}},\ and\ \bibinfo {author} {\bibfnamefont
  {Y.}~\bibnamefont {Ping}},\ }\bibfield  {title} {\bibinfo {title}
  {{Spin-Phonon Relaxation from a Universal Ab Initio Density-Matrix
  Approach}},\ }\href@noop {} {\bibfield  {journal} {\bibinfo  {journal} {Nat.
  Commun.}\ }\textbf {\bibinfo {volume} {11}},\ \bibinfo {pages} {2780}
  (\bibinfo {year} {2020})}\BibitemShut {NoStop}%
\bibitem [{\citenamefont {Xu}\ \emph {et~al.}(2021{\natexlab{c}})\citenamefont
  {Xu}, \citenamefont {Takenaka}, \citenamefont {Habib}, \citenamefont
  {Sundararaman},\ and\ \citenamefont {Ping}}]{xu2021giant}%
  \BibitemOpen
  \bibfield  {author} {\bibinfo {author} {\bibfnamefont {J.}~\bibnamefont
  {Xu}}, \bibinfo {author} {\bibfnamefont {H.}~\bibnamefont {Takenaka}},
  \bibinfo {author} {\bibfnamefont {A.}~\bibnamefont {Habib}}, \bibinfo
  {author} {\bibfnamefont {R.}~\bibnamefont {Sundararaman}},\ and\ \bibinfo
  {author} {\bibfnamefont {Y.}~\bibnamefont {Ping}},\ }\bibfield  {title}
  {\bibinfo {title} {Giant spin lifetime anisotropy and spin-valley locking in
  silicene and germanene from first-principles density-matrix dynamics},\
  }\href@noop {} {\bibfield  {journal} {\bibinfo  {journal} {Nano Lett.}\
  }\textbf {\bibinfo {volume} {21}},\ \bibinfo {pages} {9594} (\bibinfo {year}
  {2021}{\natexlab{c}})}\BibitemShut {NoStop}%
\bibitem [{\citenamefont {Xu}\ and\ \citenamefont
  {Ping}(2023{\natexlab{b}})}]{xu2023substrate}%
  \BibitemOpen
  \bibfield  {author} {\bibinfo {author} {\bibfnamefont {J.}~\bibnamefont
  {Xu}}\ and\ \bibinfo {author} {\bibfnamefont {Y.}~\bibnamefont {Ping}},\
  }\bibfield  {title} {\bibinfo {title} {Substrate effects on spin relaxation
  in two-dimensional dirac materials with strong spin-orbit coupling},\
  }\href@noop {} {\bibfield  {journal} {\bibinfo  {journal} {npj Comput.
  Mater.}\ }\textbf {\bibinfo {volume} {9}},\ \bibinfo {pages} {47} (\bibinfo
  {year} {2023}{\natexlab{b}})}\BibitemShut {NoStop}%
\bibitem [{\citenamefont {Chen}\ \emph {et~al.}(2022)\citenamefont {Chen},
  \citenamefont {Ye}, \citenamefont {Zou}, \citenamefont {Gu}, \citenamefont
  {Xu},\ and\ \citenamefont {Duan}}]{chen2022basic}%
  \BibitemOpen
  \bibfield  {author} {\bibinfo {author} {\bibfnamefont {H.}~\bibnamefont
  {Chen}}, \bibinfo {author} {\bibfnamefont {M.}~\bibnamefont {Ye}}, \bibinfo
  {author} {\bibfnamefont {N.}~\bibnamefont {Zou}}, \bibinfo {author}
  {\bibfnamefont {B.-L.}\ \bibnamefont {Gu}}, \bibinfo {author} {\bibfnamefont
  {Y.}~\bibnamefont {Xu}},\ and\ \bibinfo {author} {\bibfnamefont
  {W.}~\bibnamefont {Duan}},\ }\bibfield  {title} {\bibinfo {title} {Basic
  formulation and first-principles implementation of nonlinear magneto-optical
  effects},\ }\href@noop {} {\bibfield  {journal} {\bibinfo  {journal} {Phys.
  Rev. B}\ }\textbf {\bibinfo {volume} {105}},\ \bibinfo {pages} {075123}
  (\bibinfo {year} {2022})}\BibitemShut {NoStop}%
\bibitem [{\citenamefont {Sipe}\ and\ \citenamefont
  {Shkrebtii}(2000)}]{sipe2000second}%
  \BibitemOpen
  \bibfield  {author} {\bibinfo {author} {\bibfnamefont {J.~E.}\ \bibnamefont
  {Sipe}}\ and\ \bibinfo {author} {\bibfnamefont {A.~I.}\ \bibnamefont
  {Shkrebtii}},\ }\bibfield  {title} {\bibinfo {title} {Second-order optical
  response in semiconductors},\ }\href@noop {} {\bibfield  {journal} {\bibinfo
  {journal} {Phys. Rev. B}\ }\textbf {\bibinfo {volume} {61}},\ \bibinfo
  {pages} {5337} (\bibinfo {year} {2000})}\BibitemShut {NoStop}%
\bibitem [{\citenamefont {Nastos}\ and\ \citenamefont
  {Sipe}(2006)}]{nastos2006optical}%
  \BibitemOpen
  \bibfield  {author} {\bibinfo {author} {\bibfnamefont {F.}~\bibnamefont
  {Nastos}}\ and\ \bibinfo {author} {\bibfnamefont {J.~E.}\ \bibnamefont
  {Sipe}},\ }\bibfield  {title} {\bibinfo {title} {Optical rectification and
  shift currents in gaas and gap response: Below and above the band gap},\
  }\href@noop {} {\bibfield  {journal} {\bibinfo  {journal} {Phys. Rev. B}\
  }\textbf {\bibinfo {volume} {74}},\ \bibinfo {pages} {035201} (\bibinfo
  {year} {2006})}\BibitemShut {NoStop}%
\bibitem [{\citenamefont {Nastos}\ \emph {et~al.}(2005)\citenamefont {Nastos},
  \citenamefont {Olejnik}, \citenamefont {Schwarz},\ and\ \citenamefont
  {Sipe}}]{nastos2005scissors}%
  \BibitemOpen
  \bibfield  {author} {\bibinfo {author} {\bibfnamefont {F.}~\bibnamefont
  {Nastos}}, \bibinfo {author} {\bibfnamefont {B.}~\bibnamefont {Olejnik}},
  \bibinfo {author} {\bibfnamefont {K.}~\bibnamefont {Schwarz}},\ and\ \bibinfo
  {author} {\bibfnamefont {J.}~\bibnamefont {Sipe}},\ }\bibfield  {title}
  {\bibinfo {title} {Scissors implementation within length-gauge formulations
  of the frequency-dependent nonlinear optical response of semiconductors},\
  }\href@noop {} {\bibfield  {journal} {\bibinfo  {journal} {Phys. Rev. B}\
  }\textbf {\bibinfo {volume} {72}},\ \bibinfo {pages} {045223} (\bibinfo
  {year} {2005})}\BibitemShut {NoStop}%
\bibitem [{\citenamefont {Perdew}\ \emph {et~al.}(1996)\citenamefont {Perdew},
  \citenamefont {Burke},\ and\ \citenamefont
  {Ernzerhof}}]{perdew1996generalized}%
  \BibitemOpen
  \bibfield  {author} {\bibinfo {author} {\bibfnamefont {J.~P.}\ \bibnamefont
  {Perdew}}, \bibinfo {author} {\bibfnamefont {K.}~\bibnamefont {Burke}},\ and\
  \bibinfo {author} {\bibfnamefont {M.}~\bibnamefont {Ernzerhof}},\ }\bibfield
  {title} {\bibinfo {title} {Generalized gradient approximation made simple},\
  }\href@noop {} {\bibfield  {journal} {\bibinfo  {journal} {Phys. Rev. Lett.}\
  }\textbf {\bibinfo {volume} {77}},\ \bibinfo {pages} {3865} (\bibinfo {year}
  {1996})}\BibitemShut {NoStop}%
\bibitem [{\citenamefont {Grimme}\ \emph {et~al.}(2010)\citenamefont {Grimme},
  \citenamefont {Antony}, \citenamefont {Ehrlich},\ and\ \citenamefont
  {Krieg}}]{grimme2010consistent}%
  \BibitemOpen
  \bibfield  {author} {\bibinfo {author} {\bibfnamefont {S.}~\bibnamefont
  {Grimme}}, \bibinfo {author} {\bibfnamefont {J.}~\bibnamefont {Antony}},
  \bibinfo {author} {\bibfnamefont {S.}~\bibnamefont {Ehrlich}},\ and\ \bibinfo
  {author} {\bibfnamefont {H.}~\bibnamefont {Krieg}},\ }\bibfield  {title}
  {\bibinfo {title} {A consistent and accurate ab initio parametrization of
  density functional dispersion correction (dft-d) for the 94 elements h-pu},\
  }\href@noop {} {\bibfield  {journal} {\bibinfo  {journal} {J. Chem. Phys.}\
  }\textbf {\bibinfo {volume} {132}} (\bibinfo {year} {2010})}\BibitemShut
  {NoStop}%
\bibitem [{\citenamefont {Grimme}(2006)}]{grimme2006semiempirical}%
  \BibitemOpen
  \bibfield  {author} {\bibinfo {author} {\bibfnamefont {S.}~\bibnamefont
  {Grimme}},\ }\bibfield  {title} {\bibinfo {title} {Semiempirical gga-type
  density functional constructed with a long-range dispersion correction},\
  }\href@noop {} {\bibfield  {journal} {\bibinfo  {journal} {J. Comput. Chem.}\
  }\textbf {\bibinfo {volume} {27}},\ \bibinfo {pages} {1787} (\bibinfo {year}
  {2006})}\BibitemShut {NoStop}%
\bibitem [{\citenamefont {Habib}\ \emph {et~al.}(2022)\citenamefont {Habib},
  \citenamefont {Xu}, \citenamefont {Ping},\ and\ \citenamefont
  {Sundararaman}}]{habib2022electric}%
  \BibitemOpen
  \bibfield  {author} {\bibinfo {author} {\bibfnamefont {A.}~\bibnamefont
  {Habib}}, \bibinfo {author} {\bibfnamefont {J.}~\bibnamefont {Xu}}, \bibinfo
  {author} {\bibfnamefont {Y.}~\bibnamefont {Ping}},\ and\ \bibinfo {author}
  {\bibfnamefont {R.}~\bibnamefont {Sundararaman}},\ }\bibfield  {title}
  {\bibinfo {title} {{Electric fields and substrates dramatically accelerate
  spin relaxation in graphene}},\ }\href@noop {} {\bibfield  {journal}
  {\bibinfo  {journal} {Phys. Rev. B}\ }\textbf {\bibinfo {volume} {105}},\
  \bibinfo {pages} {115122} (\bibinfo {year} {2022})}\BibitemShut {NoStop}%
\bibitem [{\citenamefont {Geller}\ and\ \citenamefont
  {Wood}(1954)}]{geller1954crystal}%
  \BibitemOpen
  \bibfield  {author} {\bibinfo {author} {\bibfnamefont {S.}~\bibnamefont
  {Geller}}\ and\ \bibinfo {author} {\bibfnamefont {E.}~\bibnamefont {Wood}},\
  }\bibfield  {title} {\bibinfo {title} {The crystal structure of rhodium
  silicide, rhsi},\ }\href@noop {} {\bibfield  {journal} {\bibinfo  {journal}
  {Acta Crystallogr.}\ }\textbf {\bibinfo {volume} {7}},\ \bibinfo {pages}
  {441} (\bibinfo {year} {1954})}\BibitemShut {NoStop}%
\bibitem [{\citenamefont {Hamann}(2013)}]{hamann2013optimized}%
  \BibitemOpen
  \bibfield  {author} {\bibinfo {author} {\bibfnamefont {D.~R.}\ \bibnamefont
  {Hamann}},\ }\bibfield  {title} {\bibinfo {title} {Optimized norm-conserving
  vanderbilt pseudopotentials},\ }\href@noop {} {\bibfield  {journal} {\bibinfo
   {journal} {Phys. Rev. B}\ }\textbf {\bibinfo {volume} {88}},\ \bibinfo
  {pages} {085117} (\bibinfo {year} {2013})}\BibitemShut {NoStop}%
\bibitem [{\citenamefont {van Setten}\ \emph {et~al.}(2018)\citenamefont {van
  Setten}, \citenamefont {Giantomassi}, \citenamefont {Bousquet}, \citenamefont
  {Verstraete}, \citenamefont {Hamann}, \citenamefont {Gonze},\ and\
  \citenamefont {Rignanese}}]{van2018pseudodojo}%
  \BibitemOpen
  \bibfield  {author} {\bibinfo {author} {\bibfnamefont {M.~J.}\ \bibnamefont
  {van Setten}}, \bibinfo {author} {\bibfnamefont {M.}~\bibnamefont
  {Giantomassi}}, \bibinfo {author} {\bibfnamefont {E.}~\bibnamefont
  {Bousquet}}, \bibinfo {author} {\bibfnamefont {M.~J.}\ \bibnamefont
  {Verstraete}}, \bibinfo {author} {\bibfnamefont {D.~R.}\ \bibnamefont
  {Hamann}}, \bibinfo {author} {\bibfnamefont {X.}~\bibnamefont {Gonze}},\ and\
  \bibinfo {author} {\bibfnamefont {G.-M.}\ \bibnamefont {Rignanese}},\
  }\bibfield  {title} {\bibinfo {title} {The pseudodojo: Training and grading a
  85 element optimized norm-conserving pseudopotential table},\ }\href@noop {}
  {\bibfield  {journal} {\bibinfo  {journal} {Comput. Phys. Commun.}\ }\textbf
  {\bibinfo {volume} {226}},\ \bibinfo {pages} {39} (\bibinfo {year}
  {2018})}\BibitemShut {NoStop}%
\bibitem [{\citenamefont {Sundararaman}\ and\ \citenamefont
  {Arias}(2013)}]{sundararaman2013regularization}%
  \BibitemOpen
  \bibfield  {author} {\bibinfo {author} {\bibfnamefont {R.}~\bibnamefont
  {Sundararaman}}\ and\ \bibinfo {author} {\bibfnamefont {T.}~\bibnamefont
  {Arias}},\ }\bibfield  {title} {\bibinfo {title} {Regularization of the
  coulomb singularity in exact exchange by wigner-seitz truncated interactions:
  Towards chemical accuracy in nontrivial systems},\ }\href@noop {} {\bibfield
  {journal} {\bibinfo  {journal} {Phys. Rev. B}\ }\textbf {\bibinfo {volume}
  {87}},\ \bibinfo {pages} {165122} (\bibinfo {year} {2013})}\BibitemShut
  {NoStop}%
\bibitem [{\citenamefont {Sundararaman}\ \emph {et~al.}(2017)\citenamefont
  {Sundararaman}, \citenamefont {Letchworth-Weaver}, \citenamefont {Schwarz},
  \citenamefont {Gunceler}, \citenamefont {Ozhabes},\ and\ \citenamefont
  {Arias}}]{sundararaman2017jdftx}%
  \BibitemOpen
  \bibfield  {author} {\bibinfo {author} {\bibfnamefont {R.}~\bibnamefont
  {Sundararaman}}, \bibinfo {author} {\bibfnamefont {K.}~\bibnamefont
  {Letchworth-Weaver}}, \bibinfo {author} {\bibfnamefont {K.~A.}\ \bibnamefont
  {Schwarz}}, \bibinfo {author} {\bibfnamefont {D.}~\bibnamefont {Gunceler}},
  \bibinfo {author} {\bibfnamefont {Y.}~\bibnamefont {Ozhabes}},\ and\ \bibinfo
  {author} {\bibfnamefont {T.}~\bibnamefont {Arias}},\ }\bibfield  {title}
  {\bibinfo {title} {Jdftx: Software for joint density-functional theory},\
  }\href@noop {} {\bibfield  {journal} {\bibinfo  {journal} {SoftwareX}\
  }\textbf {\bibinfo {volume} {6}},\ \bibinfo {pages} {278} (\bibinfo {year}
  {2017})}\BibitemShut {NoStop}%
\bibitem [{\citenamefont {Brown}\ \emph {et~al.}(2016)\citenamefont {Brown},
  \citenamefont {Sundararaman}, \citenamefont {Narang}, \citenamefont
  {Goddard~III},\ and\ \citenamefont {Atwater}}]{brown2016nonradiative}%
  \BibitemOpen
  \bibfield  {author} {\bibinfo {author} {\bibfnamefont {A.~M.}\ \bibnamefont
  {Brown}}, \bibinfo {author} {\bibfnamefont {R.}~\bibnamefont {Sundararaman}},
  \bibinfo {author} {\bibfnamefont {P.}~\bibnamefont {Narang}}, \bibinfo
  {author} {\bibfnamefont {W.~A.}\ \bibnamefont {Goddard~III}},\ and\ \bibinfo
  {author} {\bibfnamefont {H.~A.}\ \bibnamefont {Atwater}},\ }\bibfield
  {title} {\bibinfo {title} {Nonradiative plasmon decay and hot carrier
  dynamics: effects of phonons, surfaces, and geometry},\ }\href@noop {}
  {\bibfield  {journal} {\bibinfo  {journal} {ACS nano}\ }\textbf {\bibinfo
  {volume} {10}},\ \bibinfo {pages} {957} (\bibinfo {year} {2016})}\BibitemShut
  {NoStop}%
\bibitem [{\citenamefont {Habib}\ \emph {et~al.}(2018)\citenamefont {Habib},
  \citenamefont {Florio},\ and\ \citenamefont {Sundararaman}}]{habib2018hot}%
  \BibitemOpen
  \bibfield  {author} {\bibinfo {author} {\bibfnamefont {A.}~\bibnamefont
  {Habib}}, \bibinfo {author} {\bibfnamefont {F.}~\bibnamefont {Florio}},\ and\
  \bibinfo {author} {\bibfnamefont {R.}~\bibnamefont {Sundararaman}},\
  }\bibfield  {title} {\bibinfo {title} {Hot carrier dynamics in plasmonic
  transition metal nitrides},\ }\href@noop {} {\bibfield  {journal} {\bibinfo
  {journal} {J. Opt.}\ }\textbf {\bibinfo {volume} {20}},\ \bibinfo {pages}
  {064001} (\bibinfo {year} {2018})}\BibitemShut {NoStop}%
\bibitem [{\citenamefont {Kumar}\ \emph {et~al.}(2022)\citenamefont {Kumar},
  \citenamefont {Multunas},\ and\ \citenamefont
  {Sundararaman}}]{kumar2022fermi}%
  \BibitemOpen
  \bibfield  {author} {\bibinfo {author} {\bibfnamefont {S.}~\bibnamefont
  {Kumar}}, \bibinfo {author} {\bibfnamefont {C.}~\bibnamefont {Multunas}},\
  and\ \bibinfo {author} {\bibfnamefont {R.}~\bibnamefont {Sundararaman}},\
  }\bibfield  {title} {\bibinfo {title} {Fermi surface anisotropy in plasmonic
  metals increases the potential for efficient hot carrier extraction},\
  }\href@noop {} {\bibfield  {journal} {\bibinfo  {journal} {Phys. Rev.
  Mater.}\ }\textbf {\bibinfo {volume} {6}},\ \bibinfo {pages} {125201}
  (\bibinfo {year} {2022})}\BibitemShut {NoStop}%
\bibitem [{\citenamefont {Mikhailov}(2007)}]{mikhailov2007non}%
  \BibitemOpen
  \bibfield  {author} {\bibinfo {author} {\bibfnamefont {S.~A.}\ \bibnamefont
  {Mikhailov}},\ }\bibfield  {title} {\bibinfo {title} {Non-linear
  electromagnetic response of graphene},\ }\href@noop {} {\bibfield  {journal}
  {\bibinfo  {journal} {Europhys. Lett.}\ }\textbf {\bibinfo {volume} {79}},\
  \bibinfo {pages} {27002} (\bibinfo {year} {2007})}\BibitemShut {NoStop}%
\bibitem [{\citenamefont {Higuchi}\ \emph {et~al.}(2017)\citenamefont
  {Higuchi}, \citenamefont {Heide}, \citenamefont {Ullmann}, \citenamefont
  {Weber},\ and\ \citenamefont {Hommelhoff}}]{higuchi2017light}%
  \BibitemOpen
  \bibfield  {author} {\bibinfo {author} {\bibfnamefont {T.}~\bibnamefont
  {Higuchi}}, \bibinfo {author} {\bibfnamefont {C.}~\bibnamefont {Heide}},
  \bibinfo {author} {\bibfnamefont {K.}~\bibnamefont {Ullmann}}, \bibinfo
  {author} {\bibfnamefont {H.~B.}\ \bibnamefont {Weber}},\ and\ \bibinfo
  {author} {\bibfnamefont {P.}~\bibnamefont {Hommelhoff}},\ }\bibfield  {title}
  {\bibinfo {title} {Light-field-driven currents in graphene},\ }\href@noop {}
  {\bibfield  {journal} {\bibinfo  {journal} {Nature}\ }\textbf {\bibinfo
  {volume} {550}},\ \bibinfo {pages} {224} (\bibinfo {year}
  {2017})}\BibitemShut {NoStop}%
\bibitem [{\citenamefont {Luo}\ \emph {et~al.}(2017)\citenamefont {Luo},
  \citenamefont {Xu}, \citenamefont {Zhu}, \citenamefont {Wu}, \citenamefont
  {McCormick}, \citenamefont {Zhan}, \citenamefont {Neupane},\ and\
  \citenamefont {Kawakami}}]{luo2017opto}%
  \BibitemOpen
  \bibfield  {author} {\bibinfo {author} {\bibfnamefont {Y.~K.}\ \bibnamefont
  {Luo}}, \bibinfo {author} {\bibfnamefont {J.}~\bibnamefont {Xu}}, \bibinfo
  {author} {\bibfnamefont {T.}~\bibnamefont {Zhu}}, \bibinfo {author}
  {\bibfnamefont {G.}~\bibnamefont {Wu}}, \bibinfo {author} {\bibfnamefont
  {E.~J.}\ \bibnamefont {McCormick}}, \bibinfo {author} {\bibfnamefont
  {W.}~\bibnamefont {Zhan}}, \bibinfo {author} {\bibfnamefont {M.~R.}\
  \bibnamefont {Neupane}},\ and\ \bibinfo {author} {\bibfnamefont {R.~K.}\
  \bibnamefont {Kawakami}},\ }\bibfield  {title} {\bibinfo {title}
  {Opto-valleytronic spin injection in monolayer mos2/few-layer graphene hybrid
  spin valves},\ }\href@noop {} {\bibfield  {journal} {\bibinfo  {journal}
  {Nano Lett.}\ }\textbf {\bibinfo {volume} {17}},\ \bibinfo {pages} {3877}
  (\bibinfo {year} {2017})}\BibitemShut {NoStop}%
\bibitem [{\citenamefont {Ventura}\ \emph {et~al.}(2020)\citenamefont
  {Ventura}, \citenamefont {Passos}, \citenamefont {Lopes},\ and\ \citenamefont
  {dos Santos}}]{ventura2020study}%
  \BibitemOpen
  \bibfield  {author} {\bibinfo {author} {\bibfnamefont {G.~B.}\ \bibnamefont
  {Ventura}}, \bibinfo {author} {\bibfnamefont {D.~J.}\ \bibnamefont {Passos}},
  \bibinfo {author} {\bibfnamefont {J.~M. V.~P.}\ \bibnamefont {Lopes}},\ and\
  \bibinfo {author} {\bibfnamefont {J.~M. B.~L.}\ \bibnamefont {dos Santos}},\
  }\bibfield  {title} {\bibinfo {title} {A study of the nonlinear optical
  response of the plain graphene and gapped graphene monolayers beyond the
  dirac approximation},\ }\href@noop {} {\bibfield  {journal} {\bibinfo
  {journal} {J. Phys.: Condens. Matter}\ }\textbf {\bibinfo {volume} {32}},\
  \bibinfo {pages} {185701} (\bibinfo {year} {2020})}\BibitemShut {NoStop}%
\bibitem [{\citenamefont {Iba{\~n}ez-Azpiroz}\ \emph
  {et~al.}(2022)\citenamefont {Iba{\~n}ez-Azpiroz}, \citenamefont {de~Juan},\
  and\ \citenamefont {Souza}}]{ibanez2022assessing}%
  \BibitemOpen
  \bibfield  {author} {\bibinfo {author} {\bibfnamefont {J.}~\bibnamefont
  {Iba{\~n}ez-Azpiroz}}, \bibinfo {author} {\bibfnamefont {F.}~\bibnamefont
  {de~Juan}},\ and\ \bibinfo {author} {\bibfnamefont {I.}~\bibnamefont
  {Souza}},\ }\bibfield  {title} {\bibinfo {title} {Assessing the role of
  interatomic position matrix elements in tight-binding calculations of optical
  properties},\ }\href@noop {} {\bibfield  {journal} {\bibinfo  {journal}
  {SciPost Phys.}\ }\textbf {\bibinfo {volume} {12}},\ \bibinfo {pages} {070}
  (\bibinfo {year} {2022})}\BibitemShut {NoStop}%
\bibitem [{\citenamefont {Buscema}\ \emph {et~al.}(2015)\citenamefont
  {Buscema}, \citenamefont {Island}, \citenamefont {Groenendijk}, \citenamefont
  {Blanter}, \citenamefont {Steele}, \citenamefont {van~der Zant},\ and\
  \citenamefont {Castellanos-Gomez}}]{buscema2015photocurrent}%
  \BibitemOpen
  \bibfield  {author} {\bibinfo {author} {\bibfnamefont {M.}~\bibnamefont
  {Buscema}}, \bibinfo {author} {\bibfnamefont {J.~O.}\ \bibnamefont {Island}},
  \bibinfo {author} {\bibfnamefont {D.~J.}\ \bibnamefont {Groenendijk}},
  \bibinfo {author} {\bibfnamefont {S.~I.}\ \bibnamefont {Blanter}}, \bibinfo
  {author} {\bibfnamefont {G.~A.}\ \bibnamefont {Steele}}, \bibinfo {author}
  {\bibfnamefont {H.~S.}\ \bibnamefont {van~der Zant}},\ and\ \bibinfo {author}
  {\bibfnamefont {A.}~\bibnamefont {Castellanos-Gomez}},\ }\bibfield  {title}
  {\bibinfo {title} {Photocurrent generation with two-dimensional van der waals
  semiconductors},\ }\href@noop {} {\bibfield  {journal} {\bibinfo  {journal}
  {Chem. Soc. Rev.}\ }\textbf {\bibinfo {volume} {44}},\ \bibinfo {pages}
  {3691} (\bibinfo {year} {2015})}\BibitemShut {NoStop}%
\bibitem [{\citenamefont {Eginligil}\ \emph {et~al.}(2015)\citenamefont
  {Eginligil}, \citenamefont {Cao}, \citenamefont {Wang}, \citenamefont {Shen},
  \citenamefont {Cong}, \citenamefont {Shang}, \citenamefont {Soci},\ and\
  \citenamefont {Yu}}]{eginligil2015dichroic}%
  \BibitemOpen
  \bibfield  {author} {\bibinfo {author} {\bibfnamefont {M.}~\bibnamefont
  {Eginligil}}, \bibinfo {author} {\bibfnamefont {B.}~\bibnamefont {Cao}},
  \bibinfo {author} {\bibfnamefont {Z.}~\bibnamefont {Wang}}, \bibinfo {author}
  {\bibfnamefont {X.}~\bibnamefont {Shen}}, \bibinfo {author} {\bibfnamefont
  {C.}~\bibnamefont {Cong}}, \bibinfo {author} {\bibfnamefont {J.}~\bibnamefont
  {Shang}}, \bibinfo {author} {\bibfnamefont {C.}~\bibnamefont {Soci}},\ and\
  \bibinfo {author} {\bibfnamefont {T.}~\bibnamefont {Yu}},\ }\bibfield
  {title} {\bibinfo {title} {Dichroic spin--valley photocurrent in monolayer
  molybdenum disulphide},\ }\href@noop {} {\bibfield  {journal} {\bibinfo
  {journal} {Nat. Commun.}\ }\textbf {\bibinfo {volume} {6}},\ \bibinfo {pages}
  {7636} (\bibinfo {year} {2015})}\BibitemShut {NoStop}%
\bibitem [{\citenamefont {Xie}\ and\ \citenamefont
  {Cui}(2016)}]{xie2016manipulating}%
  \BibitemOpen
  \bibfield  {author} {\bibinfo {author} {\bibfnamefont {L.}~\bibnamefont
  {Xie}}\ and\ \bibinfo {author} {\bibfnamefont {X.}~\bibnamefont {Cui}},\
  }\bibfield  {title} {\bibinfo {title} {Manipulating spin-polarized
  photocurrents in 2d transition metal dichalcogenides},\ }\href@noop {}
  {\bibfield  {journal} {\bibinfo  {journal} {Proc. Natl. Acad. Sci. USA}\
  }\textbf {\bibinfo {volume} {113}},\ \bibinfo {pages} {3746} (\bibinfo {year}
  {2016})}\BibitemShut {NoStop}%
\bibitem [{\citenamefont {Barraza-Lopez}\ \emph {et~al.}(2021)\citenamefont
  {Barraza-Lopez}, \citenamefont {Fregoso}, \citenamefont {Villanova},
  \citenamefont {Parkin},\ and\ \citenamefont {Chang}}]{barraza2021colloquium}%
  \BibitemOpen
  \bibfield  {author} {\bibinfo {author} {\bibfnamefont {S.}~\bibnamefont
  {Barraza-Lopez}}, \bibinfo {author} {\bibfnamefont {B.~M.}\ \bibnamefont
  {Fregoso}}, \bibinfo {author} {\bibfnamefont {J.~W.}\ \bibnamefont
  {Villanova}}, \bibinfo {author} {\bibfnamefont {S.~S.~P.}\ \bibnamefont
  {Parkin}},\ and\ \bibinfo {author} {\bibfnamefont {K.}~\bibnamefont
  {Chang}},\ }\bibfield  {title} {\bibinfo {title} {Colloquium: Physical
  properties of group-iv monochalcogenide monolayers},\ }\href@noop {}
  {\bibfield  {journal} {\bibinfo  {journal} {Rev. Mod. Phys.}\ }\textbf
  {\bibinfo {volume} {93}},\ \bibinfo {pages} {011001} (\bibinfo {year}
  {2021})}\BibitemShut {NoStop}%
\bibitem [{\citenamefont {Mu}\ \emph {et~al.}(2021)\citenamefont {Mu},
  \citenamefont {Pan},\ and\ \citenamefont {Zhou}}]{mu2021pure}%
  \BibitemOpen
  \bibfield  {author} {\bibinfo {author} {\bibfnamefont {X.}~\bibnamefont
  {Mu}}, \bibinfo {author} {\bibfnamefont {Y.}~\bibnamefont {Pan}},\ and\
  \bibinfo {author} {\bibfnamefont {J.}~\bibnamefont {Zhou}},\ }\bibfield
  {title} {\bibinfo {title} {Pure bulk orbital and spin photocurrent in
  two-dimensional ferroelectric materials},\ }\href@noop {} {\bibfield
  {journal} {\bibinfo  {journal} {npj Comput. Mater.}\ }\textbf {\bibinfo
  {volume} {7}},\ \bibinfo {pages} {61} (\bibinfo {year} {2021})}\BibitemShut
  {NoStop}%
\bibitem [{\citenamefont {Panday}\ \emph {et~al.}(2019)\citenamefont {Panday},
  \citenamefont {Barraza-Lopez}, \citenamefont {Rangel},\ and\ \citenamefont
  {Fregoso}}]{panday2019injection}%
  \BibitemOpen
  \bibfield  {author} {\bibinfo {author} {\bibfnamefont {S.~R.}\ \bibnamefont
  {Panday}}, \bibinfo {author} {\bibfnamefont {S.}~\bibnamefont
  {Barraza-Lopez}}, \bibinfo {author} {\bibfnamefont {T.}~\bibnamefont
  {Rangel}},\ and\ \bibinfo {author} {\bibfnamefont {B.~M.}\ \bibnamefont
  {Fregoso}},\ }\bibfield  {title} {\bibinfo {title} {Injection current in
  ferroelectric group-iv monochalcogenide monolayers},\ }\href@noop {}
  {\bibfield  {journal} {\bibinfo  {journal} {Phys. Rev. B}\ }\textbf {\bibinfo
  {volume} {100}},\ \bibinfo {pages} {195305} (\bibinfo {year}
  {2019})}\BibitemShut {NoStop}%
\bibitem [{\citenamefont {Lihm}\ and\ \citenamefont
  {Park}(2022)}]{lihm2022comprehensive}%
  \BibitemOpen
  \bibfield  {author} {\bibinfo {author} {\bibfnamefont {J.-M.}\ \bibnamefont
  {Lihm}}\ and\ \bibinfo {author} {\bibfnamefont {C.-H.}\ \bibnamefont
  {Park}},\ }\bibfield  {title} {\bibinfo {title} {Comprehensive theory of
  second-order spin photocurrents},\ }\href@noop {} {\bibfield  {journal}
  {\bibinfo  {journal} {Phys. Rev. B}\ }\textbf {\bibinfo {volume} {105}},\
  \bibinfo {pages} {045201} (\bibinfo {year} {2022})}\BibitemShut {NoStop}%
\bibitem [{\citenamefont {Fregoso}\ \emph {et~al.}(2018)\citenamefont
  {Fregoso}, \citenamefont {Muniz},\ and\ \citenamefont
  {Sipe}}]{fregoso2018jerk}%
  \BibitemOpen
  \bibfield  {author} {\bibinfo {author} {\bibfnamefont {B.~M.}\ \bibnamefont
  {Fregoso}}, \bibinfo {author} {\bibfnamefont {R.~A.}\ \bibnamefont {Muniz}},\
  and\ \bibinfo {author} {\bibfnamefont {J.~E.}\ \bibnamefont {Sipe}},\
  }\bibfield  {title} {\bibinfo {title} {Jerk current: A novel bulk
  photovoltaic effect},\ }\href@noop {} {\bibfield  {journal} {\bibinfo
  {journal} {Phys. Rev. Lett.}\ }\textbf {\bibinfo {volume} {121}},\ \bibinfo
  {pages} {176604} (\bibinfo {year} {2018})}\BibitemShut {NoStop}%
\end{thebibliography}

%

\end{document}